\documentclass[a4paper,12pt]{paper}
\usepackage[authoryear]{natbib}
\usepackage[T1]{fontenc}
\usepackage{amsmath}
\usepackage{amssymb}
\usepackage[dvips]{graphicx}
\usepackage{listings}
\usepackage{epic}
\usepackage{eepic}
\usepackage{verbatim}
\usepackage{wrapfig}
\usepackage{subfigure}
\usepackage{float}
\usepackage[obeyspaces]{url}

\newcommand{\LCfilt}[1]
 {\put(#1){\begin{picture}(33,18)
 \put(0,0){\line(1,0){33}}
 \put(30,0){\line(0,1){6}}
 \put(27,6){\line(1,0){6}}
 \put(27,9){\line(1,0){6}}
 \put(30,9){\line(0,1){6}}
 \put(0,15){\line(1,0){6} \multiput(3,0)(6,0){3}{\oval(6,6)[t]}}
 \put(24,15){\line(1,0){9}}
\end{picture}}}

\newcommand{\emwave}[1]
 {\put(#1){\begin{picture}(45,10)
 \spline(0,0)(5,-5)(10,0)(15,5)(20,0) 
 \put(20,0) {\vector(1,0){5}}
\end{picture}}}

\begin{document}

\author{Frank Borg \and Ismo Hakala \and Jukka M\"a\"att\"al\"a \thanks{Jyv\"askyl\"a University, Chydenius Institute, Finland. Emails: FB, \protect \url{borgbros@netti.fi}; IH, \protect \url{ismo.hakala@chydenius.fi}; JM, \protect \url{jukka.maattala@chydenius.fi}. }}

\title{Elements of Radio Waves}

\maketitle

\tableofcontents

\begin{abstract}
We present a summary of the basic properties of the radio wave generation, propagation and reception, with a special attention to the gigahertz bandwidth region which is of interest for wireless sensor networks.
\end{abstract}

\section{Introduction}

Over a period of several months we have made measurements with a set of transceivers with the purpose of investigating how the received power varies with the surrounding and placement of the devices. The RF-devices automatically measure a parameter called RSSI for Received Signal Strength Indicator, and thus provide a convenient means to track the power level of the signal. Since our measurements raised many issues about the behaviour of electromagnetic fields, it was decided to review some of the basic physics of electromagnetism in the style of a handbook chapter. The emphasis here is on the application of the Maxwell equations to concrete problems, not on the development of the theoretical structure (in terms of differential forms, gauge theory, etc). Of physics books on EM theory we may mention \cite{Jackson1975,Landau1975,Panofsky1962}, and the engineering style books \cite{Sihvola1996,Stratton1941,Harrington1961}. A good general physics reference including material on EM is \cite{Joos1934}. For applications to antenna theory see \cite{Lindell1995}. General reviews of EM with wireless networks in mind can be found in \cite{Ahlin2006,Bertoni2000}. Arnold Sommerfeld was a an eminent mathematical physicist who made, among other things, some significant contributions to the propagation of EM fields; these ''classical'' methods are described in \cite{Frank1935,Sommerfeld1947,Sommerfeld1949,Sommerfeld1959}. Of recent texts on antenna theory we may mention \cite{Balanis2005,Harish2007}. Balanis has also written a nice minireview \cite{Balanis1992}. For an interesting online collection of lecture notes and a selection of classic papers on EM see \cite{mcdonaldwww}.

\section{Maxwell equations}
\subsection{Quasi-stationary fields}
Electromagnetism stands for one of the four fundamental forces in physics. A static point like charge $q_1$ at the point $\mathbf{r}_1$ in an isotropic homogeneous medium exerts a force on an other charge $q_2$ at $\mathbf{r}_2$ given by (Coulomb interaction),

\begin{equation}
\label{EQ:Coul}
\mathbf{F}_{1 \rightarrow 2} = \frac{1}{4 \pi \epsilon \epsilon_0} \frac{q_1 q_2 (\mathbf{r}_2 - \mathbf{r}_1 )}{|\mathbf{r}_2 - \mathbf{r}_1|^3} .
\end{equation}

\begin{wrapfigure}{r}{50mm}
\begin{flushleft}
\unitlength = 1mm
\begin{picture}(50,50)(0,0)
\thicklines
\put(40,40){\circle*{2}}
\put(30,10){\vector(1,3){10}}
\put(30,10){\vector(-2,1){20}}
\put(40,40){\vector(-3,-2){30}}
\put(37,20){\makebox(6,6)[l]{$\mathbf{r}_1$}}
\put(15,32){\makebox(6,6)[l]{$\mathbf{r}_2 - \mathbf{r}_1$}}
\put(13,10){\makebox(6,6)[l]{$\mathbf{r}_2$}}
\put(40,40){\makebox(6,6)[l]{$q_1$}}
\put(1,20){\makebox(6,6)[l]{$\mathbf{E}(\mathbf{r}_2)$}}
\put(10,20){\vector(-3,-2){5}}
\end{picture}
\end{flushleft}
\label{FIGcoul1}
\end{wrapfigure}

The quantity $\epsilon$, relative permittivity, is a quantity characterizing the medium, while $\epsilon_0$ (the permittivity of the vacuum) is a universal constant. Eq.(\ref{EQ:Coul}) can be written as

\begin{equation}
\label{EQ:Coul2}
\mathbf{F}_{1 \rightarrow 2} = q_2 \mathbf{E}(\mathbf{r}_2),  
\end{equation}
  
where 

\begin{equation}
\label{EQ:Coul3}
\mathbf{E}(\mathbf{r}_2) =  \frac{1}{4 \pi \epsilon \epsilon_0} \frac{q_1 (\mathbf{r}_2 - \mathbf{r}_1 )}{|\mathbf{r}_2 - \mathbf{r}_1|^3}
\end{equation}

is defined as the \textit{electric field} at the point $\mathbf{r}_2$ generated by the charge $q_1$ located at $\mathbf{r}_1$.

The electric field can also be expressed in terms of a \textit{potential function} $\phi$, 

\[
\mathbf{E} = - \nabla \phi,
\] 

or vice versa,

\begin{equation}
\label{EQ:pot}
\phi(\mathbf{r}) = - \int_{\mathbf{r}_0}^{\mathbf{r}} \mathbf{E} \cdot d\mathbf{r}, 
\end{equation}

where the line integral is along a path connecting the reference point $\mathbf{r}_0$ and the point $\mathbf{r}$. In the case of Eq.(\ref{EQ:Coul3}) we have

\[
\phi(\mathbf{r}) =  \frac{1}{4 \pi \epsilon \epsilon_0} \frac{q_1} {|\mathbf{r}_2 - \mathbf{r}_1|}.
\]

When charges are in a relative motion with respect to each other then we have to include in Eq.(\ref{EQ:Coul2}) a term depending on the velocity,

\begin{equation}
\label{EQ:Lor}
\mathbf{F} = q \mathbf{E} + q \mathbf{v \times B}, 
\end{equation}

where $\mathbf{B}$ defines the \textit{magnetic field strength}. Thus, Eq.(\ref{EQ:Lor}) gives the force (''Lorentz force'') acting on a charge $q$ moving with velocity $\mathbf{v}$ in an EM field characterized by $\mathbf{E}$ and $\mathbf{B}$. From this follows the familiar fact, that a straight conductor of length $l$, with the current $I$ in a magnetic field $\mathbf{B}$, will sense a force $B I l$ in a direction perpendicular to $\mathbf{B}$ and the conductor. 

Conversely, a charge $q_1$ at $\mathbf{r}_1$ moving with the velocity $\mathbf{v}_1$ generates a magnetic field strength at the point $\mathbf{r}_2$ given by (in an isotropic homogeneous medium) 

\begin{equation}
\label{EQ:BSv}
\mathbf{B}(\mathbf{r}_2) = \frac{\mu_0 \mu}{4 \pi} \frac{q_1 \mathbf{v_1} \times (\mathbf{r}_2 - \mathbf{r}_1)}{|\mathbf{r}_2 - \mathbf{r}_1|^3}.
\end{equation}

The field strength is thus affected by the \textit{magnetic permeability} $\mu$ characterizing the medium. For most non-metallics $\mu \approx 1$, while $\mu_0$ is universal constant. From the above it follows that two charges moving with velocities $\mathbf{v}_1$ and $\mathbf{v}_2$ will interact via a magnetic force given by\footnote{An interesting observation is that the force $\mathbf{F}_{1 \rightarrow 2}$ which the particle 1 exerts on the particle 2 is no longer, in general, the opposite of the force that the particle 2 exerts on 1, as would be demanded by the principle of \textit{actio est reactio} of Newtonian mechanics; that is, we no longer have $\mathbf{F}_{1 \rightarrow 2} + \mathbf{F}_{2 \rightarrow 1} = 0$. From $\mathbf{F}_{1 \rightarrow 2} + \mathbf{F}_{2 \rightarrow 1} \neq 0$ one might conclude that the closed system of charge 1 + charge 2 may start to move without an external cause. However, the momentum of the total system is conserved if we also take into account the momentum contribution of the \textit{electromagnetic field}. Calculating the magnetic forces between two current \textit{loops}, on the other hand, we get $\mathbf{F}_{1 \rightarrow 2} + \mathbf{F}_{2 \rightarrow 1} = 0$, in accordance with Newton's third law.}

\begin{align}
\label{EQ:BSvv}
&\mathbf{F}_{1 \rightarrow 2} = q_2 \mathbf{v}_2 \times \mathbf{B}(\mathbf{r}_2) =
\frac{ q_1 q_2 \mu_0 \mu}{4 \pi} \cdot \frac{ \mathbf{v_2} \times (\mathbf{v_1} \times (\mathbf{r}_2 - \mathbf{r}_1))}{|\mathbf{r}_2 - \mathbf{r}_1|^3} =
\\
\nonumber
&\frac{ q_1 q_2 \mu_0 \mu}{4 \pi} \cdot \frac{ (\mathbf{v_2} \cdot (\mathbf{r}_2 - \mathbf{r}_1)) \mathbf{v_1} - (\mathbf{v_1} \cdot \mathbf{v_2})(\mathbf{r}_2 - \mathbf{r}_1)}{|\mathbf{r}_2 - \mathbf{r}_1|^3},
\end{align}

where we have used the rule that 

\[
\mathbf{A} \times (\mathbf{B} \times \mathbf{C}) = (\mathbf{A} \cdot \mathbf{C}) \mathbf{B} - (\mathbf{A} \cdot \mathbf{B}) \mathbf{C}. 
\]

A current $I$ in a conductor consists of many moving charges, each one contributing to the total magnetic field strength according to (\ref{EQ:BSvv}). 
If we consider a small segment $d\mathbf{l}$ of the conductor, then the sum of all terms $q \mathbf{v}$ over the charges in this segment is equal to $I \, d\mathbf{l}$. Therefore the magnetic field strength generated by this segment is given by (''Biot-Savart law'')

\begin{equation}
\label{EQ:BS}
d\mathbf{B}(\mathbf{r}_2) = \frac{\mu_0 \mu}{4 \pi} \frac{I \, d\mathbf{l} \times (\mathbf{r}_2 - \mathbf{r}_1)}{|\mathbf{r}_2 - \mathbf{r}_1|^3}  ;
\end{equation}

\begin{wrapfigure}{r}{50mm}
\begin{flushleft}
\unitlength = 1mm
\begin{picture}(50,50)(0,0)
\thicklines
\put(10,30){\circle*{2}}
\put(40,40){\oval(2,2)[r]}
\put(37,5){\vector(-1,3){10}}
\put(37,5){\vector(-4,1){18}}
\put(27,35){\vector(-1,-3){8}}
\multiput(10,29)(0,2){2}{\drawline(0,0)(30,10)}
\multiput(22,34)(6,2){2}{\oval(2,2)[r]}
\put(10,20){\makebox(6,6)[l]{$\mathbf{r}_2 - \mathbf{r}_1$}}
\put(27,8){\makebox(6,6)[l]{$\mathbf{r}_2$}}
\put(35,18){\makebox(6,6)[l]{$\mathbf{r}_1$}}
\put(21,39){\makebox(6,6)[l]{$I \; d\mathbf{l}$}}
\put(21,37) {\vector(3,1){7}}
\put(19,10){\vector(-1,3){3}}
\put(5,10){\makebox(6,6)[l]{$d\mathbf{B}(\mathbf{r}_2)$}}
\put(19,10){\circle*{2}}
\end{picture}
\end{flushleft}
\label{FIGbisa}
\end{wrapfigure}

that is, moving charges forming a current $I$ in a small conducting element $d\mathbf{l}$ at $\mathbf{r}_1$ generates a magnetic field $d\mathbf{B}(\mathbf{r}_2)$ given by Eq.(\ref{EQ:BS}) at the point $\mathbf{r}_2$. In order to obtain the effect of the whole conductor one has to sum (integrate) (\ref{EQ:BS}) over all the segments.  

The permittivity $\epsilon$ and permeability $\mu$ take into account how the medium affects the electromagnetic field. The charges in the medium are affected by the field and may become displaced, which leads to a modification of the field (''backreaction''). This explains such phenomena as polarization (charge displacements) and magnetization of a medium. From Eq.(\ref{EQ:Coul3}) and Eq.(\ref{EQ:BS}) we infer that by defining

\begin{align}
\label{EQ:DH}
\mathbf{D} = \epsilon \epsilon_0 \mathbf{E},
\\
\nonumber
\mathbf{H} = \frac{1}{\mu \mu_0} \mathbf{B},
\end{align} 

we obtain the quantities $\mathbf{D}$ (electric displacement) and $\mathbf{H}$ (magnetic field), which are apparently independent of the material factors ($\epsilon$, $\mu$). From the definitions one can show that the integral of 
$\mathbf{D}$ over a boundary $\partial V$ enclosing a volume $V$ is equal to the total charge $Q$ contained in $V$, while integrating $\mathbf{H}$ along a loop (boundary) $\partial S$ enclosing a surface $S$ one obtains the total current $I$ flowing through that surface,

\begin{align}
\label{EQ:QJ}
\oint_{\partial V} \mathbf{D} \cdot d\mathsf{S} = Q,
\\
\nonumber
\oint_{\partial S} \mathbf{H} \cdot d\mathsf{s} = I.
\end{align}

\subsection{General case -- time dependent fields}
   
If we integrate $\mathbf{B}$ over a surface $S$ we obtain a quantity

\[
\Phi = \int_S \mathbf{B} \cdot d\mathsf{S}
\]

termed the magnetic flux through the surface $S$. It is experimentally observed that when the flux enclosed by a conducting loop changes, this induces a potential difference along the loop and causes a current to flow. More precisely (Faraday's law if induction), 

\[
\frac{\partial \Phi}{\partial t} = \Delta \phi \quad \Rightarrow \quad
\int_S \frac{\partial \mathbf{B}}{\partial t} \cdot d\mathsf{S} = -
\oint_{\partial S} \mathbf{E} \cdot d\mathsf{s}.
\] 

Using the mathematical identity (Stokes' theorem) 

\[
\oint_{\partial S} \mathbf{A} \cdot d\mathsf{s} = \oint_S \nabla \times \mathbf{A} \cdot d\mathsf{S},
\]

we obtain the induction law on the form,

\[
\nabla \times \mathbf{E} = - \frac{\partial \mathbf{B}}{\partial t}.
\]

This links the time change of the magnetic field strength to the spatial variation of the electric field. The final crucial step is to find an equation for the time change of the electric field. From the second equation in (\ref{EQ:QJ}) one may infer that

\[
\nabla \times \mathbf{H} = \mathbf{J} \quad \mbox{(for static fields)}
 \] 
 
but Maxwell realized that the right hand side of this equation must be complemented with the term $\partial \mathbf{D}/\partial t$ (Maxwell's displacement term), which contains the link to the time change of the electric field. This addition is needed for maintaining charge conservation (see Eq.(\ref{EQ:Cont})). Also, without this term no electromagnetic waves would exist in the theory.

Thus, J C Maxwell was able in 1864 to synthesize the known properties of electromagnetism in his now famous equations which give, as far as we know, a complete description of the electromagnetic phenomena in the classical regime,

\begin{equation}
\label{EQ:Max}
\boxed{
\begin{aligned}
&\nabla \cdot \mathbf{D} = \varrho 
\\
&\nabla \times \mathbf{H} = \mathbf{J} + \frac{\partial \mathbf{D}}{\partial t}
\\
&\nabla \times \mathbf{E} = - \frac{\partial \mathbf{B}}{\partial t}
\\
&\nabla \cdot \mathbf{B} = 0.
\end{aligned}
}
\quad \mbox{Maxwell equations}
\end{equation}

Here $\varrho$ denotes the charge density and $\mathbf{J}$ the current density. We note that there is an asymmetry between electric and magnetic fields in the equations in that there appear no magnetic charges (no magnetic monopoles) and no magnetic currents. No magnetic charge has been ever discovered, whence all magnetic fields are assumed to be generated by moving electric charges (electric currents) as described above.  

Using the rule

\[
\nabla \times (\nabla \times \mathbf{A}) = - \nabla^2 \mathbf{A} + \nabla (\nabla \cdot \mathbf{A}), 
\]

and the relations (\ref{EQ:DH}) one can show that Maxwell equations give the equations,

\begin{align}
\label{EQ:Wav}
\nabla^2 \mathbf{E} - \frac{1}{c^2} \frac{\partial^2 \mathbf{E}}{\partial t^2}
&= \mu \mu_0 
\frac{\partial \mathbf{J}}{\partial t} + \frac{1}{\epsilon \epsilon_0}\nabla \varrho,
\\
\nabla^2 \mathbf{H} - \frac{1}{c^2} \frac{\partial^2 \mathbf{H}}{\partial t^2}
&= -\nabla \times \mathbf{J},
\end{align}

where we have set (identified with the velocity of light)

\begin{equation}
\label{EQ:c}
c = \frac{1}{\sqrt{\epsilon \epsilon_0 \mu \mu_0}}.
\end{equation}

Especially in the case of the empty space ($\mathbf{J}$ = 0, $\varrho$ = 0) we obtain the wave equation

\begin{equation}
\label{EQ:Eeq}
\nabla^2 \mathbf{E} - \frac{1}{c^2} \frac{\partial^2 \mathbf{E}}{\partial t^2} = 0,
\end{equation}

whose plane wave solutions are of the form

\begin{equation}
\label{EQ:Epl}
\mathbf{E} = \mathbf{E}_0 \cos \left(\mathbf{k} \cdot \mathbf{r} \pm \omega t \right).
\end{equation}

(Here $\mathbf{E}_0$ is a constant vector.) The magnitude of the wave-vector $\mathbf{k}$ is $2 \pi/\lambda$, where $\lambda$ denotes the wave length, while $\omega$ (circular frequency) is related to the frequency $f$ by $\omega = 2 \pi f$. Inserting (\ref{EQ:Epl}) into (\ref{EQ:Eeq}) we infer that $|\mathbf{k}|c = \omega$, which is the same as $\lambda f = c$. Eq.(\ref{EQ:Epl}) describes an oscillating field whose frequency is $f$. The solution can be interpreted as a wave moving in the direction of the wave vector $\mp \mathbf{k}$ and with the velocity $c$ (light velocity in empty space). For empty space we have $\nabla \cdot \mathbf{E} = 0$ which implies, that for the plane wave (\ref{EQ:Epl}) we must have $\mathbf{k} \cdot \mathbf{E}_0 = 0$; that is, the electrical field oscillates in a direction orthogonal to the direction of propagation. From Maxwell equations we find that the corresponding plane wave solution for the magnetic field is then given by

\begin{align}
\label{EQ:Hpl}
\mathbf{H} &= \mathbf{H}_0 \cos \left(\mathbf{k} \cdot \mathbf{r} \pm \omega t \right),
\\
\nonumber
\mathbf{H}_0 &= \mp \frac{1}{\omega \mu \mu_0} \mathbf{k} \times \mathbf{E}_0 
= \mp \frac{1}{\eta} \, \frac{\mathbf{k}}{k} \times \mathbf{E}_0, 
\end{align}

where $\eta = \sqrt{\mu \mu_0 /\epsilon \epsilon_0}$ is called the \textit{wave impedance} ($\approx$ 377 $\Omega$ for vacuum). This means that the magnetic field $\mathbf{H}_0$ is orthogonal to both $\mathbf{k}$ and the electric field $\mathbf{E}_0$. When both the electric and magnetic fields are orthogonal to the wave-vector $\mathbf{k}$ the EM-wave is said to be \textit{transversal} and of the type TEM. The direction of the electric field $\mathbf{E}_0$ defines the \textit{polarization} of the wave. For instance, if $\mathbf{E}_0 = (E_x, 0, 0)$ then the wave is polarized in the $x$-direction.

\begin{figure*}[h]
\begin{center}
\unitlength = 1mm
\begin{picture}(90,40)(0,0)
\thicklines
\drawline(13,12)(85,30)
\put(85,30){\vector(4,1){0}}
\spline(13,12)(20,27)(26,16)(32,5)(37,18)(44,31)(50,21)(54,11)(59,24)(64,34)(69,26)(70,21)
\spline(13,12)(32,11)(26,16)(21,20)(37,18)(54,18)(50,21)(45,25)(59,24)(72,23)(69,26)(66,29)(77,28)
\put(19,24){\makebox(6,6)[l]{$\mathbf{E}$}}
\put(20,6){\makebox(6,6)[l]{$\mathbf{H}$}}
\end{picture}
\end{center}
\label{FIGehwave}
\end{figure*}

A charge moving with velocity $\mathbf{v}$ in an electromagnetic field feels a force $\mathbf{F}$ given by (\ref{EQ:Lor}). This involves a work per unit time (power $P$) defined as $P = \mathbf{F} \cdot \mathbf{v}$. Because of the identity $\mathbf{v} \cdot (\mathbf{v} \times \mathbf{B}) = 0$ it follows from (\ref{EQ:Lor}) that $P = q \mathbf{E} \cdot \mathbf{v}$. If we have a current density $\mathbf{J} = \varrho \mathbf{v}$ this result is generalized to

\begin{equation}
\label{EQ:Pow}
P = \int_V \mathbf{E} \cdot \mathbf{J} dV,
\end{equation}  

defining a power density $\mathcal{P} = \mathbf{E} \cdot \mathbf{J}$. Using the vector identity

\[
\nabla \cdot (\mathbf{E} \times \mathbf{H}) =
\mathbf{H} \cdot (\nabla \times \mathbf{E}) -
\mathbf{E} \cdot (\nabla \times \mathbf{H}),
\]

one can derive from Maxwell equations the following relation,

\begin{equation}
\label{EQ:EJ}
\mathbf{E} \cdot \mathbf{J} +
\frac{\partial}{\partial t} \left(
\frac{1}{2} \mu \mu_0 H^2 + \frac{1}{2} \epsilon \epsilon_0 E^2
\right) +
\nabla \cdot (\mathbf{E} \times \mathbf{H}) = 0.   
\end{equation}

This can be interpreted as an equation of energy conservation where

\[
\mathcal{E} = \frac{1}{2} \mu \mu_0 H^2 + \frac{1}{2} \epsilon \epsilon_0 E^2
\]

represents the energy per unit volume associated with the electromagnetic field, and

\begin{equation}
\label{EQ:S}
\mathbf{S} = \mathbf{E} \times \mathbf{H} \quad \mbox{(the ''Poynting'' vector)}
\end{equation}

represents the flow of energy carried away by the electromagnetic radiation \cite{Poynting1884}. That is, given an area element $d\mathsf{A}$ then $\mathbf{S} \cdot d\mathsf{A}$ represents the energy passing through $d\mathsf{A}$ per unit time due to the electromagnetic radiation. 

If we compute the vector $\mathbf{S}$ in case of the plane wave (\ref{EQ:Epl}), (\ref{EQ:Hpl}), we obtain,

\begin{equation}
\label{EQ:Spl}
\mathbf{S} = \mp \frac{1}{\eta} E_0^2 \, \frac{\mathbf{k}}{k} \cos (\mathbf{k} \cdot \mathbf{r} \pm \omega t)^2  =
\mp \eta H_0^2 \, \frac{\mathbf{k}}{k} \cos (\mathbf{k} \cdot \mathbf{r} \pm \omega t)^2.
\end{equation}

Thus we have the important result that the radiated power is proportional to the square of the electric and magnetic field amplitudes. Also, for a wave which varies as $\cos (\mathbf{k} \cdot \mathbf{r} - \omega t)$ the power is propagated in the direction of $\mathbf{k}$. If we calculate the time average of (\ref{EQ:Spl}) over a period $T = 2 \pi /\omega$ we obtain the factor

\[
\frac{1}{2} = \frac{1}{T} \int_0^T \cos (\mathbf{k} \cdot \mathbf{r} \pm \omega t)^2 dt.
\]

Thus, if the average radiation power for a plane wave is 100 mW/m$^2$, then we obtain the corresponding electrical field amplitude $E_0$ by setting 

\[
100 \; \mbox{mW/m}^2 = \frac{1}{377 \, \Omega} \, \frac{E_0^2}{2},
\]

which yields $E_0$ = $\sqrt{2 \cdot 377 \cdot 0.1}$ V/m $\approx$ 8.7 V/m (Volt per meter).

Since electromagnetic waves carry energy, they can also carry ''information'', which of course make them useful in technology. The motion of charges at one place (transmitter) will thus interact with charges at another place (the receiver). This interaction is described in terms of the electromagnetic (EM) fields. The transmitter generates EM-waves which are intercepted by the receiver. 

As seen from the second equation in (\ref{EQ:Wav}) the magnetic field is determined by the current $\mathbf{J}$; one can write a solution of the $\mathbf{H}$-equation as

\begin{equation}
\label{EQ:Hso}
\mathbf{H}(\mathbf{r}, t) = \frac{1}{4 \pi} \int \frac{\nabla_q \times \mathbf{J}(\mathbf{r}_q, \bar{t})}{|\mathbf{r}_q - \mathbf{r}|} d^3 \mathbf{r}_q = \nabla \times \left\{
\frac{1}{4 \pi} \int \frac{\mathbf{J}(\mathbf{r}_q, \bar{t})}{|\mathbf{r}_q - \mathbf{r}|} d^3 \mathbf{r}_q
\right\}.
\end{equation}

Here $\bar{t} = t - |\mathbf{r}_q - \mathbf{r}|/c$ is the \textit{retarded time} which takes into account that it takes time for the field contribution generated at the point $\mathbf{r}_q$ to reach the point $\mathbf{r}$. This form (\ref{EQ:Hso}) suggests introducing an auxillary quantity $\mathbf{A}$ called the \textit{vector potential} related to $\mathbf{E}$ and $\mathbf{B}$ by

\begin{align}
\label{EQ:Vpo}
\mathbf{B} &= \nabla \times \mathbf{A},
\\
\nonumber
\mathbf{E} &= -\nabla \phi - \frac{\partial \mathbf{A}}{\partial t}.
\end{align} 

The first equation in (\ref{EQ:Vpo}) implies that $\nabla \cdot \mathbf{B} = 0$ holds identically; also it implies in conjunction with Maxwell equations that $\nabla \times \left( \mathbf{E} + \frac{\partial \mathbf{A}}{\partial t} \right) = 0$ which suggests the second equation in (\ref{EQ:Vpo}). There is some freedom in choosing $\mathbf{A}$ and $\phi$; indeed, if we use the ''tilded'' versions 

\begin{equation}
\label{EQ:Gau}
\left.
\begin{aligned}
\tilde{{\mathbf{A}}} &= \mathbf{A} + \nabla \xi
\\
\tilde{\phi} &= \phi - \frac{\partial \xi}{\partial t}
\end{aligned}
\right\} \quad \mbox{''Gauge transformation''} 
\end{equation}

(for some function $\xi$) then the fields $\mathbf{E}$ and $\mathbf{B}$ remain unchanged in (\ref{EQ:Vpo}). If we use as the supplementary condition for $\mathbf{A}$ that (fixing the gauge to the so called Lorenz gauge\footnote{This was indeed introduced by the Danish physicist Ludvig Lorenz (1829-1891) and not by the more famous Dutch physicist Henrik Lorentz (1853-1928) to whom it is often attributed.})

\begin{equation}
\label{EQ:Con}
\nabla \cdot \mathbf{A} + \frac{1}{c^2} \frac{\partial \phi }{\partial t} = 0, 
\end{equation}

we get from Maxwell equations 

\begin{equation}
\label{EQ:Aeq}
\nabla^2 \mathbf{A} - \frac{1}{c^2} \frac{\partial^2 \mathbf{A}}{\partial t^2} = -\mu \mu_0 \mathbf{J}.
\end{equation}

This has a solution of the form (consistent with (\ref{EQ:Hso}))

\begin{equation}
\label{EQ:Aso}
\mathbf{A}(\mathbf{r},t) =
\frac{\mu \mu_0}{4 \pi} \int \frac{\mathbf{J}\left(\mathbf{r}_q,\bar{t}\right)}{|\mathbf{r}_q - \mathbf{r}|} d^3 \mathbf{r}_q.
\end{equation}

In principle, if we know the current distribution $\mathbf{J}\left(\mathbf{r}_q,\bar{t}\right)$ in the transmitting antenna we can calculate the radiated field using (\ref{EQ:Aso}) and (\ref{EQ:Vpo}). The problem thus reduces to determining the current $\mathbf{J}\left(\mathbf{r}_q,\bar{t}\right)$ which often is a very hard problem to solve analytically. However, in many cases simple approximations will do quite well. 

Finally we observe two important consequences of the above equations. If we differentiate (\ref{EQ:Con}) with respect to the time and use the second equation in (\ref{EQ:Vpo}) together with the first equation in (\ref{EQ:Max}), then we obtain the relativistic form of \textit{Poisson equation}

\begin{equation}
\label{EQ:Pos}
\nabla^2 \phi - \frac{1}{c^2} \frac{\partial^2 \phi}{\partial t^2} = - \frac{\varrho}{\epsilon \epsilon_0}.
\end{equation}

This equation determines the potential $\phi$ when the charge distribution $\varrho$ is known. A second observation is that the identity $\nabla \cdot (\nabla \times \mathbf{H}) = 0$ applied to the second of the Maxwell equations (\ref{EQ:Max}) leads to the \textit{continuity equation},

\begin{equation}
\label{EQ:Cont}
\nabla \cdot \mathbf{J} + \frac{\partial \varrho}{\partial t} = 0,
\end{equation}

which expresses the law of the conservation of electric charge. This is of importance when e.g.\ determining the current/charge distributions in antennas.  

\section{Dielectrics and conductors}
\subsection{Electric susceptibility}

Common electromagnetic phenomena are due to the interaction of electron and protons, the basic elementary particles of ordinary matter. Electromagnetic forces hold atoms and molecules together. Since matter is thus made up of electrons and protons (and neutrons) we expect matter to affect electromagnetic fields and vice versa. Although atoms and molecules may be electrically neutral (contain an equal number of electrons and protons), the charges may be shifted so that one region is dominantly negative while another region is dominantly positive. The matter is then said to be polarized. The polarization can be understood in terms of electric dipoles. Suppose we have a positive charge $q$ at the point $\mathbf{r}_1 + \mathbf{l}$, and a negative charge $-q$ at the point $\mathbf{r}_1$, then the potential of the system measured at the point $\mathbf{r}_2$ becomes

\[
\phi(\mathbf{r}_2) = \frac{1}{4 \pi \epsilon_0} \frac{q}{|\mathbf{r}_1 + \mathbf{l} - \mathbf{r}_2|}
- \frac{1}{4 \pi \epsilon_0} \frac{q}{|\mathbf{r}_1 - \mathbf{r}_2|}.
\]

When $\mathbf{l}$ approaches zero such that $q \mathbf{l}$ remains a finite vector $\mathbf{p}$ (the electric dipole moment), the potential becomes

\[
\phi(\mathbf{r}_2) = \frac{1}{4 \pi \epsilon_0} \frac{\mathbf{p} \cdot \mathbf{r}}{r^3} \quad (\mathbf{r} = \mathbf{r}_2 - \mathbf{r}_1). 
\]

The corresponding electric field is given by

\begin{equation}
\label{EQ:Edip}
\mathbf{E}(\mathbf{r}_2) = -\nabla \phi (\mathbf{r}_2) =
\frac{1}{4 \pi \epsilon_0} 
\left(
\frac{3 \mathbf{r} (\mathbf{r} \cdot \mathbf{p})}{r^5} - \frac{\mathbf{p}}{r^3}
\right) \quad (\mathbf{r} = \mathbf{r}_2 - \mathbf{r}_1).
\end{equation}

The point is that even though the dipole is electrically neutral it generates a non-zero electric field which depends the orientation of the dipole. If a dipole is placed in an (homogeneous) external electric field $\mathbf{E}$, then a force $q\mathbf{E}$ acts on the positive end and a force $-q\mathbf{E}$ acts on the negative end creating a torque $\mathbf{N} = \mathbf{p} \times \mathbf{E}$ trying to line up the dipole along the direction of the field. This in turn affects the field generated by the dipole.

The charges associated with dipoles are called \textit{bound charges} since they cannot move freely. Thus, the total charge density can be written as $\varrho = \varrho_{\text{bound}} + \varrho_{\text{free}}$. While the displacement field $\mathbf{D}$ is defined such that $\nabla \cdot \mathbf{D} = \varrho_{\text{free}}$, one can define the \textit{polarization density} $\mathbf{P}$ such that $\nabla \cdot \mathbf{P} = -\varrho_{\text{bound}}$. Since ~$\epsilon_0 \nabla \cdot \mathbf{E} = \varrho$, we obtain

\[
\epsilon_0 \nabla \cdot \mathbf{E} = \nabla \cdot \mathbf{D} - \nabla \cdot \mathbf{P}
\]

which suggests that

\begin{equation}
\label{EQ:EDP}
\epsilon_0 \mathbf{E} = \mathbf{D} - \mathbf{P}.
\end{equation} 

Experimentally it is found, under not too extreme conditions, that there is a linear relation between the external field $\mathbf{E}$ and the induced polarization, $\mathbf{E} = \epsilon_0 \chi_e \mathbf{P}$. This finally gives using (\ref{EQ:EDP})

\begin{equation}
\label{EQ:ED}
\mathbf{D} = \epsilon_0 (1 + \chi_e) \mathbf{E} = \epsilon \epsilon_0 \mathbf{E},
\end{equation}

which is equation (\ref{EQ:DH}) with the relative permittivity given by $\epsilon = 1 + \chi_e$, where $\chi_e$ is the \textit{electric susceptibility}. 

\subsection{Magnetic susceptibility}

In the magnetic case we do not have magnetic dipoles formed by magnetic charges, because, as pointed out earlier, there appears not to exist any magnetic charges in the nature. Instead the magnetic fields are generated entirely by electric currents. On the atomic and molecular scales we have electric currents due the electrons ''circling'' around the atoms. Also the ''spin'' of the electrons contribute to magnetism. An external magnetic field may deflect the atomic currents and thus change the corresponding field generated by the currents in an analogy with the electric polarization. The atomic (bound) currents $\mathbf{J}_\text{bound}$ generate a \textit{magnetization} $\mathbf{M}$ defined by

\[
\nabla \times \mathbf{M} = \mathbf{J}_\text{bound}.
\]

\begin{wrapfigure}{r}{50mm}
\begin{flushleft}
\unitlength = 1mm
\begin{picture}(50,50)(0,0)
\thicklines
\put(25,25){\ellipse{40}{20}}
\put(25,25){\vector(0,1){15}}
\put(28,37){\makebox(6,6)[l]{$\mathbf{a}$}}
\put(20,12){\vector(1,0){8}}
\put(23,15){\makebox(6,6)[l]{$I$}}
\end{picture}
\end{flushleft}
\label{FIGmagmom}
\end{wrapfigure}

Besides the bounded currents $\mathbf{J}_\text{bound}$ which average to zero, we might in conductors have a free (macroscopic) current $\mathbf{J}_\text{free}$ related to the magnetic field by $\nabla \times \mathbf{H} = \mathbf{J}_\text{free}$. Since $\mu_0^{-1} \nabla \times \mathbf{B} = \mathbf{J} = \mathbf{J}_\text{bound} + \mathbf{J}_\text{free}$ we conclude that

\[
\mu_0^{-1} \mathbf{B} = \mathbf{H} + \mathbf{M}.
\]

For para- and diamagnetic substances the magnetization and the magnetic field are, under ''normal'' circumstances, linearly related, $\mathbf{M} = \chi_m \mathbf{H}$, where $\chi_m$ is the \textit{magnetic susceptibility}. Thus, we get $\mathbf{B} = \mu_0 (1 + \chi_m) \mathbf{H} = \mu_0 \mu \mathbf{H}$, with the \textit{magnetic permeability} given by $\mu = 1 + \chi_m$, which is the second equation in (\ref{EQ:DH}). Whereas the electric polarization was analyzed in terms of electric dipoles, the magnetization may be analyzed in terms of small current loops. If one consider such a small current loop in a magnetic field characterized by $\mathbf{B}$, then the total force acting on it is zero whereas the torque becomes (applying (\ref{EQ:Lor})),

\[
\mathbf{N} = \oint I \mathbf{r} \times (d\mathbf{r} \times \mathbf{B}) = \mathbf{m} \times \mathbf{B},
\] 

with the magnetic moment $\mathbf{m}$ defined by $\mathbf{m} = I \mathbf{a}$, where 

\[
\mathbf{a} = \frac{1}{2} \oint \mathbf{r} \times d\mathbf{r}
\]

is the area enclosed by the loop and $I$ its current. The magnetization $\mathbf{M}$ corresponds to the density of magnetic moments.

\subsection{Ohm's law}

A ''free'' charge $q$ in an electric field $\mathbf{E}$ feels a force $q \mathbf{E}$ which causes it to move. Thus, electrons in the conduction band in metals (conductors) can form a current when a potential difference is applied over a piece of a metal. The electrons though meet resistance caused e.g.\ by the thermal motion of the atoms. This is manifested in the well known ''law'' of Ohm according to which one needs a potential difference $U = RI$ in order to drive a current $I$ through a conductor with the resistance $R$. (It is conventional to use $U$ for the potential in the theory of circuits.) In terms of the current density $\mathbf{J}$ and the electrical field $\mathbf{E}$ driving the current, the law of Ohm can be written as

\begin{equation}
\label{EQ:Ohm}
\mathbf{J} = \sigma \mathbf{E}
\end{equation}

where the \textit{conductivity} $\sigma$ is inversely related to the resistance.
More precisely, for a conductor of length $L$ and cross section $A$ we have for the resistance

\[
R = \frac{L}{\sigma A} = \rho \frac{L}{A}.
\]

where $\rho = 1/\sigma$ defines the \textit{resistivity} (typically of the order of 10$^{-7}$ $\Omega\,$m for metals). When treating electromagnetic waves in conductors we thus have to use the relation (\ref{EQ:Ohm}) in Maxwell equation (\ref{EQ:Max}).

\subsection{Shielding}

\subsubsection{Skin effect}

It is well known that metallic enclosures (''Faraday cages'') protect against external electromagnetic fields. This shielding is caused by the fact that the charge carriers generate an opposing field. Suppose we have an incident plane wave along the $z$-direction on a metallic surface with the normal direction in the $-z$-direction. The equation (\ref{EQ:Wav}) becomes\footnote{The $\varrho$-term vanishes in this special case. Combining the continuity equation (\ref{EQ:Cont}) with $\mathbf{J} = \sigma \mathbf{E}$ we obtain for an harmonic plane field parallel with surface, $\mathbf{E} = (E_x, 0, 0)$, 

\[
\varrho = -i\frac{\sigma}{\omega} \nabla \cdot \mathbf{E},
\]

which is 0 since $E_x$ depends only on $z$ as the plane field travels in the $z$-direction.}

\begin{equation}
\label{EQ:ESJ}
\nabla^2 \mathbf{E} - \frac{1}{c^2} \frac{\partial^2 \mathbf{E}}{\partial t^2}
- \sigma \mu \mu_0 
\frac{\partial \mathbf{E}}{\partial t} = 0.
\end{equation}

Inserting a solution of the form (in the metallic medium $z > 0$; here we consider only the transmitted component, not the incident and reflected components in $z < 0$) 

\[
E_x = E_{0x} e^{i (k z - \omega t) }
\]

we obtain for $k$ the equation

\begin{equation}
\label{EQ:kequ}
k^2 = \frac{\omega^2}{c^2} + i \sigma \mu_0 \mu \omega.
\end{equation}

The imaginary part leads to exponentially decaying factor in $\exp(i k z)$. For instance, if the second term in (\ref{EQ:kequ}) dominates then we have

\[
k \approx \frac{1 + i}{\sqrt{2}} \sqrt{\sigma \mu_0 \mu \omega}, 
\]

which leads to 

\[
e^{ikz} \approx \exp\left(i\frac{1}{\sqrt{2}}\sqrt{\sigma \mu_0 \mu \omega } z\right) \cdot
\exp\left(-\frac{1}{\sqrt{2}}\sqrt{\sigma \mu_0 \mu \omega } z\right).
\]

The second decay factor shows that we have characteristic \textit{penetration depth} of

\begin{equation}
\label{EQ:pen}
\delta = \sqrt{\frac{2}{\sigma \mu \mu_0 \omega}}.
\quad \mbox{(Penetration depth, for high $\sigma$)}
\end{equation}

As an example, for copper we have $\sigma = 6\cdot 10^7 (\Omega \mbox{m})^{-1}$, $\mu \approx 1$, which give at $\omega$ = 2 $\pi f$ = 2 $\pi$ 2.4 GHz a penetration depth of $\delta = 1.3 \cdot 10^{-6}$m; that is, about 1 $\mu$m. From this it follows that a 1 mm Cu-sheet will practically stop the field completely. For an aluminum ($\delta = 3.8 \cdot 10^{-6}$m) foil of thickness 0.01 mm we would get a suppression factor around $\exp(-0.01 \, \mbox{mm}/\delta) \approx 2.5 \cdot 10^{-3}$, or -52 dB.   

The shielding (absorption of radiation) can also be interpreted in terms of a complex permittivity. From the second of Maxwell equations (\ref{EQ:Max}) we see that the current density $\mathbf{J}$ and the field $\mathbf{D}$ occur in form of the combination 

\[
\mathbf{J} + \frac{\partial \mathbf{D}}{\partial t}.
\]

Assuming harmonic fields depending on time as $\exp(-i \omega t)$ the time derivative above can be replaced by the factor $-i \omega$, and if we replace $\mathbf{J}$ by $ \sigma \mathbf{E}$, the above expression becomes

\[
\sigma \mathbf{E} + \frac{\partial \mathbf{D}}{\partial t} = -i \omega \left( i \frac{\sigma}{\omega} + \epsilon \epsilon_0 \right) \mathbf{E}. 
\]

This means that for conductors the effect of the current on the fields can be taken into account by using a complex relative permittivity given by

\begin{equation}
\label{EQ:perm}
\epsilon_r = \epsilon + i \frac{\sigma}{\omega \epsilon_0} \equiv
\epsilon' + i \epsilon''. 
\end{equation}

It is conventional to denote the real part by $\epsilon'$ and the imaginary part by $\epsilon''$. In many texts they write the complex permittivity as $\epsilon' - i \epsilon''$, which follows from assuming a time dependence $\exp(i \omega t)$ instead of $\exp(-i \omega t)$; thus, it is purely a matter of convention. The effect of the conductor is also to make the wave impedance $\eta$ defined in (\ref{EQ:Hpl}) imaginary. Indeed, using Maxwell equations for the plane waves we find that the E- and H-amplitudes are related by

\begin{equation}
\label{EQ:wimp}
H = \frac{E}{\eta}, \quad \eta = \frac{k}{\epsilon \epsilon_0 \omega + i\sigma},
\end{equation}

where $k$ is given by (\ref{EQ:kequ}). The above relation reduces to the one in
(\ref{EQ:Hpl}) when $\sigma$ = 0. A material is called a good conductor if the dielectric imaginary part dominates, $\epsilon'' \gg \epsilon'$, which translates into

\begin{equation}
\label{EQ:cond}
\tag{Good conductor criterion}
\frac{\sigma}{\epsilon \epsilon_0 \omega} \gg 1. 
\end{equation}

As an example, for $\omega$ = 2 $\pi f$ = 2 $\pi$ 2.4 GHz we get 

\[
\epsilon_0 \omega \approx 0.133 \; (\Omega \mbox{m})^{-1} \quad (f = 2.4 \; \mbox{GHz}).
\]

This can be compared with the conductivity of copper, $\sigma$ = 5.8 $\cdot 10^7$ ($\Omega$m)$^{-1}$, which thus, as all metals, qualifies as a ''good conductor'' by a safe margin. The human body has a conductivity around 0.2 ($\Omega$m)$^{-1}$ and is thus a poor conductor at this frequency. Sea water is a borderline case at this frequency having a conductivity around 4 ($\Omega$m)$^{-1}$ and $\epsilon \approx$ 80. The penetration depth is an important characteristic for food that is heated in microwave ovens; $\delta$ must be of the order of centimeters for the radiation to heat the food thoroughly. 

Because of the small penetration depth for good conductors it is typically assumed that the electric field is zero in the main part of the conductor, and that it is, like the current, confined to a thin layer of thickness about $\delta$ near the surface. This has important consequences. Suppose the surface lies along the $xy$-plane, and that the electric field is tangential in the $x$-direction. Then we have from the third equation in (\ref{EQ:Max}),

\[
\left| \Delta E_x \right| = |\Delta z| \left| \frac{\partial B_y}{\partial t} \right| .
\]  

If take the difference $\Delta E_x$ to be over the interface, and let $|\Delta z| \rightarrow 0$, then we obtain that

\begin{equation}
\label{EQ:Etan}
\mathbf{E}_t^{(II)} = \mathbf{E}_t^{(I)},
\end{equation} 

that is, the tangential component $\mathbf{E}_t$ of the electric field changes continuously across an interface between two mediums (I) and (II). Thus, if the electric field is zero inside the conductor ($\mathbf{E}_t^{(I)} = 0$) it must also be zero at the outside surface ($\mathbf{E}_t^{(II)} = 0$). It follows that when an oscillating electric field $\mathbf{E}^{i}$ impinges on a conducting surface, it generates a surface current $\mathbf{J}$ causing magnetic field $\mathbf{H}$ which in turn, according to Maxwell equations, causes a reflected field $\mathbf{E}^{r}$, such that the tangent component of the total field is zero at the surface (this is a simplification valid only on a scale large compared with the skin depth)

\begin{equation}
\label{EQ:Ezer}
\mathbf{E}^\text{tot}_t = \mathbf{E}^{i}_t + \mathbf{E}^{r}_t = 0. \quad \mbox{(At the surface of a conductor.)}
\end{equation}

This is used as a boundary condition when treating radiation in cavities and antenna radiation. Thus, the current $\mathbf{J}$ caused by an impinging field in an antenna will generate a field outside the antenna from which one may, for example, calculate the gapfield and corresponding potential difference in case of a dipole antenna (see sec. \ref{SEC:Antennaex}). Another consequence of the penetration layer is that electromagnetic waves will loose energy due to ohmic losses; the waves induce surface currents which encounter a resistance given by

\begin{equation}
\label{EQ:Rs}
R_s = \frac{\rho}{\delta} = \frac{1}{\sigma} \sqrt{\frac{\sigma \mu \mu_0 \omega}{2}} = \sqrt{\frac{ \mu \mu_0 \omega}{2 \sigma}}. \quad \mbox{(Surface resistance.)}
\end{equation}

Using previous data on copper we get for its surface resistance $R_s \approx$  1/78 $\Omega$ (at 2.4 GHz).

\subsubsection{Leakage through slots}
\label{SEC:leak}

If there are holes in the shield then there will be no counteracting current at that place and radiation can leak through. This effect can be used to an advantage when constructing slot antennas, but for shielding purposes the leakage is of course a nuisance. In order to get an estimate of the leakage through a slot one may consider a rectangular hole in a conducting sheet. We suppose the sheet is in the $xy$-plane with the normal along the $-z$-axis and has thickness $d$. Further we take the hole to have the corners (0,0), (0,$b$), ($a$,0), ($a$,$b$) in the $xy$-plane.

We will treat the hole as a wave guide on which impinges a planar EM-wave along the $z$-axis. Intuitively it seems clear that waves with a wavelength $\lambda \gg a, b$ will have difficulty in passing through the hole; that is, the hole acts as a high-pass filter damping waves with wavelengths exceeding the dimension of the hole, but letting smaller wavelengths through (the high frequency part). We will consider a TE-wave passing the wave guide; thus, the electric field is transversal while the magnetic field may also have a longitudinal component along the $z$-axis\footnote{The TEM case leads to a trivial solution of zero fields inside an empty wave guide. Indeed, in this special case one obtains from Maxwell equation that ($T$ means here that the operators are restricted to the transversal $xy$-plane)

\[
\nabla_T \times \mathbf{E} = 0,
\]

from which one may posit that there is a function $\phi$ such that

\[
\mathbf{E} = -\nabla_T \phi.
\]

Combining this with the equation $\nabla_T \times \mathbf{E} = 0$ gives the equation $\nabla_T^2 \phi = 0$. This is the Laplace equation in two dimensions, and for a simple region (such as the cross-section of the wave guide) this has the trivial solution $\phi$ = constant. Indeed, $E_\text{tangent}$ = 0 implies that $\phi$ is constant along the boundary of the cross-section, and therefore that $\nabla_T^2 \phi = 0$ has the trivial solution $\phi$ = constant in the cross-section. In the TE case, where we allow for a magnetic longitudinal component, $\nabla_T \times \mathbf{E} = 0$ is replaced by $\nabla_T \times \mathbf{E} = i \omega \mu_0 \mu H_z$, and non-trivial solutions become possible.}. We will therefore assume that

\begin{align*}
&\mathbf{E}(x,y,z,t) = (E_x (x,y), E_y (x,y), 0) \cdot e^{ i(\beta z - \omega t)} ,
\\
&\mathbf{H}(x,y,z,t) = (H_x (x,y), H_y (x,y), H_z (x,y)) \cdot e^{i (\beta z - \omega t)}.
\end{align*}

The dependence on $z$ is thus factored out as $\exp(i \beta z)$ since we are interested in wave solutions progressing in the $z$-direction. If we insert the above ansatz into the wave-equation (\ref{EQ:Wav}) for the E-field we obtain ($\mathbf{J} = 0$ and $\varrho$ = 0 in empty space)

\begin{align}
\label{EQ:wguide}
&\nabla_T^2 E_x = (\beta^2 - k^2) E_x, \quad \mbox{(Helmholz equation)}
\\
\nonumber
&\quad \mbox{with}
\\
\nonumber
&\nabla_T^2 \equiv \frac{\partial^2}{\partial x^2} + \frac{\partial^2}{\partial y^2} \quad \mbox{and} \quad k = \frac{\omega}{c} = \frac{2 \pi}{\lambda}, 
\end{align}

\begin{wrapfigure}{r}{60mm}
\begin{flushleft}
\unitlength = 1mm
\begin{picture}(60,45)(-10,0)
\thicklines
\multiput(0,20)(20,-20){2}{\drawline(0,0)(0,25)}
\multiput(0,20)(0,25){2}{\drawline(0,0)(20,-20)}
\multiput(20,0)(0,25){2}{\dottedline{1}(0,0)(30,0)}
\multiput(30,20)(20,-20){2}{\dottedline{1}(0,0)(0,25)}
\multiput(30,20)(0,25){2}{\dottedline{1}(0,0)(20,-20)}
\dottedline{1}(0,45)(30,45)
\drawline(0,20)(20,20)
\dottedline{1}(20,20)(30,20)
\put(0,20){\vector(1,-1){7}}
\put(7,20){\makebox(6,6)[l]{$z$}} 
\put(0,20){\vector(1,0){10}}
\put(7,12){\makebox(6,6)[l]{$y$}} 
\put(0,20){\vector(0,1){10}}
\put(1,27){\makebox(6,6)[l]{$x$}}
\put(22,10){\makebox(6,6)[l]{$a$}}
\put(11,35){\makebox(6,6)[l]{$b$}} 
\emwave{-6,8}
\end{picture}
\end{flushleft}
\label{FIGwellg}
\end{wrapfigure}

with a similar equation for the $y$-component. Here $\nabla_T^2$ refers to the Laplacian operator restricted to the transversal plane. Since the tangential components of the electric fields vanish at the surface of the walls, they will be of the form

\begin{align*}
&E_x (x,y) = g(x) \sin \left( \frac{n \pi y}{b}\right),
\\
&E_y (x,y) = h(y) \sin \left( \frac{m \pi x}{a}\right). 
\end{align*}

Using the Maxwell equation $\partial_x E_x + \partial_y E_y = 0$ we can determine the functions $h$ and $g$, obtaining finally ($m, n \neq 0$),

\begin{align}
\label{EQ:Tmn}
&E_x (x,y) = E_{mn} \frac{a}{m} \cos \left( \frac{m \pi x}{a}\right) \sin \left( \frac{n \pi y}{b}\right),
\\
\nonumber
&E_y (x,y) = - E_{mn} \frac{b}{n} \sin \left( \frac{m \pi x}{a}\right) \cos \left( \frac{n \pi y}{b}\right). 
\end{align}

The case $m$ = 0 corresponds to a solution of the form

\begin{align*}
E_x (x,y) = a_n \sin \left( \frac{n \pi y}{b}\right),
\\
E_y (x,y) = b_n \sin \left( \frac{n \pi x}{a}\right).
\end{align*} 

The general solutions may be constructed as superpositions of these ($m, n$)-mode solutions. Inserting an ($m, n$)-mode solution into (\ref{EQ:wguide}) we obtain the relation

\begin{equation}
\label{EQ:beta}
\beta^2 = k^2 - \left(\frac{n \pi}{a} \right)^2 - \left(\frac{m \pi}{b} \right)^2 = \left(\frac{ 2 \pi}{\lambda} \right)^2 - \left(\frac{n \pi}{a} \right)^2 - \left(\frac{m \pi}{b} \right)^2.
\end{equation}

From this we see that if $a, b < \lambda$ then $\beta$ must be necessarily imaginary leading to an exponential decay factor $\exp(i \beta z)$. Suppose we have a narrow slot $a$ = 1 mm in a $d$ = 2 mm thick conducting plate, then we may estimate the radiation through the slot to be damped by factor of the order

\[
\exp \left\{ -\sqrt{\left(\frac{\pi}{ a} \right)^2 - \left(\frac{ 2 \pi}{\lambda} \right)^2} \cdot d \right\} \approx \exp(-3141 \cdot 0.002) \approx 0.0019,
\]  

which corresponds to a -54 dB damping (power) at 2.4 GHz ($\lambda$ = 0.125 m).
For $a \ll \lambda$ the above formula for damping can be approximated by (in dB) 

\[
\text{damping [dB]} \; = \; 20 \cdot \log\left(e^{-\frac{\pi d}{a}}\right) \approx -27.3 \cdot \frac{d}{a}.
\]

  
\begin{wrapfigure}{r}{60mm}
\begin{flushleft}
\unitlength = 1mm
\begin{picture}(60,30)(0,0)
\thicklines
\drawline(5,20)(30,20)
\multiput(30,18)(0,4){2}{\drawline(0,0)(25,0)}
\multiput(30,18)(2,0){11}{\drawline(0,0)(4,4)}
\drawline(30,20)(32,22)
\drawline(30,18)(30,22)
\put(10,30){\makebox(6,6)[l]{(I)}}
\put(40,30){\makebox(6,6)[l]{(II)}}
\put(10,25){\vector(1,0){7}}
\put(17,15){\vector(-1,0){7}}
\put(40,25){\vector(1,0){7}}
\end{picture}
\end{flushleft}
\label{FIGreflline}
\end{wrapfigure}

\subsection{Reflexion and refraction}
\label{SEC:reflex}

Reflexion and refraction of waves is a familiar phenomenon from our daily experiences with light, sound and water. Reflexion is a basic effect when a wave hits an inhomogeneity in the medium, typically the interface between two different mediums such as air and water. We will discuss this fundamental feature in terms of a very simple model. Consider a ''wave'' traveling along the $x$-axis in a medium (I, $x$ < 0) whose propagation velocity is $u_1$. At $x$ = 0 starts another medium (II, $x$ > 0) with a different propagation velocity $u_2$. (In case of EM-fields $u = c/\sqrt{\epsilon \epsilon_0}$ where $c$ is the light velocity in vacuum.) We use $\phi$ for a propagating field satisfying the following equations: 

\begin{align}
\label{EQ:refl1}
&\frac{\partial^2 \phi }{\partial x^2} - \frac{1}{u_1^2} \frac{\partial^2 \phi }{\partial t^2} = 0, \quad \mbox{(I)} \; (x < 0) 
\\
\nonumber
&\frac{\partial^2 \phi }{\partial x^2} - \frac{1}{u_2^2} \frac{\partial^2 \phi }{\partial t^2} = 0. \quad \mbox{(II)} \; (x > 0)
\end{align}

We write the basic harmonic solutions for regions (I) and (II) as:

\begin{align}
&\label{EQ:refl2}
\phi_1(x,t) = e^{i(k_1 x - \omega t)} + R e^{i(-k_1 x - \omega t)}, &\quad \mbox{(I)}
\\
\nonumber
&\phi_2(x,t) = T e^{i(k_2 x - \omega t)}.  &\quad \mbox{(II)}
\end{align}

\begin{wrapfigure}{r}{50mm}
\begin{flushleft}
\unitlength = 1mm
\begin{picture}(50,50)(0,0)
\thicklines
\drawline(0,25)(50,25)
\drawline(5,45)(25,25)(45,45)
\put(15,35){\vector(1,1){5}}
\put(35,35){\vector(-1,1){5}}
\drawline(25,25)(32,4)
\put(30,10){\vector(3,1){6}}
\put(12,27){\makebox(6,6)[l]{$\theta$}}
\put(37,27){\makebox(6,6)[l]{$\theta_r$}}
\put(37,14){\makebox(6,6)[l]{$\theta_t$}}
\dashline{3}(25,50)(25,0)
\put(25,45){\vector(0,1){5}}
\put(27,45){\makebox(6,6)[l]{$z$}}
\put(50,25){\vector(1,0){0}}
\put(45,26){\makebox(6,6)[l]{$y$}}
\multiput(0,23)(2,0){13}{\line(1,1){2}}
\multiput(32,23)(2,0){8}{\line(1,1){2}}
\put(25,25){\arc{8}{3.1416}{3.927}}
\put(25,25){\arc{8}{5.498}{0}}
\put(25,25){\arc{10}{5.498}{0}}
\put(25,25){\arc{8}{0}{1.249}}
\put(25,25){\arc{10}{0}{1.249}}
\put(25,25){\arc{12}{0}{1.249}}
\put(12,40){\makebox(6,6)[l]{(I)}}
\put(12,10){\makebox(6,6)[l]{(II)}}
\end{picture}
\end{flushleft}
\label{FIGrefl}
\end{wrapfigure}

Here the wavenumbers $k_i$ are given by $k_i = \omega/u_i$. The interpretation of the solution (\ref{EQ:refl2}) is the $R$-term represent the reflected part propagating in the $-x$-direction, and the $T$-term the transmitted part. The coefficients $R, T$ can be determined from the requirement that $\phi$ and ~$\partial \phi/\partial x$ be continuous at the boundary $x$ = 0; that is, $\phi_1(0-) = \phi_2(0+)$, $\partial \phi_1(0-)/\partial x = \partial \phi_2(0+)/\partial x$. This yields the equations

\begin{align}
\label{EQ:refl3}
1 + R &= T,
\\
k_1 (1 - R) &= k_2 T,
\end{align}

from which obtain $R$ and $T$,

\begin{align}
\label{EQ:reflRT}
&R = \frac{k_1 - k_2}{k_1 + k_2} = \frac{u_2 - u_1}{u_1 + u_2},
\\
&T = \frac{2 k_1}{k_1 + k_2} = \frac{2 u_2}{u_1 + u_2}.
\end{align}

The above model may e.g.\ be used to describe the effect of connecting two cables with different impedances; the discontinuity at the connection gives rise to reflexions. We may also note that the sign of the reflexion coefficient $R$ depends on whether the wave travels faster or slower in region II than in region I.

Next we will consider an EM plane wave in open space ($z$ > 0) impinging on a dielectric surface in the $xy$-plane at $z$ = 0. The total electric field on the side I ($z$ > 0) will consist of the incoming and the reflected part ($R$), while on the side II ($z$ < 0) we will have the transmitted (refracted) part ($T$), 

\begin{align}
\label{EQ:Erefl}
&\mathbf{E} e^{i(\mathbf{k} \cdot \mathbf{r} - \omega t)}
+ \mathbf{E}^R e^{i(\mathbf{k}^R \cdot \mathbf{r} - \omega t)},
&\quad \mbox{(I)}
\\
\nonumber
&\mathbf{E}^T e^{i(\mathbf{k}^T \cdot \mathbf{r} - \omega t)}. 
&\quad \mbox{(II)}
\end{align}

Here the magnitudes of the wavevectors are given by

\begin{align*}
&k_1 = |\mathbf{k}| = |\mathbf{k}^R| = \omega \sqrt{\mu_1 \mu_0 \epsilon_1 \epsilon_0},
\\
&k_2 = |\mathbf{k}^T| = \omega \sqrt{\mu_2 \mu_0 \epsilon_2 \epsilon_0}.
\end{align*}
 
As demonstrated in connection with (\ref{EQ:Etan}) the tangential component of the electric field does not change across the interface. Hence, choosing the coordinate system so that the tangent component is along the $y$-axis we obtain
at the I-II interface, $z$ = 0,

\begin{equation}
\label{EQ:Exy}
E_y e^{i(\mathbf{k} \cdot \mathbf{r} - \omega t)}
+ E_y^R e^{i(\mathbf{k}^R \cdot \mathbf{r} - \omega t)} =
E_y^T e^{i(\mathbf{k}^T \cdot \mathbf{r} - \omega t)}. 
\end{equation}

This equality is only possible if (at $z$ = 0) 

\[
\mathbf{k} \cdot \mathbf{r} = \mathbf{k}^R \cdot \mathbf{r} = \mathbf{k}^T \cdot \mathbf{r}
\]

\begin{wrapfigure}{r}{50mm}
\begin{flushleft}
\unitlength = 1mm
\begin{picture}(50,45)(0,0)
\thicklines
\shade\put(25,30){\ellipse{30}{20}}
\put(25,24){\ellipse{30}{20}}
\put(25,27){\ellipse{30}{20}}
\path(5,10)(10,45)(40,45)(45,10)(5,10)
\multiput(10,24)(30,0){2}{\drawline(0,0)(0,6)}
\put(25,30){\vector(0,1){20}}
\put(27,39){\makebox(6,6)[l]{$D_z$}}
\end{picture}
\end{flushleft}
\label{FIGpillbx}
\end{wrapfigure}

from which one deduces (set $\mathbf{r}$ = $\mathbf{\hat{y}}$~) that the angle of reflexion $\theta_r$ is equal to the incident angle $\theta$ (angles are here measured as those made by the directions of propagation with the interface), while the angle of refraction $\theta_t$ on the other hand is related by ''Snellius' law'' 

\begin{equation}
\label{EQ:Snell}
n_1 \cos \theta = n_2 \cos \theta_t \quad \mbox{(Snellius)}
\end{equation}

where $n_i$ are the indexes of refraction of the mediums given by $n_i = \sqrt{\epsilon_i \mu_i}$. From (\ref{EQ:Exy}) we also obtain that

\begin{equation}
\label{EQ:Eqtan}
E_y + E_y^R =
E_y^T. 
\end{equation}

If we consider the $E$-field to be polarized in the $yz$-plane (vertical, V-polarization case; H-field will be along the $y$-axis) then the components of the fields can be expressed in terms of the amplitudes,

\begin{align}
\label{EQ:Ecomp} 
E_y &= E \sin \theta, &\quad H_y &= 0, 
\\
\nonumber
E_z &= E \cos \theta,  &\quad H_x &= \frac{1}{\eta_1} E,
\\
\nonumber
E_y^R &= - E^R \sin \theta, &\quad H_y^R &= 0, 
\\
\nonumber
E_z^R &= E^R \cos \theta, &\quad H_x^R &= \frac{1}{\eta_1} E^R,
\\
\nonumber 
E_y^T &= E^T \sin \theta_t, &\quad H_y^T &= 0,
\\
\nonumber
E_z^T &= E^T \cos \theta_t, &\quad H_x^T &= \frac{1}{\eta_2} E^T.
\end{align}

In order to determine the amplitudes $E^R$, $E^T$, we need one further equation besides (\ref{EQ:Exy}). This can be found by applying (\ref{EQ:QJ}) to a very thin ''pill-box'' which contains the I-II interface, then one obtains that

\[
\oint \mathbf{D} \cdot d\mathsf{S} = \left(D_z^{I} - D_z^{II} \right) \mathsf{S} = Q,
\]   

where $\mathsf{S}$ is the top (bottom) surface area of the pill-box, and $Q$ the surface charge contained  by it. Going to the infinitesimal thin pill-box limit we obtain the general result on the normal component of the displacement vector, 

\begin{equation}
\label{EQ:Dnorm}
\mathbf{D}_n^{I} = \mathbf{D}_n^{II} + \varrho_s,
\end{equation}

where $\varrho_s$ is the surface charge density ($Q/S$). In our particular case we can assume that there are no extra surface charges ($\varrho_s = 0$), whence, using $\mathbf{D} = \epsilon \epsilon_0 \mathbf{E}$,

\begin{equation}
\label{EQ:Enorm}
\epsilon_1 (E_z + E_z^R)= \epsilon_2 E_z^T.
\end{equation}  

Another alternative is to use the boundary condition that the tangent component of the H-field is continuous across the interface, $\mathbf{H}^I_t = \mathbf{H}^{II}_t$, which can be derived in a similar way as in the case of the E-field. Note that the wave impedances $\eta_i$ in (\ref{EQ:Ecomp}) are given by (\ref{EQ:wimp}) which cover the case of conductive media too.

Combining (\ref{EQ:Enorm}), (\ref{EQ:Eqtan}) and (\ref{EQ:Snell}), we can obtain after some algebra,

\begin{align}
\label{EQ:Fresnelv}
&\rho_v \equiv \frac{E^R}{E} = \frac{\epsilon_r \sin\theta - \sqrt{\epsilon_r - (\cos\theta)^2}}{\epsilon_r \sin\theta + \sqrt{\epsilon_r - (\cos\theta)^2}},
\\
\nonumber
&\tau_v \equiv \frac{E^T}{E} = \frac{ 2 \sin\theta \sqrt{\epsilon_r}}{\epsilon_r \sin\theta + \sqrt{\epsilon_r - (\cos\theta)^2}},
\\
\nonumber
&\mbox{(V-polarization case.)}
\end{align}  

where we have used the notation $\epsilon_r = \epsilon_2/\epsilon_1$. Similar considerations can be applied in the horizontal (H) polarization case, with the end result,

\begin{align}
\label{EQ:Fresnelh}
&\rho_h \equiv \frac{E^R}{E} = \frac{\sin\theta - \sqrt{\epsilon_r - (\cos\theta)^2}}{\sin\theta + \sqrt{\epsilon_r - (\cos\theta)^2}},
\\
\nonumber
&\tau_h \equiv \frac{E^T}{E} = \frac{ 2 \sin\theta}{\sin\theta + \sqrt{\epsilon_r - (\cos\theta)^2}}.
\\
\nonumber
&\mbox{(H-polarization case.)}
\end{align}

The equations (\ref{EQ:Fresnelv}), (\ref{EQ:Fresnelh}), are known as Fresnel [FRA-nel] equations. These equations also apply when interfacing a conducting material by replacing $\epsilon$ by a complex number as explained in connection with (\ref{EQ:perm}). Thus, if the medium II is a perfect conductor this corresponds to letting $|\epsilon_r| \rightarrow \infty$, and we have then $\rho_v = 1$ for V-polarization and $\rho_h = -1$ for H-polarization. For dielectrics there is a special angle, the \textit{Brewster angle} $\theta_B$, 

\begin{equation}
\label{EQ:Brew}
\sin \theta_B = \frac{1}{\sqrt{1 + \epsilon_r}}
\end{equation} 

at which the reflected vertical component goes to zero, $\rho_v = 0$. Thus, if the incoming field is vertically polarized, none of it will be reflected at the Brewster angle (for a planar interface). This means that if the transmitter (using vertical antenna) and the receiver are placed such that the Brewster angle condition is satisfied then the reflecting component of the radiation is eliminated from the transmission, and only the direct field is received. This configuration may be used to measure how the inclination of the receiver antenna affects the reception; that is, to measure the function $G(\theta)$ for varying $\theta$. The above analysis can be generalized to the case where we have two mediums I, III, with a second medium II of thickness $d$ sliced between them. One example might be air (I), and ice sheet (II) with water (III) below (which we have investigated experimentally). The reflected field in I is thus reflected both from the interface I-II and from the interface II-III. We may treat the problem with the methods used above, pasting together plane wave solutions in the regions at the interfaces. We may also use the methods of geometrical optics and sum the contributions from all the additional reflexions from the intermediary layer interface II-III. We denote by (and similarily for transmission coefficient) $\rho\left(
\frac{\epsilon_2}{\epsilon_1}, \theta \right)$ the reflexion coefficient (subindexes will indicate the polarization states) for an EM-wave in a medium I impinging on a surface of a medium II at the angle $\theta$.

\begin{wrapfigure}{r}{50mm}
\begin{flushleft}
\unitlength = 1mm
\begin{picture}(50,50)(0,0)
\thicklines
\multiput(5,20)(0,10){2}{\drawline(0,0)(40,0)}
\path(5,45)(20,30)(35,45)
\put(35,45){\vector(1,1){0}}
\path(20,30)(25,20)(30,30)(45,45)
\put(45,45){\vector(1,1){0}}
\drawline(25,20)(30,5)
\put(30,5){\vector(1,-3){0}}
\path(30,30)(35,20)(40,5)
\put(40,5){\vector(1,-3){0}}
\path(35,20)(40,30)(45,35)
\put(45,35){\vector(1,1){0}}
\put(5,33){\makebox(6,6)[l]{(I)}}
\put(39,22){\makebox(6,6)[l]{(II)}}
\put(5,10){\makebox(6,6)[l]{(III)}}
\drawline(10,20)(10,30)
\put(10,20){\vector(0,-1){0}}
\put(10,30){\vector(0,1){0}}
\put(12,23){\makebox(6,6)[l]{$d$}}
\put(12,30){\makebox(6,6)[l]{$\theta$}}
\put(19,20){\makebox(6,6)[l]{$\theta_t$}}
\put(28,13){\makebox(6,6)[l]{$\theta_{t'}$}}
\end{picture}
\end{flushleft}
\label{FIGrefl2lay}
\end{wrapfigure}

The total reflexion will be a sum of the primary reflection at point $A$ (see figure), the next contribution comes from the transmitted part which reflects from point $B$ and then exits the surface at point $C$, and so on. It is important to note that the parts that bounce through the intermediary layer II pick up additional phase differences due to the factor \[\exp(i \mathbf{k}^{\text{II}} \cdot \mathbf{r}).\] The phase contribution due to a a given optical path is $k s$ where $s$ is the length of the path. Thus, after some trigonometrical exercises, the phase difference between the paths $ABC$ and $AD$ will turn out to be,

\begin{equation*}
\Delta = \frac{n_2 2 \pi 2 d}{\lambda \sin \theta_t} -
\frac{n_1 2 \pi }{\lambda} 2 d \cot \theta_t \cos \theta = 
\frac{2 n_2 2 \pi d}{\lambda}
\left(
\frac{1}{\sin \theta_t} -
\frac{\cos \theta_t^2}{\sin \theta_t}
\right) = 2 k n_2 d \sin \theta_t,
\end{equation*}

where we have used the law of Snellius, and  $k = 2 \pi/\lambda$ for the  wavevector magnitude in vacuum, and $n_i$ for the indexes of refraction. Summing all the reflexion contributions we get,

\begin{align}
\label{EQ:Rtotal}
&\rho_\text{total} =
\rho\left(
\frac{\epsilon_2}{\epsilon_1}, \theta
\right)   
+
e^{i\Delta}
\tau\left(
\frac{\epsilon_2}{\epsilon_1}, \theta
\right) 
\rho\left(
\frac{\epsilon_3}{\epsilon_2}, \theta_t
\right)   
\tau\left(
\frac{\epsilon_1}{\epsilon_2}, \theta_t
\right) + 
\\
\nonumber
&e^{i2\Delta}
\tau\left(
\frac{\epsilon_2}{\epsilon_1}, \theta
\right) 
\rho\left(
\frac{\epsilon_3}{\epsilon_2}, \theta_t
\right) 
\rho\left(
\frac{\epsilon_1}{\epsilon_2}, \theta_t
\right) 
\rho\left(
\frac{\epsilon_3}{\epsilon_2}, \theta_t
\right) 
\tau\left(
\frac{\epsilon_1}{\epsilon_2}, \theta_t
\right) + \cdots =
\\
\nonumber 
&\rho\left(
\frac{\epsilon_2}{\epsilon_1}, \theta
\right)   
+ e^{i\Delta} 
\frac{
\rho\left(\frac{\epsilon_3}{\epsilon_2}, \theta_t\right)
\tau\left(\frac{\epsilon_2}{\epsilon_1}, \theta\right)
\tau\left(\frac{\epsilon_1}{\epsilon_2}, \theta_t\right)   
}
{
1 - 
e^{i\Delta}
\rho\left(\frac{\epsilon_3}{\epsilon_2}, \theta_t\right)
\rho\left(\frac{\epsilon_1}{\epsilon_2}, \theta_t\right)
}.
\end{align}

\begin{wrapfigure}{r}{50mm}
\begin{flushleft}
\unitlength = 1mm
\begin{picture}(50,45)(0,0)
\thicklines
\multiput(5,10)(0,20){2}{\drawline(0,0)(40,0)}
\path(5,40)(15,30)(30,45)
\put(30,45){\vector(1,1){0}}
\path(15,30)(25,10)(35,30)(45,40)
\put(45,40){\vector(1,1){0}}
\drawline(25,40)(35,30)
\put(35,37){\makebox(6,6)[l]{(I)}}
\put(35,17){\makebox(6,6)[l]{(II)}}
\drawline(10,10)(10,30)
\put(10,10){\vector(0,-1){0}}
\put(10,30){\vector(0,1){0}}
\put(12,19){\makebox(6,6)[l]{$d$}}
\put(8,30){\makebox(6,6)[l]{$\theta$}}
\put(18,11){\makebox(6,6)[l]{$\theta_t$}}
\put(13,32){\makebox(6,6)[l]{$A$}}
\put(24,4){\makebox(6,6)[l]{$B$}}
\put(20,38){\makebox(6,6)[l]{$D$}}
\put(35,24){\makebox(6,6)[l]{$C$}}
\end{picture}
\end{flushleft}
\label{FIGphased}
\end{wrapfigure}

Here we have used the geometric summation rule $1 + x + x^2 + \cdots = 1/(1-x)$.
By a similar calculation we obtain for the transmission coefficient $\tau$ for the radiation that enters into the medium III,

\begin{equation}
\label{EQ:Ttotal}
\tau =
\frac{e^{i\Delta/2}
\tau\left(\frac{\epsilon_2}{\epsilon_1}, \theta_t\right)
\tau\left(\frac{\epsilon_3}{\epsilon_2}, \theta_{t'}\right)
}
{
1 - 
e^{i\Delta}
\rho\left(\frac{\epsilon_3}{\epsilon_2}, \theta_t\right)
\rho\left(\frac{\epsilon_1}{\epsilon_2}, \theta_t\right)
}.
\end{equation}
 
As an example we can calculate the transmission coefficient for radiation impinging normally on a brick wall of thickness $d$ = 10 cm assuming $\epsilon_1 = \epsilon_3 = 1$ and $\epsilon_2 = 4$, leading to (setting $\epsilon_r = \epsilon_2/\epsilon_1$)

\begin{equation}
\label{EQ:tauT} 
\tau = \frac{4 \sqrt{\epsilon_r} e^{i k \sqrt{\epsilon_r} d}}{\left(1 + \sqrt{\epsilon_r}\right)^2 + 
e^{i 2 k \sqrt{\epsilon_r} d} \left(\sqrt{\epsilon_r} - 1 \right)^2}  = -0.74 + i 0.43,
\end{equation}

whose absolute value is 0.86 (@ 2.4 GHz) corresponding to a reduction of power by the factor 0.86$^2$ = 0.74 (-1.3 dB). Since $\epsilon_r$ is assumed to be real there is no absorption in the wall. If we use $\epsilon''$ = 0.07 for the brick wall then we have for a 1 m wall $|\tau|$ = 0.36 thus showing already significant absorption (-8.9 dB). In reality there would be a further loss of power due to scattering caused by the inhomogeneities in the wall. Brick walls are seldom 1 m thick, instead the radiation may have to pass several brick walls which are say 10 cm thick. Then a quick estimate would be that the power decreases with a factor about 0.74 per wall. For a more detailed treatment one can extend the above methods to an arbitrary numbers of dielectric layers \cite{Rulf1987}. One can also apply the transmission theory for cables using for impedance the wave impedances $Z = \sqrt{\mu \mu_0 /\epsilon \epsilon_0}$ (see Appendix \ref{SEC:Acab}).

\subsection{Image charges}
\label{SEC:image}

We consider a charge $q$ above a plane perfect conductor which we take to be in the $xy$-plane at $z$ = 0. The plane will now affect the electric field. As explained earlier the electric field must have a zero tangential component on the surface of the conductor. An equivalent formulation is that the potential $\varphi$ is constant along the surface. Thus, given this boundary condition, one has to solve the Laplace equation $\nabla^2 \varphi = 0$ which is valid for $z > 0$, except at the place of the charge which we may suppose is at the point $\mathbf{r}_0$ = ($0, 0, h$). One can convince oneself that the solution must be

\begin{equation}
\label{EQ:mirr1}
\varphi(\mathbf{r}) = \frac{q}{4 \pi \epsilon \epsilon_0} \frac{\mathbf{r} - \mathbf{r}_0}{|\mathbf{r} - \mathbf{r}_0|^2}       
-
\frac{q}{4 \pi \epsilon \epsilon_0} \frac{\mathbf{r} + \mathbf{r}_0}{|\mathbf{r} + \mathbf{r}_0|^2}.       
\end{equation} 

It satisfies the Laplace equation for $z > 0$ except at the point $\mathbf{r}_0$, and it vanishes for $z = 0$; that is, $z = 0$ is an equi\-potential surface. Furthermore, if we integrate $\mathbf{E} = - \nabla \varphi$ over a small sphere containing $q$ we obtain $q/\epsilon \epsilon_0$ proving that it is indeed the potential of the charge $q$. The solution (\ref{EQ:mirr1}) means that the effect of the conducting plane is the same as if we had an additional extra charge of the opposite sign at the place of its mirror image, $\mathbf{r}_0 - 2(\mathbf{n} \cdot \mathbf{r}_0) =-\mathbf{r}_0$ ($\mathbf{n}$ is the normal of the surface), in an empty space. One consequence of the mirror effect is that the charge is attracted toward the conducting plane by the apparent opposite image charge. Physically the effect of the conducting plane is that the charge $q$ polarizes the free charges in the plane by attracting them if of opposite sign, and repelling them otherwise. In fact, the \textit{induced} surface charge at $z = 0$ can be calculated from $\varrho_s = - \epsilon \epsilon_0 \, \partial \varphi/\partial z$ by inserting the solution (\ref{EQ:mirr1}). If we integrate $\varrho_s$ over the surface we get in fact for the total induced charge the result $-q$.

This mirroring method can be generalized to other surface that can be construed as equi\-potential surfaces for some distribution of charges. Consider the case where we have two conducting planes meeting along the $z$-axis. We may take the conducting planes to be the $xz$-plane and the $yz$-plane (see part (b) in the figure). The potential in the open space region is obtained adding three image charges as shown in the figure.  

\begin{figure}[h]
\centering
\subfigure[]
{
\setlength{\unitlength}{1mm}
\begin{picture}(50,50)(0,0)
\thicklines
\dottedline(25,10)(25,40)
\drawline(5,25)(45,25)
\put(25,40){\circle*{1}}
\put(25,10){\circle{1}}
\drawline(20,25)(15,10)
\drawline(30,25)(35,10)
\drawline(15,25)(5,10)
\drawline(35,25)(45,10)
\spline(25,40)(22,33)(20,25)
\spline(25,40)(28,33)(30,25)
\qbezier(25,40)(16,30)(15,25)
\qbezier(25,40)(34,30)(35,25)
\put(7,40){\makebox(6,6)[l]{(I)}}
\put(7,5){\makebox(6,6)[l]{(II)}}
\put(22,41){\makebox(6,6)[l]{$q$}}
\put(26,41){\makebox(6,6)[l]{$q''$}}
\put(24,4){\makebox(6,6)[l]{$q'$}}
\end{picture}
}
\hspace{1cm}
\subfigure[]
{
\setlength{\unitlength}{1mm}
\begin{picture}(50,50)(0,0)
\thicklines
\dottedline(5,25)(25,25)(25,45)
\dashline{2}(35,15)(15,15)(15,35)(35,35)(35,15)
\drawline(25,5)(25,25)(45,25)
\put(35,15){\circle*{1}}
\put(35,35){\circle{1}}
\put(15,15){\circle{1}}
\put(15,35){\circle*{1}}
\put(38,15){\makebox(6,6)[l]{$q$}}
\put(38,33){\makebox(6,6)[l]{$-q$}}
\put(5,15){\makebox(6,6)[l]{$-q$}}
\put(7,33){\makebox(6,6)[l]{$q$}}
\multiput(23,5)(0,2){10}{\drawline(0,0)(2,2)}
\multiput(25,25)(2,0){10}{\drawline(0,0)(2,2)}
\end{picture}
}
\label{FIGimageq}
\end{figure}

Indeed, one sees that this arrangement makes the total potential zero along the planes. This example has application in the case we use 90$^\circ$-degree bent sheet as an antenna reflector. A somewhat more involved case is that of placing a charge between to parallel conducting planes which requires an infinite number of image charges (as one can ''see''  from the analogous case of placing a candle between two parallel mirrors). 

The imaging principle can also be applied to the case of a dielectric instead of a conducting plane. Thus consider the case where we have a homogeneous dielectric I in the region $z > 0$, and a different dielectric II in the region $z < 0$. We place a charge $q$ at the point $\mathbf{r}_0$ = ($0, 0, h$) in I, and the problem is to determine the resulting potential in I and II. We make the ansatz that the potential in I is the sum of the potential generated by $q$ and an imaginary charge $q''$ in II, and that the potential in II is generated by a charge $q''$ at $\mathbf{r}_0$ possible different from $q$ due to ''screening''.

\begin{align}
\label{EQ:imdiel}     
&\phi^I(\mathbf{r}) = \frac{q}{4 \pi \epsilon_1 \epsilon_0} \frac{1}{|\mathbf{r} - \mathbf{r}_0|} + \frac{q'}{4 \pi \epsilon_1 \epsilon_0} \frac{1}{|\mathbf{r} + \mathbf{r}_0|}, \quad &\mbox{(In I.)} 
\\
\nonumber
&\phi^{II}(\mathbf{r}) = \frac{q''}{4 \pi \epsilon_2 \epsilon_0} \frac{1}{|\mathbf{r} - \mathbf{r}_0|}. \quad &\mbox{(In II.)}
\end{align}

We can determine the unknown charges $q'$ and $q''$ from the boundary conditions at $z = 0$ where we have the continuity of the tangent electric field, $\mathbf{E}^I_t = \mathbf{E}^{II}_t$, and the normal electric displacement, $\mathbf{D}^I_n = \mathbf{D}^{II}_n$ (since no free surface charges are expected for the dielectrics). Expressing these conditions in terms of the potential (\ref{EQ:imdiel}) we obtain the equations

\begin{align}
\label{EQ:dielq}
\frac{q}{\epsilon_1} + \frac{q'}{\epsilon_1} &= \frac{q''}{\epsilon_2},
\\
\nonumber
-q + q' &= -q''.
\end{align}

These equations have the solution

\begin{align}
\label{EQ:dielq1}
&q' = q \frac{\epsilon_1 - \epsilon_2}{\epsilon_1 + \epsilon_2},
\\
\nonumber
&q'' = q \frac{2\epsilon_2}{\epsilon_1 + \epsilon_2}.
\end{align}

The conducting plane corresponds to the limit $\epsilon_2 \rightarrow \infty$ and we see that in this case we indeed recover the solution $q' = -q$.

\section{Interference and diffraction}

\subsection{Geometric optics} 

As is well known, light, which is EM-wave of very short wavelengths (around 0.1--1 $\mu$m), can in many problems be treated as consisting of ''rays''. This is the method of \textit{geometric optics}. The plane waves discussed above do not exist in reality as they would be of infinite extent. But locally EM-waves may often quite well be approximated by plane waves. The geometric optical methods give good approximations far away from the source (transmitter) and when we consider spatial dimension large in comparison with the wavelength $\lambda$.

\begin{wrapfigure}{r}{50mm}
\begin{flushleft}
\setlength{\unitlength}{1mm}
\begin{picture}(50,50)(0,0)
\thicklines
\drawline(5,25)(45,25)
\put(45,25){\vector(1,0){0}}
\drawline(5,35)(30,45)
\put(30,45){\vector(2,1){0}}
\drawline(5,15)(30,5)
\put(30,5){\vector(2,-1){0}}
\put(-10,25){\arc{50}{5.64}{0.6435}}
\put(-10,25){\arc{72}{5.695}{0.588}}
\put(-10,25){\arc{98}{5.865}{0.4182}}
\put(9,5){\makebox(6,6)[l]{$S$}}
\end{picture}
\end{flushleft}
\label{FIGcphase}
\end{wrapfigure}

If we write $\psi$ to represent some component of the EM-field, then one may set

\begin{equation}
\label{EQ:psi}   
\psi(\mathbf{r},t) = A \cdot e^{iS(\mathbf{r},t)},
\end{equation}

where $A$ (amplitude) and $S$ (phase) are real functions. In dielectric medium  $\psi$ satisfies the wave equation

\begin{equation}
\label{EQ:psiwe}
\nabla^2 \psi - \frac{n^2}{c^2} \frac{\partial^2 \psi }{\partial t^2} = 0
\end{equation}

where $n = \sqrt{\epsilon \mu}$~ is the index of refraction.
   
A plane wave corresponds to the case where $A$ = constant and $S(\mathbf{r},t)$ = $\mathbf{k} \cdot \mathbf{r} -\omega t$. For a fixed time $t$ the sets of points $\mathbf{r}$ satisfying 

\[
S(\mathbf{r},t) = \text{constant}
\]

form surfaces (wavefronts) of constant \textit{phase}. If we draw the lines that are everywhere orthogonal to the wavefronts we obtain the \textit{rays}. The concept of rays is useful only when the characteristic dimensions of the regions considered are large in comparison with the wavelength $\lambda$. Then one make the assumption that the amplitude $A$ changes only a little over regions of the dimension $\lambda$. Mathematically this condition may be expressed as $ |{\nabla A}| \lambda \ll |A| $. If the phase locally is close to that of the plane wave we may assume that $ |{\nabla S}| \lambda \sim 2 \pi $. Using these assumptions one obtain from (\ref{EQ:psi}) the approximate equation

\begin{equation}
\label{EQ:eikon}
\left( \nabla S \right)^2 = \frac{n^2}{c^2} \, \left(\frac{\partial S }{\partial t}\right)^2 = n^2 k^2.
\quad \mbox{(Eikonal equation.)}
\end{equation}

The last equality follows if we assume an harmonic wave for which $S(\mathbf{r},t)$ is of the form $S_0(\mathbf{r}) - \omega t$. In (\ref{EQ:eikon}) $k = 2 \pi/\lambda$ is the magnitude of the wave vector in vacuum. The eikonal equation is equal to Fermat's principle\footnote{This follows from the fact the eikonal equation is analogous to the so called Hamilton-Jacobi equation of classical mechanics for a particle moving in a potential proportional to $-n(\mathbf{r})^2$. This in turn is related to the least action principle (Maupertuis) which finally leads to the Fermat's principle.} according to which the rays are paths $\Gamma$ which minimizes the traveling time defined by

\begin{equation}
\label{EQ:Fermat}
t_{\Gamma} = \int_{\Gamma} \frac{n ds}{c}. 
\end{equation}

\begin{wrapfigure}{r}{50mm}
\begin{flushleft}
\setlength{\unitlength}{1mm}
\begin{picture}(50,40)(0,0)
\thicklines
\drawline(2,10)(48,10)
\multiput(5,10)(40,0){2}{\drawline(0,0)(0,15)}
\dashline{3}(5,25)(45,25)
\dashline{3}(5,25)(25,10)(45,25)
\multiput(3,8)(2,0){22}{\drawline(0,0)(2,2)}
\put(25,25){\vector(1,0){1}}
\put(35,17.5){\vector(4,3){1}}
\put(24,26){\makebox(6,6)[l]{$r$}}
\put(4,26){\makebox(6,6)[l]{$T$}}
\put(44,26){\makebox(6,6)[l]{$R$}}
\put(6,14){\makebox(6,6)[l]{$h$}}
\put(15,10){\makebox(6,6)[l]{$\theta$}}
\end{picture}
\end{flushleft}
\label{FIGinterfer}
\end{wrapfigure}

From this it follows that rays are straight lines in homogeneous regions ($n$ independent of position). If we are far away from an antenna ($r \gg \lambda$) then we may think of the EM radiation as arriving from the antenna along rays. If the antenna is close (in relation to the distance to the observation point) to the ground, buildings etc, then, besides the contribution coming along the straight line between the antenna and the observation point, we may also have rays reflected from the ground, buildings etc, arriving at the observation point. Thus, the field at a point $\mathbf{r}$ may be a sum of contributions due to many paths,

\begin{equation}
\label{EQ:paths}
\psi(\mathbf{r}) = A_1 e^{iS_1(\mathbf{r})} + A_2 e^{iS_2(\mathbf{r})} + \cdots.
\end{equation}

For a flat ground this sum reduces to just two parts (''two-ray model''): the direct contribution  and the reflection from the ground. In case of reflexions the terms in (\ref{EQ:paths}) will be of the simple form $A \exp(i k s)$ where $s$ is the total path length.

Because the different paths may have different lengths $s_i$ the sum (\ref{EQ:paths}) may lead to either constructive or destructive interference. We have maximum destructive interference if the path difference is $\lambda/2 + n\lambda$ (phase difference 180$^\circ$) and maximum constructive interference if the path difference is $n\lambda$ (phase difference 0$^\circ$) in terms of the wave length $\lambda$. As a simple example we consider the interference  between the direct and reflected ray from a transmitter (T) to a receiver (R), both at the height $h$ above a perfectly conducting ground. When the distance $r$ between R and T is considerably larger than $h$ we may assume that $A_1 \sim A_2$ and the magnitude of the interference becomes proportional to (this corresponds to the case of vertically polarized EM waves)

\begin{equation} 
\label{EQ:interf}
\left| 
e^{i k r} + e^{i k 2 \sqrt{h^2 + r^2/4}} 
\right|^2
\approx 
\left|
1 + e^{i k h^2/r}
\right|^2 = 4 \left\{\cos \left(\frac{\pi h^2}{\lambda r} \right)
\right\}^2.
\end{equation}

This leads to a characteristic interference pattern with changing distance $r$ with a separation $\Delta r$ between the positions of maximum amplitude approximately given by

\[
\Delta r = \frac{2 \lambda r^2}{h^2}.
\]

\begin{wrapfigure}{r}{50mm}
\begin{flushleft}
\setlength{\unitlength}{1mm}
\begin{picture}(50,50)(0,0)
\thicklines
\put(0,5){\arc{60}{4.8}{0}}
\put(0,5){\arc{80}{4.9}{0}}
\put(28.53,14.27){\circle{20}}
\put(24.27,22.63){\circle{20}}
\put(17.63,29.27){\circle{20}}
\put(9.27,33.53){\circle{20}}
\put(31,7){\makebox(6,6)[l]{$S_1$}}
\put(41,7){\makebox(6,6)[l]{$S_2$}}
\end{picture}
\end{flushleft}
\label{FIGhuygens}
\end{wrapfigure}

For $r > 2 h^2/\lambda$ the cosines term approaches 1. With reference to the Fresnel equations (\ref{EQ:Fresnelv}), (\ref{EQ:Fresnelh}), we make the observation that, for a reflexion from a perfectly conducting surface, the reflexion coefficient $\rho_h$ is negative for the horizontal polarization case. This is also true for the vertical polarization reflexion coefficient $\rho_v$ in case of a dielectric ground when the distance $r$ (see Eq.~(\ref{EQ:BrewsterD})) is large enough to make $\theta$ smaller than the Brewster angle. In these cases the ''$+$''-sign in (\ref{EQ:interf}) must be replaced by a ''$-$''-sign, and the cosines term becomes instead a sinus-term, \[4 \left\{\sin \left(\frac{\pi h^2}{\lambda r} \right) \right\}^2.\] This has the interesting property of approaching \[4 \left(\frac{\pi h^2}{\lambda r} \right)^2\] as $r \rightarrow \infty$, meaning that the interference reduces the power (which is proportional to the square of the amplitude) with an additional $r^{-2}$-factor. Since the amplitude $A$ falls of as $r^{-1}$ the power will fall off as $A^2 r^{-2} \propto r^{-4}$ in this case. Thus the long-range behaviour is vastly different depending on whether we have a summation or a subtraction in (\ref{EQ:interf}).

One way to understand the propagation of the wavefronts is to imagine that every point on the wavefront is the source of an expanding spherical wavefront (a ''Huygens wavelet''), which together with the other spherical wavefronts form the new wavefront. This principle was advanced by Huygens (1678), and it gives a nice explanation of why reflection angle is equal the incidence angle, and for the law of Snellius (noting that the velocity of propagation of the wavefronts is $c/n$ where $n$ is the index of refraction). It also provides a picture of the diffraction of EM waves e.g.\ through a hole in a screen. 

\subsection{Fraunhofer and Fresnel diffractions}

Specifically consider a rectangular hole extending from $-a/2$ to $a/2$ in the $x$-direction and from $y$ = $-b/2$ to $y$ = $b/2$ in the $y$-direction of a an opaque screen. We suppose that a plane wave $\psi$ propagates in the $z$-direction and impinges on the hole at $z$ = 0. The field at the observation point $\mathbf{r}_0 = (z_0, y_0, x_0)$, will be obtained by summing the phase factors $\exp(iks)$ over rays from the surface of the hole to the observation point 

\begin{equation}
\label{EQ:diffh}
\psi(\mathbf{r}_0) = A \cdot \int_{-a/2}^{a/2} \int_{-b/2}^{b/2} e^{i k s(x,y)} dx dy.
\end{equation}

If we take $r$ to be the distance from the center of the hole to the observation $\mathbf{r}_0$ point then the path length $s(x,y)$ can be evaluated as

\begin{equation}
\label{EQ:paths2}
s = r \sqrt{1 + \left(\frac{x - x_0}{r}\right)^2 + \left(\frac{y - y_0}{r}\right)^2} \approx r + \frac{(x - x_0)^2}{2r} + \frac{(y - y_0)^2}{2r}.
\end{equation}

Here we have assumed that $r \gg |a|, |b|$. If $x_0 \gg a$ and $y_0 \gg b$ (Fraunhofer case) one can retain only the linear term in $(x - x_0)^2 = x^2 - 2x x_0 + x_0^2$ (and similarly for $y$) which simplifies the integrals (\ref{EQ:diffh}) to the form

\[
\int_{-a/2}^{a/2} \int_{-b/2}^{b/2} e^{- i k x_0 x/r - i k y_0 y/r } dx dy 
= \frac{2r}{k x_0} \sin \left( \frac{k x_0 a}{2r} \right)
\frac{2r}{k y_0} \sin \left( \frac{k y_0 a}{2r} \right).
\]

This means that the intensity at the point of observation will be proportional to the factor

\begin{equation}
\label{EQ:fhof}
\left(\frac{\sin(k a x_0/2r)}{k x_0/r}\right)^2
\left(\frac{\sin(k a y_0/2r)}{k y_0/r}\right)^2.
\end{equation}

\begin{wrapfigure}{r}{50mm}
\begin{flushleft}
\setlength{\unitlength}{1mm}
\begin{picture}(50,50)(0,0)
\thicklines
\multiput(5,5)(5,0){3}{\drawline(0,0)(0,40)}
\multiput(20,5)(0,25){2}{\shade\path(0,0)(2,0)(2,15)(0,15)(0,0)}
\multiput(22,20)(0,5){3}{\arc{10}{4.71}{1.57}}
\spline(22,10)(25,11)(30,15)(32,25)(30,35)(25,39)(22,40)
\dottedline(21,20)(21,30)
\multiput(5,10)(0,15){3}{\vector(1,0){8}}
\put(24,32){\vector(1,1){8}}
\put(24,18){\vector(1,-1){8}}
\end{picture}
\end{flushleft}
\label{FIGdiffr}
\end{wrapfigure}

Closer to the hole we may no longer have $x_0 \gg a$ and $y_0 \gg b$. In this case (Fresnel diffraction) we have to retain the full quadratic expression, which leads to integrals of the form

\[
\int_{-a/2}^{a/2} e^{ik (x - x_0)^2/2r} dx.
\]

This integral cannot be solved analytically but is easily evaluated numerically using computers. Traditionally the integral has been analyzed in terms of the Fresnel integrals

\begin{align*}
\label{EQ:fint}
\text{C}(\sigma) = \int_0^{\sigma} \cos(\pi s^2/2) ds ,
\\
\text{S}(\sigma) = \int_0^{\sigma} \sin(\pi s^2/2) ds,
\end{align*}

and the so called \textit{Cornu spiral} defined by $x = \text{C}(\sigma)$, $y = \text{S}(\sigma)$. 

We may consider the example of diffraction by an edge at $y$ = 0 in the $xy$-plane with the open space for $y$ > 0, and the screen in the half $y$ < 0. Suppose the observation point is at $\mathbf{r}_0 = (0,d,D_p)$; that is, $D_p$
is the distance from the $xy$-plane to the observation point. We have the geometric shadow for $d$ < 0, and the illuminated region for $d$ > 0.
Repeating the same arguments as above the calculation of the field at $\mathbf{r}_0$ leads to an expression involving the integral

\begin{equation}
\label{EQ:edge1}
\int_0^{\infty} e^{i k (y - d)^2/2D_p} dy = \sqrt{\frac{2 D_p}{k}}
\int_{-w}^{\infty} e^{i \eta^2} d\eta,
\end{equation}

where the \textit{diffraction parameter} $w$ is defined by

\begin{equation}
\label{EQ:edgew}
w = d \sqrt{\frac{k}{2 D_p}}.
\end{equation}

We consider the shadow region $d$ < 0. Then it is possible to estimate the integral (\ref{EQ:edge1}) by partial integration \cite[p.152]{Landau1975} -- doing it twice yields,

\begin{equation}
\label{EQ:edge2}
\int_{|w|}^{\infty} e^{i \eta^2} d\eta = e^{i w^2} 
\left\{
- \frac{1}{2 i |w|} + \frac{1}{4 |w|^3}
\right\}
- \frac{3}{4} \int_{|w|}^{\infty} \eta^{-4} e^{i \eta^2} d\eta
\end{equation}

For large $|w|$ the first term will dominate and the intensity of the field $|\psi|^2$ will therefore be proportional to $1/4w^{2}$. Comparing with the corresponding calculation for $d \rightarrow \infty$ (far into the illuminated region)\footnote{For this it is useful to know that 
\[
\int_{-\infty}^{\infty} \exp(i \eta^2) d\eta = \sqrt{\pi/2} \cdot (1 + i) .
\]
}, it follows that the intensity $P$ in the shadow region varies as

\begin{equation}
\label{EQ:edge3} 
P = \frac{P_0}{4 \pi w^2},
\end{equation}

where $P_0$ is the intensity in the illuminated region. We may apply this to an example discussed in \cite[p.132]{Bertoni2000} and treated by other methods there. Thus consider a transmitter at the height of 12 meter and a receiver 37.3 meters away at the height 2 m. Between them is a house with 12 m to the notch, which in turn is 20 m from the transmitter. For a rough estimate of the reduction of signal strength due to the building, which is assumed to act as a screen, we can use (\ref{EQ:edge3}) and (\ref{EQ:edgew}) with $D_p \sim$ 17 m and $d$ = -10 m. In the case of $f$ = 900 MHz we find the reduction to be about -28 dB, which can be compared with the value -25.8 dB given by Bertoni. From (\ref{EQ:edgew}) we observe that the boundary of the shadow region $w$ = constant is given by the lower part ($y$ < 0) of the parabola

\begin{equation}
\label{EQ:shad}
z = \frac{k}{w^2} y^2.
\end{equation}

Above we assumed that the waves from the transmitter are plane waves (parallel rays), but if we take into account that it is a finite distance $D_q$ from the screen we need only to modify the expression for diffraction parameter $w$ (\ref{EQ:edgew}) according to

\begin{equation}
\label{EQ:edgew2}
w = d \cdot \sqrt{\frac{k D_q}{2 D_p (D_p + D_q)}}.
\end{equation}

The Fresnel and Fraunhofer cases above be roughly distinguished by that the Fresnel approach deals with the situation where we have diverging rays while the Fraunhofer approximation treats the case with parallel rays. Since the Fraunhofer approximation uses linear terms in the argument of the $\exp$-function it lends itself readily to Fourier methods, and this has lead to the development of \textit{Fourier optics}. 

\subsection{Fresnel zones}

Consider a transmitter ($T$) and receiver ($R$) whose line of sight (LOS) distance is $D = d_1 + d_2$. Let $U$ be a point on the line of sight with $\overline{TU} = d_1$, $\overline{UR} = d_2$, and let $UV$ be perpendicular line to the line of sight with length $s$. Denoting $s_1 = \overline{TV}$ and $s_2 = \overline{VR}$ the difference between lengths of the paths $TR$ and $TVR$ becomes

\begin{equation}
\label{EQ:Fresnz1}
F = s_1 + s_2 - d_1 -d_2 = \sqrt{d_1^2 + d^2} + \sqrt{d_2^2 + d^2} - d_1 -d_2.
\end{equation}

The $n$:th \textit{Fresnel zone} is the region bounded by $F = n \lambda/2$ which forms an ellipsoid with the transmitter and receiver at the focuses. The first Fresnel zone $F_1$ thus corresponds to maximum phase shift of 180$^\circ$ due to a reflection. As a rule an undisturbed path between the transmitter and receiver requires that there be no obstacles in the Fresnel zone $F_1$. If $D \gg d$ then one can, for $d_1 = d_2 = D/2$, approximate (\ref{EQ:Fresnz1}) by

\[
F_n = \frac{2 d^2}{D},
\]

from which we obtain the width of the $n$:th Fresnel zone,

\[
r_n = d = \sqrt{\frac{D F_n}{2}} = \sqrt{\frac{D n \lambda}{4}}. 
\]

Inserting this into (\ref{EQ:edgew}) and (\ref{EQ:edge3}) we find that the reduction of the intensity of the field due to a shadow of depth $r_n$ would be
$1/\pi n$. Roughly, if the Fresnel zone $F_n$ is free then the transmission power is disturbed at most by a factor of ($1 - 1/\pi n$). In decibels $1/\pi$ corresponds to $10 \, \log(1/\pi) \approx$ -5 dB. The parameter

\[
\frac{d}{r_n}
\]

\begin{wrapfigure}{r}{60mm}
\begin{flushleft}
\setlength{\unitlength}{1mm}
\begin{picture}(60,35)(0,0)
\thicklines
\multiput(7.09,5)(45.83,0){2}{\drawline(0,0)(0,15) \put(0,15){\circle{1}}}
\put(30,20){\ellipse{50}{20}}
\drawline(7.09,20)(40.57,29.07)(52.91,20)
\drawline(40.57,20)(40.57,29.07)
\dottedline(0,20)(60,20)
\dottedline(0,5)(60,5)
\put(20,23){\makebox(6,6)[l]{$s_1$}}
\put(20,14){\makebox(6,6)[l]{$d_1$}}
\put(47,22){\makebox(6,6)[l]{$s_2$}}
\put(47,14){\makebox(6,6)[l]{$d_2$}}
\put(37,22){\makebox(6,6)[l]{$d$}}
\end{picture}
\end{flushleft}
\label{FIGFresnelz}
\end{wrapfigure}

where $d$ is shortest distance from LOS to obstructing objects is called the \textit{Fresnel zone clearance} and is used in investigating the path clearance. While objects protruding into the Fresnel zones cause scatterings,   
one draws a series of Fresnel zones also in order to see whether there are some surfaces close to the tangent of the Fresnel zones; such surfaces may cause reflexions and thus interference effects.

\subsection{Kirchoff equation}

Above we presented the Fresnel-Huygens principle in an intuitive fashion. It can also be justified starting from the wave-equation (\ref{EQ:psiwe}) which in the harmonic case $\psi(\mathbf{r}, t) = \psi(\mathbf{r}) \exp(-i \omega t)$ becomes,

\begin{equation}
\label{EQ:psik}
\nabla^2 \psi(\mathbf{r}) + k^2 \psi(\mathbf{r}) = 0. \quad (k = \omega/c.) 
\end{equation}

Suppose $\psi(\mathbf{r})$ is known on some surface $S_1$, the idea is to try to express the value of $\psi(\mathbf{r})$ at an observation point $\mathbf{r}_0$ in terms of its values at the surface $S_1$. First we note that the function (a Green function)

\begin{equation}
\label{EQ:Gfun}
G(\mathbf{r}) = -\frac{1}{4\pi r} e^{ikr}   
\end{equation}

satisfies

\begin{equation}
\label{EQ:Gequ}
\nabla^2 G(\mathbf{r}) + k^2 G(\mathbf{r}) = \delta(\mathbf{r}).
\end{equation}

From that follows (using Green's theorem)

\begin{align}
\label{EQ:Kirch1}
&\oint_{\partial V} (\psi(\mathbf{r}) \nabla G(\mathbf{r} - \mathbf{r}_0) -
G(\mathbf{r} - \mathbf{r}_0) \nabla \psi(\mathbf{r})) \cdot d\mathsf{S} =
\\
\nonumber
&\int_{V} (\psi(\mathbf{r}) \nabla^2 G(\mathbf{r} - \mathbf{r}_0) -
G(\mathbf{r} - \mathbf{r}_0) \nabla^2 \psi(\mathbf{r}))  d\mathsf{V} =
\\
\nonumber
&\int_{V} (\psi(\mathbf{r}) (\nabla^2 + k^2) G(\mathbf{r} - \mathbf{r}_0) -
G(\mathbf{r} - \mathbf{r}_0) (\nabla^2 + k^2) \psi(\mathbf{r})) d\mathsf{V} =
\\
\nonumber
&\int_{V} \psi(\mathbf{r}) \delta(\mathbf{r} - \mathbf{r}_0)  d\mathsf{V}
= \psi(\mathbf{r}_0).
\end{align}

We can take $V$ to be a volume enclosed by a screen $S_1$ (and some hole in it) and an infinitely far away surface $S_2$ where the field disappears. It may assumed that the field is zero at the screen and is approximately a plane wave at the hole area $A$, where we thus have $\nabla \psi = i \mathbf{k}_\text{in} \psi$. Hence (\ref{EQ:Kirch1}) becomes (Kirchoff equation, 1882)

\begin{equation}
\label{EQ:Kirch2}
\psi(\mathbf{r}_0) = \int_A \psi(\mathbf{r})( \nabla G(\mathbf{r} - \mathbf{r}_0) - i \mathbf{k}_\text{in} G(\mathbf{r} - \mathbf{r}_0)) \cdot d\mathsf{S}.
\end{equation}

Since 

\begin{equation*}
\nabla G(\mathbf{r} - \mathbf{r}_0) = G(\mathbf{r} - \mathbf{r}_0)
\left(
- \frac{\mathbf{r} - \mathbf{r}_0}{|\mathbf{r} - \mathbf{r}_0|^2} +
k\frac{\mathbf{r} - \mathbf{r}_0}{|\mathbf{r} - \mathbf{r}_0|} 
\right) \approx
G(\mathbf{r} - \mathbf{r}_0)
k\frac{\mathbf{r} - \mathbf{r}_0}{|\mathbf{r} - \mathbf{r}_0|}, 
\end{equation*}

when $k|\mathbf{r} - \mathbf{r}_0| \gg 1$ (that is, $|\mathbf{r} - \mathbf{r}_0| \gg \lambda$), we get finally,

\begin{equation}
\label{EQ:Kirch3}
\psi(\mathbf{r}_0) = -i \int_A \psi(\mathbf{r}) G(\mathbf{r} - \mathbf{r}_0)\left(k\frac{\mathbf{r} - \mathbf{r}_0}{|\mathbf{r} - \mathbf{r}_0|} - \mathbf{k}_\text{in}\right) \cdot d\mathsf{S}.
\end{equation}

Noting that the normal of the surface $A$ is pointing away from the point $\mathbf{r}_0$ we infer that

\begin{equation*}
\left(k\frac{\mathbf{r} - \mathbf{r}_0}{|\mathbf{r} - \mathbf{r}_0|} - \mathbf{k}_\text{in}\right) \cdot d\mathsf{S} = 
k (\cos \theta + \cos \theta_\text{in}) \; d \mathsf{S},
\end{equation*}

where $\theta$ is the angle between $\mathbf{r} - \mathbf{r}_0$ and the normal of the surface $d\mathsf{S}$ at the hole. The new thing obtained here beyond the Fresnel-Huygens principle used above is the geometrical factor $\cos\theta + \cos \theta_\text{in}$. The Kirchoff method gives often good approximations though it is not entirely mathematically consistent \cite[sect. 9.8]{Jackson1975}. Indeed, the assumption of the vanishing of $\psi$ and $\partial \psi/\partial n$ at the screen is no longer necessarily valid for the Kirchoff solution itself. Sommerfeld and Rayleigh have introduced slight modifications (employing the mirror principle) that remove these inconsistencies. There are also a few exact solutions of the díffraction problem -- the half plane screen and the wedge shaped screen -- presented by Arnold Sommerfeld and others, which are of great theoretical interest though approximate methods will typically suffice in practical problems.

\section{Radiation from antennas}

\subsection{The Hertz dipole}

If we consider the situation of an antenna in free space, then $\mathbf{J} \neq 0$ only in the region of the antenna (and the RF-circuit of course); elsewhere the magnetic fields satisfy the free space equations. It is useful to consider a simplified case where we have short piece $L$ of a wire as an antenna at $\mathbf{r}_0$, directed along the $z$-direction \cite{Hertz1889}. We suppose that the current density in this antenna may be described as

\begin{equation}
\label{EQ:Her}
\mathbf{J}(\mathbf{r}, t) = \delta (\mathbf{r} - \mathbf{r}_0) L \mathbf{I}_0 \cos \omega t ,
\end{equation}

using the Dirac (generalized) delta-function $\delta(\dots)$ in three dimensions, $\delta(\mathbf{r}) = \delta(x) \delta(y) \delta(z)$. $\mathbf{I}_0$ denotes the constant current amplitude directed along the $z$-direction. Inserting (\ref{EQ:Her}) into (\ref{EQ:Aso}) gives

\begin{equation}
\label{EQ:Her2}
\mathbf{A}(\mathbf{r}, t) = \frac{\mu \mu_0}{4 \pi} \frac{ L\mathbf{I}_0}{|\mathbf{r} -\mathbf{r}_0|} \cos \omega(t - |\mathbf{r} -\mathbf{r}_0|/c),
\end{equation}

from which we can calculate the magnetic field $\mathbf{H} = \nabla \times \mathbf{A}/ \mu \mu_0$,

\begin{align}
\label{EQ:Her3}
&\mathbf{H}(\mathbf{r}, t) = \frac{1}{4 \pi}
\nabla \left(
\frac{\cos \omega(t - |\mathbf{r} -\mathbf{r}_0|/c)}{|\mathbf{r} -\mathbf{r}_0|}
\right) \times (L\mathbf{I}_0) =
\\
\nonumber
&\left(
-\frac{\cos \omega(t - |\mathbf{r} -\mathbf{r}_0|/c)}{|\mathbf{r} -\mathbf{r}_0|^3} + \frac{\omega}{c}\frac{\sin \omega(t - |\mathbf{r} -\mathbf{r}_0|/c)}{|\mathbf{r} -\mathbf{r}_0|^2}
\right) (\mathbf{r} -\mathbf{r}_0) \times \frac{(L\mathbf{I}_0)}{4 \pi}.
\end{align}

If we use $\mathbf{R} \equiv \mathbf{r} -\mathbf{r}_0$, we can see that the $\cos$-term in (\ref{EQ:Her3}) on the right hand side falls off as $R^{-2}$ while the $\sin$-term falls off only as $R^{-1}$ with growing $R$. Thus, for $R^{-2} < (\omega/c) R^{-1}$; that is, for (the ''radiation field region'')

\begin{equation}
\label{EQ:Ff}
R > \frac{\lambda}{2 \pi}
\end{equation}
 
the second term (radiation field) will dominate giving\footnote{The near-field region is of interest when one considers cases of energy transfer not by radiation but through inductive couplings \cite{Kurs2007}.}

\begin{align}
\label{EQ:Her4}
\mathbf{H}(\mathbf{r}, t) \approx 
\frac{\omega}{c}\frac{\sin \omega(t - R/c)}{R^2}
\mathbf{R} \times \frac{L\mathbf{I}_0}{4 \pi}.
\end{align}

This can be interpreted as a planar field propagating in the direction $\mathbf{\hat{R}} = \mathbf{R}/R$ with an amplitude 

\begin{equation}
\label{EQ:Her5}
H_0 = \frac{\omega}{c} \frac{L I_0 \sin \theta}{4 \pi R},
\end{equation}

where $\theta$ is the angle between $\mathbf{\hat{R}}$ and $\mathbf{I}_0$ (= the $z$-direction). As the antenna is along the $\theta = 0$ direction we can see that there is no radiation in this direction.

The interpretation of our results is that the oscillating current in a short wire antenna generates a field which, far away form the source, approximates a plane wave with an amplitude which falls off with distance as $1/R$. Since the electric and magnetic fields are related by (\ref{EQ:Hpl}) (second equation), it follows that the electric field amplitude $\mathbf{E}_0$ will be orthogonal to both $\mathbf{H}_0$ and $\mathbf{\hat{R}}$; in fact, we have $\mathbf{E}_0 = \eta \mathbf{H}_0 \times \mathbf{\hat{R}}$.\footnote{One can work out the electrical field directly (see e.g. \cite[sec. II]{Jackson2006}) from the relation 

\[
\mathbf{E} = -\nabla \phi - \frac{\partial \mathbf{A}}{\partial t}
\]

by calculating $\phi$ from the charge distribution $\varrho$ which in turn is obtained from the continuity equation 

\[
\nabla \cdot \mathbf{J} + \frac{\partial \varrho}{\partial t} = 0.
\]

Another procedure is to first determine the magnetic field from $\mathbf{H} = \nabla \times \mathbf{A}/ \mu \mu_0$ and then the electric field from $\mathbf{E} = i \nabla \times \mathbf{H}/\omega \epsilon \epsilon_0$, which follows from the Maxwell equation $\nabla \times \mathbf{H} = \epsilon \epsilon_0 \partial \mathbf{E}/\partial t$ in the current-free region ($\mathbf{J} = 0$).}

Inserting (\ref{EQ:Her5}) into the expression for the Poynting vector $\mathbf{S}$ (\ref{EQ:Spl}), and integrating over the surface of a sphere of radius $R$, centered at the antenna ($\mathbf{r}_0$), we obtain,

\begin{equation}
\label{EQ:Her6}
P = \int_{|\mathbf{r}-\mathbf{r}_0| = R} \mathbf{S} \cdot d\mathsf{S} = \frac{\eta}{12 \pi} 
\left(
\frac{\omega L I_0}{c}
\right)^2,
\end{equation}

for the average (over time) power radiated by the antenna. If we compare this to the average power $R I_0^2/2$ dissipated by a resistor $R$ through which flows an oscillating current $I_0 \cos (\omega t)$, then (\ref{EQ:Her6}) suggests that the antenna can be associated with a \textit{radiation resistance}

\begin{equation}
\label{EQ:Res}
R_s = \frac{2 \eta}{12 \pi} 
\left(
\frac{\omega L} {c}
\right)^2 = \eta \frac{2 \pi}{3} 
\left(
\frac{L} {\lambda}
\right)^2 \approx 790 \left(
\frac{L} {\lambda}
\right)^2 \Omega.
\end{equation}

One implication of this analysis is that, if an AC-circuit contains wires whose lengths approaches the order of $\lambda = c/f$, where $f$ is the frequency of the AC-current, then one has to take into account that the wires may radiate a significant power and have to be treated as antennas. However, if the typical dimensions $D$ are much less than $\lambda$, then the system can be treated as an ''ordinary'' point-to-point circuit where Kirchoff's laws can still be applied \cite{Zozaya2007}.  

The above model based on (\ref{EQ:Her}), and the assumption $L \ll \lambda$, is referred to as the Hertz dipole. It already shows some general features. Thus, in the ''far away region'' the magnetic and electric field amplitudes fall off as $R^{-1}$, and since the radiated power per unit area is proportional to the square of the field amplitude, it falls off as $R^{-2}$. This is logical since the power flowing through the spherical surface area $4 \pi R^2$ must be constant (in empty space) and independent of $R$ since energy is conserved. We also found that the direction of the electrical field (direction of polarization) lies in the plane of the antenna and the radius vector $\mathbf{R}$; more exactly, $\mathbf{E} = E \, \mathbf{\hat{\theta}}$.

If the antenna is of the size of the wave length, $L \sim \lambda$, then the variation of the current along the antenna becomes an important problem. Basically one would have to solve the field equations for the propagation of the field along the antenna. The conductors act as waveguides directing the fields along the surface of the conductors. For good conductors there is no field inside the conductor and thus no energy is transported in or out the conductor. If there is slight resistance in the conductor then some of the energy transported by fields will flow \textit{into} the conductor and will be dissipated as heat. In this sense \textit{the antenna does not as such radiate, instead it guides the fields along its surface}. This is vivildly illustrated by the diagrams computed by Landstorfer and coworkers \cite{Landstorfer1972} which show how the electromagnetic energy flows along the surface of antennas and is spread into the surrounding space.

If an oscillating voltage source is connected to two parallel wires it will generate a traveling wave between the wires which will detach from the free ends of the wires and radiate into the surrounding space. By bending the wires at the free end forming a T-dipole the part of energy radiated into space may be increased. The problem of determining the radiating field thus becomes a boundary value problem, with excitation voltage given at the feedpoint, and with the boundary condition that the tangential electric field vanishes at the surface of the conductors. However, linear wire antennas are often treated by the simpler method of making some more or less well justified assumptions about the current distribution in the antenna, from which the fields are calculated as exemplified by the Hertz dipole case. The problem of determining the antenna current given the exciting voltage is called the \textit{antenna excitation problem}. Mathematically it leads to integral equations that are numerically solved using various discretization procedures, such as the moment method and the method of Galerkin \cite{Fikioris2001}. For a freely available software (\textsc{4Nec2}, ''numerical electromagnetic code'') for computing antenna fields and parameters see \cite{4nec2,Burke1981}.

\subsection{Dipole antennas}
\subsubsection{Antenna excitation}
\label{SEC:Antennaex}

The dipole antenna consists of two wires extending in opposite directions with the feeding point at the meeting ends. A simple dipole can be made from a coaxial cable by exposing a $\lambda/4$ end of the inner wire and make a sleeve of the outer conductor as shown in the adjoining figure. The added advantage of the sleeve is that it acts as a so called \textit{balun}. In order to study the current distribution in a common dipole antenna we assume that the wires are cylinders of radius $a \ll \lambda$, oriented along the $z$-direction, and separated by a small gap where an oscillating electrical field $E_z$ drives the antenna current. For a thin wire dipole-antenna it is typically assumed (based on the thin-wire approximation; see below) that the current distribution is sinusoidal along the center of the antenna ($x$ = 0, $y$ = 0),

\begin{equation}
\label{EQ:Dip1}
\mathbf{J}(z) = I_m \sin\left(\frac{kL}{2} - k|z|\right) \delta(x) \delta(y) \mathbf{\hat{z}}.  
\end{equation}

(Note that the feedpoint current amplitude $I(z = 0) = I_0$ is related to the maximum current amplitude $I_m$ by $I_0 = I_m \cdot \sin(kL/2)$.) Here the antenna extends from $z$ = -$L/2$ to $z$ = $L/2$, and (\ref{EQ:Dip1}) is consistent with the requirement that the current must vanish at the endpoints. (As usual $k$ denotes the wave-number $2 \pi/\lambda$.) The approximation (\ref{EQ:Dip1}) for the current however is inconsistent with the fact that the current is concentrated on the surface of the conductor. Still, for calculating fields far away from the antenna (in terms of the radius $a$ of the wire) the difference will be small.

\begin{wrapfigure}{r}{50mm}
\begin{flushleft}
\setlength{\unitlength}{1mm}
\begin{picture}(50,50)(0,0)
\thicklines
\put(17,10){\drawline(0,0)(0,20)}
\put(17,30){\drawline(0,0)(0,20)}
\put(17,10){\vector(0,-1){0}}
\put(17,30){\vector(0,1){0}}
\put(17,50){\vector(0,1){0}}
\put(17,30){\vector(0,-1){0}}
\put(20,18){\makebox(6,6)[l]{$\frac{\lambda}{4}$}}
\put(20,38){\makebox(6,6)[l]{$\frac{\lambda}{4}$}}
\multiput(32,0)(0,1){29}{\drawline(0,0)(2,2)}
\multiput(36,2)(0,1){29}{\drawline(2,-2)(0,0)}
\Thicklines
\put(30,0){\drawline(0,0)(0,28)(-2,28)(-2,10)(-4,10)(-4,30)(2,30)(2,0)}
\put(38,0){\drawline(0,0)(0,30)(6,30)(6,10)(4,10)(4,28)(2,28)(2,0)}
\put(34,0){\drawline(0,0)(0,50)(2,50)(2,0)}
\end{picture}
\end{flushleft}
\label{FIGDipSleeve}
\end{wrapfigure}

The mathematical form of the antenna excitation problem for wire antennas is usually developed as follows. Differentiating the second equation in (\ref{EQ:Vpo}) with respect to time, and using (\ref{EQ:Con}) to eliminate $\phi$ we obtain

\begin{equation}
\label{EQ:Dip2}
\nabla (\nabla \cdot \mathbf{A}) - \frac{1}{c^2} \frac{\partial^2 \mathbf{A}}{\partial t^2} = \frac{1}{c^2} \frac{\partial \mathbf{E}}{\partial t}.
\end{equation}

Because of the cylindrical symmetry we may assume that $\mathbf{A} = (0, 0, A_z)$. Secondly we express $\mathbf{A}$ as a function of the current $\mathbf{J}$ using (\ref{EQ:Aso}), leading to the equation 

\begin{align}
\label{EQ:Dip3}
\frac{\partial^2 A_z(z, t)}{\partial z^2} - \frac{1}{c^2} \frac{\partial^2 A_z(z,t)}{\partial t^2} = 
\\
\nonumber
\frac{\mu \mu_0}{4 \pi}
\int_{-L/2}^{L/2} 
\left\{ 
\frac{\partial^2} {\partial z^2} - \frac{1}{c^2} \frac{\partial^2}{\partial t^2}
\right\} 
\frac{ I(u, \bar{t})}{\sqrt{(u-z)^2 + a^2}} du = 
\\
\nonumber
\frac{1}{c^2} \frac{\partial E_z(z, t)}{\partial t} 
\end{align}

on the cylinder (wire) surface (to repeat this is strictly speaking inconsistent, see e.g. \cite{Jackson2006}). Finally, using the time-harmonic form 

\begin{align}
\label{EQ:Har}
I(u, \bar{t}) &= I(u) e^{-i\omega (t - R/c)} \quad (R = \sqrt{(u-z)^2 + a^2}\, ), 
\\
E_z(z, t) &=  E_z(z) e^{-i\omega t}, 
\end{align}

we end up with \textit{Pocklington equation} (which does not appear in this form in \cite{Pocklington1897})

\begin{align}
\label{EQ:Poc}
\int_{-L/2}^{L/2}  I(u) 
\left\{
\frac{\partial^2}{\partial z^2} + k^2
\right\}
\frac{e^{ik\sqrt{(u-z)^2 + a^2}}}{4 \pi \sqrt{(u-z)^2 + a^2}} du =
\\
\nonumber
- i \epsilon \epsilon_0 \omega E_z(z)
\end{align}

where we have used (\ref{EQ:c}) and the fact that $\omega / c = k$. This is an integral equation in the unknown current $I$ in terms of the excitation field $E_z$. Since it solves for the current $I$, given the input potential $V$, it gives the impedance $Z = V/I$ of the antenna. If the antenna is assumed to be a perfect conductor then $E_z = 0$ at the surface of the antenna since a non-zero tangential field would generate an infinite current (more precisely, in a perfect conductor it generates a field that cancels it such that the total tangential field becomes zero -- more on this in the section on shielding). $E_z$ differs from zero only at the feedgap where it may be assumed to be $-V/\Delta$, where $V$ is the voltage fed into the antenna, and $\Delta$ is the gap size. The same equation applies also to the receiving antenna in case which $E_z = -E_z^{(in)}$ (in the right hand side of Eq.(\ref{EQ:Poc})) at the surface of the conductor; i.e., the scattered field offsets the incoming field such that the total tangential component vanishes. Pocklington developed a somewhat more general theory of wire antennas in \cite{Pocklington1897} as he considered wire antennas of arbitrary shapes, such as circular rings and helixes. His basic idea was to consider the wire antenna as being composed of an array of infinitesimal Hertz dipoles. The total electric field can then be obtained by summing (or integrating) the contributions from the elementary Hertz dipoles, each characterized by a current $I(\mathbf{s})$ and direction $d \mathbf{s}$,

\begin{equation}
\label{EQ:Poc1}
\mathbf{E}(\mathbf{r}) = -\nabla \int \frac{\partial G(\mathbf{r} -\mathbf{s}(u))}{\partial u} I(u) du + k^2 \int I(u) G(\mathbf{r} -\mathbf{s}(u)) \frac{\partial \mathbf{s}}{\partial u} du,
\end{equation}

where

\begin{equation}
\label{EQ:Poc2}
G(\mathbf{r}) = \mu \mu_0 \frac{e^{ikr}}{4 \pi r}.
\end{equation}

The integration in Eq.(\ref{EQ:Poc1}), which can be derived from Eq.(\ref{EQ:Hzpi}) and Eq.(\ref{EQ:Hertzfree}), is along the wire antenna ($u$ is the length parameter along the curve defined by $du = |d\mathbf{s}|$). Finally Pocklington imposed the boundary condition that the tangential component of the electric field (\ref{EQ:Poc1}) vanishes at the surface of the conductor. In the limit of a vanishing small radius this leads to the equation\footnote{For closed antenna loops, or where current is assumed to be zero at the ends of the antenna, the first term in (\ref{EQ:Poc1}) becomes by partial integration

\[
\nabla \int G(\mathbf{r} - \mathbf{s}(u)) \frac{\partial I(u)}{ \partial u} du.
\]

Pocklington then evaluates (\ref{EQ:Poc1}) at surface of the conductor of a small radius $\delta$. For small $\delta$ the leading contribution to the field at a point on the surface is from the nearby section of the conductor yielding a term proportional to $\ln(M / \delta)$ (for some constant $M$). Indeed, note that $\int_0^C dz/\sqrt{z^2 + \delta^2} \rightarrow \ln(2C/\delta)$ for small $\delta$. The electric field parallel to the conductor at its surface thus becomes proportional to

\[
\left( \frac{\partial^2 I(u)}{\partial u^2}  + k^2 I(u)  \right) \ln(M/\delta)
\]   

when neglecting terms small compared with $\ln(M/\delta)$ as $\delta \rightarrow 0$. The vanishing of the tangential field component therefore leads in the limit of a thin wire to (\ref{EQ:Poc3}) excepting the end points and feedpoints of the antenna. It is to be emphasized that the above argument \textit{does not} demonstrate that sinusoidal currents do produce zero tangential fields for nonzero radius. In fact, for sinusoidal current the tangent field will in general look sinusoidal too.  
}

\begin{equation}
\label{EQ:Poc3}
\frac{\partial^2 I(u)}{\partial u^2} + k^2 I(u) = 0,
\end{equation}

i.e., the current is a sinusoidal function of the wire length parameter $u$. (This equation follows from the approximation of (\ref{EQ:Aso}), that $A_z$ on the conductor is proportional to the current at that point, in combination with Eq.(\ref{EQ:Hallen1}). For a pedagogic discussion of various approximation schemes for the antenna current see \cite{Mcdonald2007}.)

If we return to the version (\ref{EQ:Poc}) and let $\Delta \rightarrow 0$ we arrive at the  deltagap case where the feedgap is assumed to be infinitesimally small at $z = 0$. Specializing the earlier considerations of the dipole antenna to this case (usually associated with the name \textit{Hall\'en} \cite{Hallen1938}) we get the equation

\begin{equation}
\label{EQ:Hallen1}
\frac{\partial^2 A_z}{\partial z^2} +k^2 A_z = 0 \quad (z \neq 0),
\end{equation}

which suggests that $A_z$ is of the form

\begin{equation}
\label{EQ:Hallen2}
A_z = \frac{1}{c} \left(B \cos(kz) + C \sin(k|z|) \right),
\end{equation}

where $B$ and $C$ are constants to be determined. This form ensures the symmetry $A_z(z) = A_z(-z)$ and $I(z) = I(-z)$. When differentiating the $|z|$-argument this introduces a discontinuity at $z = 0$. In fact, using the equation

\begin{equation}
\label{EQ:Hallen3}
- \frac{i \omega}{c^2} E_z = \frac{\partial^2 A_z}{\partial z^2} +k^2 A_z,
\end{equation}
  
in order to compute $-V = \int E_z dz$ we obtain the value $2 C i$ which determines $C$ in terms of $V$. $B$ finally is determined by the condition that the current $I$, solved (numerically) from

\begin{equation}
\label{EQ:Hallen4}
A_z = \frac{1}{c} \left(B \cos(kz) + C \sin(k|z|) \right) = \mu \mu_0 \int I(z') \frac{e^{i k R}}{4 \pi R} dz', 
\end{equation}

should vanish at the ends of the antenna, $I(\pm L/2) = 0$. Having solved for $I(z)$ we obtain the impedance $Z$ of the antenna from $Z = V/I(0)$. (If we use the approximation that $A_z$ is proportional to the current $I$ at the wire surface, then the condition $I(\pm L/2) = 0$ would simply become $B \cos(kL/2) + C \sin(kL/2) = 0$, but such an approximation is too crude in order to yield a reliable estimate of the impedance of the antenna.)

In the integral in Eq.(\ref{EQ:Hallen4}) it is assumed that the current $I$ is at the axis of the antenna leading to the simplified ''kernel'' $e^{i k R} / 4 \pi R$ used above (called the \textit{reduced kernel approximation}), with $R$ denoting the distance from a surface point to a point on the axis (see Eq.(\ref{EQ:Har})). In the cylindrical antenna theory the reduced kernel

\[
\frac{e^{ik\sqrt{(u-z)^2 + a^2}}}{4 \pi \sqrt{(u-z)^2 + a^2}}
\]

in Eq.(\ref{EQ:Poc}) becomes replaced with

\begin{equation}
\label{EQ:kernelcyl}
K(u-z) = \frac{1}{2 \pi} \int_0^{2 \pi} \frac{e^{ik\sqrt{(u-z)^2 + 4 a^2 \sin^2 \frac{\varphi}{2} }}}{4 \pi \sqrt{(u-z)^2 + 4 a^2 \sin^2 \frac{\varphi}{2}}} d \varphi.
\end{equation}

This kernel is based on the replacing an axial current $I$ at the cross section $z$ with a surface current $I/2 \pi a$ uniformly distributed on the circumference at the section $z$. One sees that (\ref{EQ:Hallen4}) is, in fact, mathematically inconsistent if one uses the reduced kernel, in the sense that the left hand side is continuously differentiable at $z = 0$ (supposing the current $I$ is ''well behaved'') while the right hand side is not \cite{Fikioris2001}. The cylinder kernel fares better on this point since it has a singularity $R = 0$ (for the reduced kernel one has always $R \geq a$). Indeed, the kernel $K(z)$ grows like $-\ln(z)$ when $z$ goes to zero. However, integrals $\int K(z-u) I(u) du$ still remain finite because $\int \ln(z) dz = z \ln z - z$ stays finite when $z$ approaches 0. 

Returning to Eq.(\ref{EQ:Hallen4}) we can discretize it by dividing the antenna into $N = 2Q+1$ slices of length $\Delta = L/N$ with the feedgap at $z = 0$ (corresponding to the index $n = Q$). The midpoints of the slices are given by $z_n = (n-Q)\Delta$. A discretized version of (\ref{EQ:Hallen4}) can thus be written as the matrix equation

\begin{equation}
\label{EQ:Hallen5} 
V_n = \sum_{k = 0}^{N-1} M_{n,k} I_k
\end{equation}

where

\begin{align}
\label{EQ:Hallen6}
&M_{n,k} = \frac{\mu \mu_0}{4 \pi^2}
\int_0^{\pi} \int_{-\Delta/2}^{\Delta/2} \frac{e^{ik\sqrt{((n-k)\Delta-z)^2 + 4 a^2 \sin^2 \frac{\varphi}{2} }}}{\sqrt{((n-k)\Delta-z)^2 + 4 a^2 \sin^2 \frac{\varphi}{2}}} d\varphi dz,\\
&V_k = \frac{1}{c}\left(B \cos(k z_k) + C \sin(k |z_k|) \right),\\
&C = i\mathcal{U}/2 \quad (\mathcal{U} \, \text{is the applied feedgap voltage}).
\end{align}

\begin{figure}[H]
\centering
\includegraphics[width = 0.8\textwidth, angle = -90]{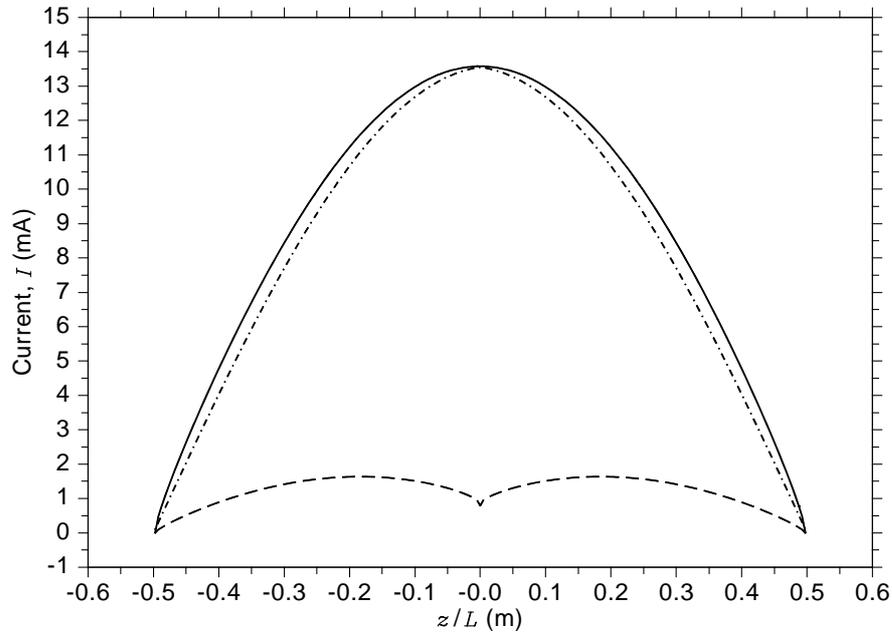}
\caption{The figure shows the distribution of the current along a dipole antenna, with an applied deltagap potential of 1 V at the center. The current has been calculated by solving the discretized Hall{\'e}n equation (\ref{EQ:Hallen5}). The solid line represents the real part of the current, the dashed line the imaginary part, while the dotted line represents a pure sinusoidal current. Parameters: $L = 0.48 \lambda$, antenna radius $a = 0.001 \lambda$; the calculated input impedance becomes $Z = 73.4 - i 4.3 \, \Omega$. 
Note that in the literature it is often the absolute magnitude of the current that is display instead of the real and imaginary part separately.}
\label{FIG5}
\end{figure}

The matrix equation $V = MI$ is easily implemented with mathematics software such as \textsc{Mathcad}, \textsc{Matlab}, etc. One solves for $I = M^{-1}V$ and adjusts the coefficient $B$ such that $I_0 = I_{N-1} = 0$. Writing the vector $V$ in (\ref{EQ:Hallen6}) as $V = (B/c) V^{(1)} + (C/c) V^{(2)}$ one obtains $B$ from $B (M^{-1} V^{(1)})_0 + C (M^{-1} V^{(2)})_0 = 0$ (which is thus generally a complex quantity). The adjoining figure (Fig.\ref{FIG5}) shows the results for imaginary and real parts of the current in the case $Q = 100$, $a = 0.001 \lambda$ and $L = 0.48 \lambda$ ($\lambda$ = 12.5 cm). As can be seen the calculated current deviates slightly from the sinusoidal current. The impedance was obtained as $Z = \mathcal{U}/I(z=0) = 73.4 - i4.3 \, \Omega$ whose real part may be compared to the radiation resistance (73 $\Omega$) for the half-wave dipole to be calculated below assuming a sinusoidal current. This result indicates that by making the dipole a bit shorter than $\lambda/2$ one can make the complex part (reactance) of the input impedance disappear, which simplifies the impedance matching of the antenna (such an antenna is called a ''resonant antenna'').

Another approach to solving the Hall{\'e}n equation (\ref{EQ:Hallen5}) is to express the current $I$ as trigonometric sum \cite{Neff1969}

\begin{equation}
\label{EQ:Hallen7}
I(z) = \sum_{n=1}^N D_n \sin \left(\frac{n \pi}{L}\left(\frac{L}{2} - |z|\right)\right).
\end{equation}   
  
The idea is thus to try to approximate the current using basis functions that are expected to be similar to the true current. We note that (\ref{EQ:Hallen7}) automatically satisfies the condition $I(\pm L/2) = 0$. If the expression (\ref{EQ:Hallen7}) is inserted into Eq.(\ref{EQ:Hallen4}) we obtain an equation of the form

\begin{equation}
\label{EQ:Hallen8}
\int_{-L/2}^{L/2} \sum_{n=1}^N D_n \sin \left(\frac{n \pi}{L}\left(\frac{L}{2} - |u|\right)\right) K(z-u) du = B \cos(kz) + C \sin(k|z|).
\end{equation}

Choosing $N$ + 1 points in the interval $(-L/2, L/2)$ we can write (\ref{EQ:Hallen8}) as a matrix equation

\begin{equation}
\label{EQ:Hallen9}
\sum_{n=1}^{N} K_{m,n} D_n + K_{m, N+1} B = C \sin(k |z_m|),
\end{equation}

where

\begin{align}
&K_{m,n} = \frac{\mu \mu_0 c}{4 \pi} \int_{-L/2}^{L/2} K(z_m - u) \sin \left(\frac{n \pi}{L}\left(\frac{L}{2} - |u|\right)\right) du \quad (n \leq N), \\
&K_{m, N+1} = - \cos(k z_m).
\end{align}
 
As before $C$ is related to the applied feedgap voltage $\mathcal{U}$ by $C = i\mathcal{U}/2$, and $K(z)$ denotes the kernel (\ref{EQ:kernelcyl}). When selecting the points $z_n$ one should take care not to make the $(N+1) \times (N+1)$ matrix $K$ singular. (This may happen if the points are selected symmetrically along the $z$-axis with respect to $z = 0$.) Using $N +1 = 6$ points we obtain a five term trigonometrical approximation of the current which hardly improves much going to higher values of $N$ -- the computational cost increases rapidly due to the integration over the interval $(-L/2, L/2)$.  
  

\subsubsection{Receiving dipole antenna}
\label{SEC:Receiving}

For simplicity assume that the incoming field $\mathbf{E}^{(in)}$ is parallel with the dipole antenna, chosen again to be along the $z$-axis. For a perfect conductor the incoming field will induce a current and a reflecting field $\mathbf{E}$ such that the tangential component of the total field $\mathbf{E} + \mathbf{E}^{(in)}$ is zero. From this we obtain that $E_z = - E^{(in)}_z$. For an incoming plane wave we can assume that $E^{(in)}_z$ is constant along the antenna.\footnote{In the more general case the last term in (\ref{EQ:Hallenrec}) is replaced by 
\[
\frac{ik}{\omega} \int_0^z E^{(in)}_z(\zeta) \sin(k(z-\zeta)) d\zeta.
\]

} We can therefore write the solution to Eq.(\ref{EQ:Hallen3}) as,

\begin{equation}
\label{EQ:Hallenrec}
A_z = \frac{1}{c} \left(B \cos(kz) + C \sin(k|z|) \right) + \frac{i}{\omega} E^{(in)}_z \quad ( \Delta/2 \leq |z| \leq L/2).
\end{equation}

Making use of the Lorenz gauge condition 

\[
\nabla \cdot \mathbf{A} + \frac{1}{c^2} \frac{\partial \phi}{\partial t} = 
\frac{\partial A_z}{\partial z} - \frac{i \omega}{c^2} \phi = 0,
\]

the potential $\phi$ can be expressed in terms of $A_z$,

\[
\phi = \frac{c^2}{i\omega} \frac{\partial A_z}{\partial z}.
\] 

For a thin wire dipole antenna we can use the ''Pocklington approximation'' and assume that $A_z$ is proportional to the current \cite{Mcdonald2007a}. In the open circuit case we have the condition that the current is zero at $|z| = \Delta/2, L/2$. The consequent condition $A_z(z) = 0$ for $|z| = \Delta/2, L/2$, then leads in the limit $\Delta = 0$ to the solution,

\begin{equation}
\label{EQ:BCrec}
B = \frac{i c}{\omega} E^{(in)}_z , \quad C = \frac{i c E^{(in)}_z}{\omega}
\cdot \frac{1 - \cos\left( kL/2 \right)}{\sin \left( kL/2\right) }.
\end{equation}

In the same limit $\Delta \rightarrow 0$ we then get for the induced open circuit voltage,

\begin{align}
\label{EQ:RecU}
\mathcal{U} = \phi(\Delta/2) - \phi(-\Delta/2) &= \frac{c^2}{i \omega} \left( \frac{\partial A_z(z=\Delta/2)}{\partial z} - \frac{\partial A_z(z=-\Delta/2)}{\partial z} \right) \\
\nonumber
&= \frac{2 E^{(in)}_z}{k} \cdot \frac{1 - \cos\left( kL/2 \right)}{\sin \left( kL/2\right) } .
\end{align}

Especially we obtain for the half-wave dipole ($L = \lambda/2$): $\mathcal{U} = E^{(in)}_z \lambda/\pi$.

\subsubsection{Dipole field}

In order to be able to derive some analytical results for the dipole antenna of length $L = 2 \, l$ we will assume a sinusoidal current (\ref{EQ:Dip1}) (which, as pointed out above, is typically sufficiently accurate for estimating the far field.). As in the case of the Hertz dipole we can then calculate the radiation field. For an antenna oriented in the $z$-direction the electric component of the radiation field (in complex representation) in the far-field region ($r > \lambda$, $r$ is the distance from the antenna and $\lambda$ is the wavelength) thus becomes

\begin{equation}
\label{EQ:Dipole}
\mathbf{E}(\mathbf{r}) = \mathbf{\hat{\theta}} 2 \eta I_m \frac{\cos(kl \cos \theta)- \cos(kl)}{\sin \theta} \cdot \frac{e^{ikr}}{4 \pi r}.
\end{equation}

Here $I_0$ is the amplitude of the sinusoidal antenna current, $\mathbf{\hat{\theta}}$ is the unit vector in the direction of the $\theta$-rotation, $k = 2 \pi/\lambda$ is the wave-number, and $\eta$ is the ''wave impedance''. In Eq.(\ref{EQ:Dipole}) we have ignored the effects of the surrounding, such as the circuit board (PCB) of the transmitter. In our measurements it was found that the PCB contributed to an anisotropic form of the radiation field in the horizontal plane which can be cured enclosing the PCB in a symmetrical (cylindrical) shielding box. Eq.(\ref{EQ:Dipole}) is a special case of the more general form 
 
\begin{equation}
\label{EQ:Efar}
\mathbf{E}(\mathbf{r}) = \frac{e^{ikr}}{r} \cdot \mathbf{G}(\theta, \phi)
\end{equation}

of the far-field electric component. Since the power flux of the radiation field is proportional to $E^2$, Eqs.(\ref{EQ:Dipole}, \ref{EQ:Efar}) predict that the power depends on the distance $r$ as $r^{-2}$. Power $P$ is usually measured in the units of decibel (dB) relative to a standard power $P_0$ (such as 100 mW),

\begin{equation}
\label{EQ:dB}
P_\text{dB} = 10 \cdot \log_{10} \left( \frac{P}{P_0} \right) .
\end{equation}

which is equivalent to the field strength.

Thus, if we have a $r^{-2}$-dependence of the power then it is reduced by 20 dB for every decade of distance and by ca 6 dB for every doubling of distance. From this relation one could deduce that, if the power at $r$ = 4 m is -10 dB, and we measure a power -40 dB at an unknown distance $r$, then $r$ must be 4 m $\times$ 10$^{(40-10)/20} \approx$ 126.5 m. Besides the distance $r$ the received signal strength depends also on the antenna orientations. 

\begin{wrapfigure}{r}{50mm}
\begin{flushleft}
\setlength{\unitlength}{1mm}
\begin{picture}(55,50)(0,0)
\Thicklines
\dottedline(10,25)(25,35)
\drawline(5,23)(10,23)(10,5)
\drawline(5,27)(10,27)(10,45)
\drawline(15,5)(15,23)
\put(15,5){\vector(0,-1){0}}
\put(15,23){\vector(0,1){0}}
\put(13,30){\makebox(6,6)[l]{$\theta$}}
\put(17,12){\makebox(6,6)[l]{$l$}}
\drawline(30,23)(35,23)(35,5)(40,5)(40,45)(35,45)(35,27)(30,27)
\end{picture}
\end{flushleft}
\label{FIGdipole}
\end{wrapfigure}

A common dipole antenna is the half-wave ($\lambda/2$) T-dipole with $l = \lambda/4$ and $kl = \pi/2$. In this case the field (\ref{EQ:Efar}) component $E_\theta$ becomes,

\begin{equation}
\label{EQ:Efar1}   
E_\theta = \frac{\eta I_m}{2 \pi r} \frac{\cos\left(\frac{\pi}{2} \cos \theta\right)}{\sin \theta} e^{ikr}.
\end{equation}

The total radiated power is obtained by integrating $E_\theta^2/2 \eta$ over a spherical surface $r$ = constant,

\begin{equation*}
\label{EQ:Ptotdip} 
P_{\text{tot}} = \frac{1}{2 \eta}  \left(\frac{\eta I_m}{2 \pi}\right)^2
4 \pi \int_0^{\pi/2}\left(\frac{\cos\left(\frac{\pi}{2} \cos \theta\right)}{\sin \theta}\right)^2 \sin \theta d\theta.
\end{equation*}

The integral cannot be solved analytically but can be evaluated numerically on the computer. The end result is that,

\begin{equation}
\label{EQ:Ptotdip1}
P_{\text{tot}} \approx 0.097 \cdot \eta \cdot I_m^2,
\end{equation}

and identifying this with $R_s I_0^2/2$ (note that for an $\lambda/2$-dipole $I_m = I_0$) we obtain for the radiation resistance $R_s$ of the $\lambda/2$-dipole, $R_s \approx 2 \cdot 0.097 \cdot \eta \approx 73 \,\Omega$. We can generalize these calculations using (\ref{EQ:Efar1}) to antennas of lengths $L_n = (2n -1) \lambda/2$ (the calculations are simplified for these special lengths) which yield for the corresponding antenna resistances the expression

\begin{align}
\label{EQ:Res2}
R_n = 30 \, \Omega \cdot \mathrm{Cin}((4n -2)\pi), \\
\intertext{where the cosine integral is defined by}
\mathrm{Cin}(x) = \int_0^x \frac{1 - \cos y}{y}dy.
\end{align}

A variant of the dipole antenna is the folded $\lambda/2$-dipole where the endpoints of the T-dipole are connected by a wire. Thus the current will counted twice when evaluating the radiation field which therefore will be twice compared to the T-dipole. The radiated power will consequently be four-fold. Hence the radiation resistance of the folded dipole is also four-fold, or about 4 $\times$ 73 $\Omega$ = 292 $\Omega$. 

\begin{wrapfigure}{r}{50mm}
\begin{flushleft}
\setlength{\unitlength}{1mm}
\begin{picture}(55,50)(0,0)
\Thicklines
\multiput(0,0)(25,0){2}{\drawline(10,23)(10,5)(15,5)(15,23)}
\multiput(0,0)(25,0){2}{\drawline(15,27)(15,45)(10,45)(10,27)}
\multiput(10,35)(5,0){2}{\vector(0,1){6}}
\put(35,35){\vector(0,1){6}}
\put(40,40){\vector(0,-1){6}}
\put(4,23){\makebox(6,6)[l]{$\frac{\mathcal{U}}{2}$}}
\put(4,35){\makebox(6,6)[l]{$I_s$}}
\put(4,5){\makebox(6,6)[l]{(a)}}

\put(17,23){\makebox(6,6)[l]{$\frac{\mathcal{U}}{2}$}}
\put(28,23){\makebox(6,6)[l]{$\frac{\mathcal{U}}{2}$}}
\put(28,35){\makebox(6,6)[l]{$I_a$}}
\put(28,5){\makebox(6,6)[l]{(b)}}
\put(41,23){\makebox(6,6)[l]{$-\frac{\mathcal{U}}{2}$}}
\end{picture}
\end{flushleft}
\label{FIGFold}
\end{wrapfigure}

In a bit more careful treatment of the folded dipole case \cite[section 9.5]{Balanis2005} the conventional analysis is based on the trick of viewing the system as consisting of two parallel coupled dipoles separated by a small distance $d$ (see figure). These are excited by voltages $\mathcal{U}/2$ and $\mathcal{U}/2$ (symmetrical case (a)), and $\mathcal{U}/2$ and $-\mathcal{U}/2$ (asymmetrical case (b)). The current $I$ of the original problem is then obtained as the sum of the currents $I_a$ and $I_b$ (at the feedpoints) for the separate cases. In case (a) we have the equation

\[
\mathcal{U}/2 = Z_{11} I_a + Z_{12} I_a,
\]

and assuming that for small separation $d$ the mutual impedance $Z_{12}$ is close to the self-impedance $Z = Z_{11}$ of the single dipole we obtain,

\[
I_a = \frac{\mathcal{U}}{4 Z}.
\]

The asymmetrical case (b) can be handled with the help of the transmission line theory (see Appendix \ref{SEC:Acab}). The input impedance of two-wire cable of length $L/2$ short circuited at the end is $Z_b = -i Z_c \tan(kL/2)$ where $Z_c$ is cable impedance. Thus the current $I_b$ becomes

\[
I_b = \frac{\mathcal{U}}{2} \div -i Z_c \tan(kL/2).
\]

The impedance of the $Z_f$ of the folded dipole is therefore given by

\[
\frac{1}{Z_f} = \frac{I_a+I_b}{\mathcal{U}} = \frac{1}{4Z} + \frac{i}{ 2 Z_c \tan(kL/2)}.
\]

When $L = \lambda/2$ we obtain that the folded dipole impedance $Z_f$ is four times the simple dipole impedance, $Z_f = 4 Z$. In this case only $I_a$ contributes to the total current since the short circuited $\lambda/4$-length transmission line has infinite impedance which suppresses the asymmetrical mode (b).

In the cases considered we have assumed that the feedpoint of the dipole antennas is centered at the midpoint (corresponding to $z = 0$). If the feedpoint is off-center by an amount $h$, then the current at the feedpoint -- assuming a sinusoidal current $I_m \sin(kL/2 - k|z|)$ for the antenna -- will be $I_m \sin(kL/2 - k|h|)$ which for a $\lambda/2$-dipole becomes $I_h = I_m \cos(kh)$. The radiated power can therefore be written $P = 1/2 \cdot \Re (Z I_m^2) = 1/2 \cdot \Re (Z/\cos(kh)^2) I_h^2$ implying that the off-center input impedance is $Z_h = Z/\cos(kh)^2$.

\subsection{Mutual impedance}

A current in a conductor $1$ generates an electric field which induces a current and its reaction field in any nearby conductor. This mutual influence is the basis of the use of antennas. The mutual influence is measured in terms of the mutual impedance. If we consider two (unloaded) dipole antennas and denote the voltages and the currents at the feedgaps (antenna terminals) as $V_i, I_i$, then we have the relations,

\begin{align}
\label{EQ:Imped1}
V_1 = Z_{11} I_1 + Z_{12} I_2,\\
V_2 = Z_{21} I_1 + Z_{22} I_2,
\end{align}

where $Z_{11}, Z_{22}$ are the antenna impedances, and $Z_{12} = Z_{21}$ is the mutual impedance (see Appendix and the Reciprocity Theorem). Following Bechmann \cite{Bechmann1931} one can compute the impedance for a set of $N$ conductors by starting from the Poynting's energy theorem (\ref{EQ:EJ}). Taking the time average for harmonic fields the term $\partial / \partial t (\dotsm) $ disappears, and we obtain using phasors,

\begin{equation}
\label{EQ:Imped2}
P_{av} = -\frac{1}{2} \int \Re \left( \mathbf{E} \cdot \mathbf{J}^\star  \right) dV.
\end{equation} 

Denote by $\mathbf{E}_i$ the electric field generated by the current density $\mathbf{J}_i$ of the $i$th conductor. Then the equation (\ref{EQ:Imped2}) becomes

\begin{equation}
\label{EQ:Imped3}
P_{av} = -\frac{1}{2} \sum_{i,j} \int \Re \left( \mathbf{E}_i \cdot \mathbf{J}_j^\star  \right) dV.
\end{equation} 
 
The term $\int \left( \mathbf{E}_i \cdot \mathbf{J}_j^\star + \mathbf{E}_j \cdot \mathbf{J}_i^\star \right) dV$ measures the mutual influence between conductors $i$ and $j$. Since $\mathbf{E}_i$ can be supposed to scale with the amplitude of the current density $\mathbf{J}_i$, due to the linearity of the Maxwell equations, one can define a current-independent quantity called (mutual) impedance by

\begin{equation}
\label{EQ:Imped4}
Z_{ij} = \frac{1}{2 I_i I_j}  \int \left( \mathbf{E}_i \cdot \mathbf{J}_j^\star + \mathbf{E}_j \cdot \mathbf{J}_i^\star \right) dV,
\end{equation} 

where $I_i$ denotes the characteristic current amplitude of the $i$th conductor (one may refer to the current amplitude at the antenna terminals, or the maximum current amplitude of the conductor).

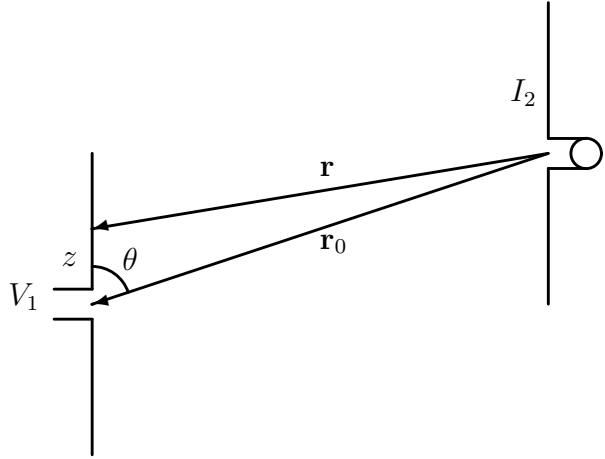
\begin{figure}
\centering
\setlength{\unitlength}{1mm}
\begin{picture}(100,70)(0,0)
\Thicklines
\put(15,27){\drawline(0,0)(5,0)(5,18)}
\put(15,23){\drawline(0,0)(5,0)(5,-18)}
\put(85,47){\drawline(0,0)(-5,0)(-5,18)}
\put(85,43){\drawline(0,0)(-5,0)(-5,-18)}
\drawline(80,45)(20,25)
\put(20,25){\vector(-4,-1){0}}
\drawline(80,45)(20,35)
\put(20,35){\vector(-4,-1){0}}
\put(85,45){\circle{4}}
\put(50,30){\makebox(6,6)[l]{$\mathbf{r}_0$}}
\put(50,40){\makebox(6,6)[l]{$\mathbf{r}$}}
\put(16,28){\makebox(6,6)[l]{$z$}}
\put(24,28){\makebox(6,6)[l]{$\theta$}}
\put(9,23){\makebox(6,6)[l]{$V_1$}}
\put(75,50){\makebox(6,6)[l]{$I_2$}}
\put(20,25){\arc{10}{-1.57}{-0.32}}
\end{picture}
\caption{Field by antenna 2 induces an open-circuit voltage $V_1$ over the terminals of antenna 1.}
\label{FIG2antennas}
\end{figure}

In the Appendix on the Reciprocity relations we consider the transmission between two antennas. If the antenna 2 is a transmitting dipole while antenna 1 is the receiving dipole then it was demonstrated that the open circuit voltage $V_1$ induced by the field generated by 2 is given by $V_1 = Z_{12} I_2$ (see Eq.(\ref{EQ:Volt1}) in the Appendix) 

\begin{equation}
\label{EQ:Imped5}
V_1 = \frac{1}{I_1} \int \mathbf{E}_2 \cdot \mathbf{J}_1 dV.
\end{equation}

This is typically calculated by inserting a sinusoidal current for $\mathbf{J}_1$, letting $\mathbf{E}_2$ be the incident field and integrating over the antenna.

In the situation considered by Bechmann $\mathbf{E}_2$ would refer to the incoming field and $\mathbf{J}_1$ to the induced current, and the total field (satisfying the BC) would be $\mathbf{E}_1 + \mathbf{E}_2$. Then the integral in Eq.(\ref{EQ:Imped5}) over the antenna can be interpreted as an electromotive force generated in the antenna.\footnote{Carter \cite{Carter1932} argues that the impressed field $E_z(z)$ at $z$ of the antenna induces a voltage $E_z(z) dz$ at the point $z$ over the section $dz$, and a current $I(z)$, and that $E_z(z) dz \, I(z)$ by reciprocity is equal to $I(0) dV$ where $dV$ is the corresponding induced voltage at the antenna terminal $z=0$. Summing over all sections gives the induced total voltage $V(0)$.

In \cite{Stratton1941I} the authors adopt the approach to let $\mathbf{E}_2$ be an arbitrarily prescribed incoming field which generates a current $\sigma \mathbf{E}_2$ in the antenna which is assumed to have finite conductivity $\sigma$. This current is then added to the Maxwell equations for the ''reaction'' fields $\mathbf{E}_1, \mathbf{H}_1$ in the antenna (conductor):
\begin{align*}
&\nabla \times \mathbf{E}_1 + i \mu \mu_0 \omega \mathbf{H}_1 = 0,\\
&\nabla \times \mathbf{H}_1 - \sigma \mathbf{E}_1 = \sigma \mathbf{E}_2.
\end{align*}

(In the second equation the displacement term has been dropped as negligible for a good conductor.) Combining this with the equations for the fields outside the conductor (free space or a dielectric) leads to a system of equations describing waves traveling along conductors which was treated by Sommerfeld already in 1899 (a more recent reference is \cite{Sommerfeld1949}).} This voltage can furthermore be expressed as $V_1 = \mathbf{h}_1(\theta) \cdot \mathbf{E}_2$ where $\mathbf{h}_1(\theta)$ defined by (for a linear antenna)

\begin{equation}
\label{EQ:Imped6}
\mathbf{h}_1(\theta) = \frac{1}{I_1} \int_{-L/2}^{L/2} e^{i k z \cos\theta} \mathbf{I}_1(z) dz
\end{equation}

is called the \textit{effective antenna length} ($\theta$ is the angle between the wave direction and the linear antenna -- see illustration). The exponential factor comes from the fact that the far field $\mathbf{E}_2(\mathbf{r})$ varies as $\exp(i \mathbf{k} \cdot \mathbf{r}) = \exp(i \mathbf{k} \cdot \mathbf{r}_0) \, \exp(i k z\cos\theta)$ along the receiving antenna. Again using sinusoidal currents as an approximation for a $\lambda/2$-dipole we obtain

\begin{align}
\label{EQ:Imped7} 
h(\theta) = \int_{-L/2}^{L/2} e^{i k z \cos\theta} \cos(kz) dz &= 
\frac{2}{k} \int_{0}^{\pi/2} \cos\left(u\cos\theta\right) \cos u \, du
\\ \nonumber
&= \frac{2}{k} \frac{\cos\left(\frac{\pi}{2}\cos\theta\right)}{(\sin\theta)^2}.
\end{align}

Hence for a $\lambda/2$-dipole at $\theta = \pi/2$ we obtain $h = \lambda/\pi$. This agrees with the result in section \ref{SEC:Receiving}. As discussed in an earlier example, a near planar EM-field with power flux of 100 mW/m$^2$ corresponds to an electric field amplitude of $E_0$ = 8.7 V/m. If this is aligned with a receiving $\lambda/2$-antenna (i.e. the antenna is parallel with the polarization), and given $\lambda$ = 0.125 m, the field will induce an open circuit voltage of amplitude $h \times 8.7$ V/m = $0.125/\pi \times 8.7$ V = 0.35 V between the antenna terminals. 

\subsection{Antenna above ground}

We have discussed the antenna radiation field in the open space. However, antennas are usually situated near the ground which will thus affect the radiation field. As shown in section \ref{SEC:image} the static field of a charge in presence of a dielectric or conducting plane can be handled using the image method; that is, the total field above the ground is the sum of the direct field and the one generated by an opposite ''mirror'' charge. An analogous situation prevails also in the case of moving charges in antennas. Thus, if the transmitter-antenna $T$ is situated at $\mathbf{r}$ = 0, and the ground is on the level $z = -h$, then the radiation field is the same as in the free space case with a second transmitter $T_M$ situated at $\mathbf{r} = (0, 0, -2h)$. $T_M$ is a mirror image of $T$ reflected in the plane $z = -h$ and is fed with the current $I_M = -I$. For a vertically oriented dipole antenna above the ground the resulting far field $\mathbf{E}_{\text{tot}}(\mathbf{r})$ becomes,

\begin{equation}
\label{EQ:Ground}
\mathbf{E}_{\text{tot}}(\mathbf{r}) = \mathbf{E}(\mathbf{r}) + \mathbf{E}(\mathbf{r - D}),
\end{equation}

where $\mathbf{D} = (0, 0, -2h)$ and $\mathbf{E}$ is the dipole field of Eq.(\ref{EQ:Dipole}). This equation for the total field is justified in the case of a perfectly conducting ground since it satisfy the condition of a zero tangential total field at the interface.

\begin{figure}[H]
\centering
\setlength{\unitlength}{1mm}
\begin{picture}(120,60)
\Thicklines
\drawline(10,30)(10,50)
\drawline(10,45)(7,50)
\drawline(10,45)(13,50)
\drawline(110,30)(110,50)
\drawline(110,45)(107,50)
\drawline(110,45)(113,50)
\drawline(5,30)(115,30)
\dashline{2}(10,10)(110,50)
\dashline{2}(10,50)(110,50)
\dottedline{1}(10,30)(10,10)
\dottedline{1}(10,15)(7,10)
\dottedline{1}(10,15)(13,10)
\put(15,40){\makebox(5,5)[l]{$T$}}
\put(15,15){\makebox(5,5)[l]{$T_M$}}
\put(102,40){\makebox(5,5)[l]{$R$}}
\put(65,25){\makebox(20,5)[l]{\text{Ground}}}
\put(73,30){\makebox(5,5)[l]{$\theta$}}
\multiput(5,28)(2,0){30}{\drawline(0,0)(2,2)}
\multiput(80,28)(2,0){17}{\drawline(0,0)(2,2)}
\end{picture}
\label{FIG00}
\end{figure}

\begin{figure}[htb]
\centering
\includegraphics[width = 0.8\textwidth, angle = -90]{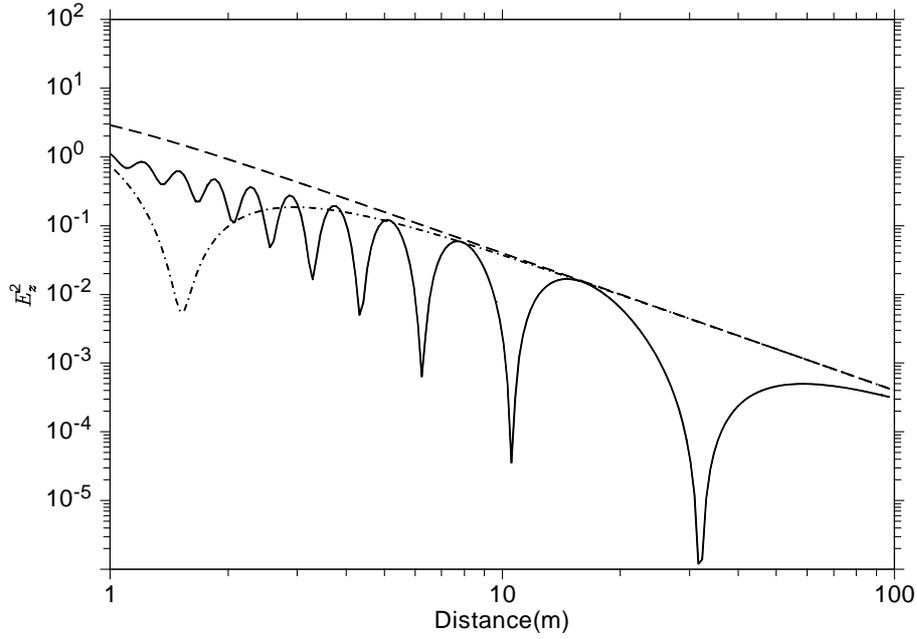}
\caption{Variation of power with distance due to interference calculated according to the model of Eq.(\ref{EQ:Ground}). Antennas (transmitter and receiver) are supposed to be vertically oriented. Solid line corresponds to antennas at the height $h$ = 1 m; dashed line to $h$ = 0.1 m; dotted line to $h_\text{sender}$ = 0.5 m and $h_\text{receiver}$ = 0.1 m.}
\label{FIG0}
\end{figure}

The field $\mathbf{E}_{\text{tot}}$ represents the sum of the direct field and the reflected field. One thus speaks of a ''two ray model''. If we use Eq.(\ref{EQ:Ground}) and Eq.(\ref{EQ:Dipole}) in order to calculate the power (which is set proportional to $E_z^2$) as function of the distance from the transmitter we obtain Fig.\ref{FIG0}. The solid curve represents the case when the receiver and the transmitter are at the height $h$ = 1 m above the ground. As can be seen the interference has a marked effect and the power does not decrease monotonically with distance as it would do in the free space. In contrast, the dashed curve representing the case $h$ = 0.1 m shows a monotonic fading with distance.

Our measurements in an outdoor soccer field revealed that the ground (sand, gravel) was best described as a dielectric with the relative permittivity $\epsilon \approx$ 3. In this case we need to employ the Fresnel relations for the reflected part of the horizontally (H) and vertically (V) polarized components of the field, (\ref{EQ:Fresnelv}), (\ref{EQ:Fresnelh}). Eq.(\ref{EQ:Ground}) has then to be modified by adding the reflexion factor,

\begin{equation}
\label{EQ:GroundR}
\mathbf{E}_{\text{tot}}(\mathbf{r}) = \mathbf{E}(\mathbf{r}) + \rho_v(\theta)\mathbf{E_v}(\mathbf{r - D}) + \rho_h(\theta)\mathbf{E_h}(\mathbf{r - D}),
\end{equation}

for the horizontally and vertically polarized components. (For vertical antennas $\mathbf{E_v} = \mathbf{E}$ .) Eq.(\ref{EQ:GroundR}) can be justified by the boundary conditions at the interface when the field can be assumed to be approximately planar. A.\ Sommerfeld has obtained an exact (for the idealized model) solution for the Hertz dipole above the ground \cite{Sommerfeld1909,Sommerfeld1947}. An alternative (but mathematically equivalent) approach has been presented by Weyl \cite{Weyl1919} where he decomposes the dipole field into a sum (integral) of planar waves. Thus one may apply the Fresnel relations to each planar wave component an then sum the components which yield the solution.

\subsubsection{Sommerfeld's analysis of the antenna above ground}
\label{SEC:Sommerfeld}

We consider a vertical Hertz dipole at a height $h$ above a planar ground characterized by parameters $\varepsilon$ and $\sigma$. Above ground the radiation field can be considered as a sum of the primary (open space solution) excitation plus a reflected part, while the field in the ground consists of the transmitted part. The task is to is to determine the reflected and transmitted fields by invoking the boundary conditions at the interface. The difference here compared to the earlier reflection problems is that we no longer assume that the primary field is a planar wave but is given by the full Hertz dipole field (in complex notion),

\begin{equation}
\label{EQ:Ahertz}
\mathbf{A}(\mathbf{r},t) = \mu \mu_0 l I_0 \frac{e^{i\omega(r/c - t)}}{4 \pi r},
\end{equation}

where $r = |\mathbf{r} - \mathbf{r}_0|$ is the distance from the antenna. The electric field is then given by $\mathbf{E} = - \nabla \phi - \partial \mathbf{A} / \partial t$. However, the $\phi$-part can be avoided if we use the Maxwell equations and the Lorenz gauge to derive the equation

\begin{equation}
\label{EQ:AEwav}
\nabla (\nabla \cdot \mathbf{A}) - \frac{1}{c^2} \frac{\partial^2 \mathbf{A}}{\partial t^2} = \frac{1}{c^2} \frac{\partial \mathbf{E}}{\partial t}.
\end{equation}

Then for harmonious waves we can replace $\partial / \partial t (\dotsm) $ by $-i \omega (\dotsm)$, and introducing the Hertz function $\vec{\Pi}$ through

\begin{equation}
\label{EQ:Hertzpi}
\mathbf{A} = \frac{1}{c^2} \frac{\partial \vec{\Pi}}{\partial t},
\end{equation}

we obtain

\begin{equation}
\label{EQ:Hzpi}
\nabla (\nabla \cdot \vec{\Pi}) + k^2 \vec{\Pi} 
 =  \mathbf{E}.
\end{equation}

(As usual $kc = \omega$.) In open space we have the Hertz dipole solution which we write simply as (the important part here is only how it depends on $r$)

\begin{equation}
\label{EQ:Hertzfree}
\vec{\Pi} = \hat{\mathbf{z}}\frac{e^{ikr}}{r},
\end{equation}

for a dipole oriented along the vertical direction $\hat{\mathbf{z}}$. It satisfies the free space wave equation

\begin{equation}
\label{EQ:Hertzfree_eq}
\nabla^2 \vec{\Pi} + k^2 \vec{\Pi} = 0.
\end{equation}

It will be assumed that the dipole is at $z = h$ while the surface of the ground corresponds to $z = 0$ (and air for $z > 0$). Eq.(\ref{EQ:Hertzfree}) represents the primary field $\vec{\Pi}_\mathrm{prim}$
excited by the antenna. With no ground this would constitute the total field. However, with the ground present the total field will become a sum 

\begin{equation}
\label{EQ:Hertztotal}
\vec{\Pi} = \vec{\Pi}_\mathrm{prim} + \vec{\Pi}_\mathrm{sec},
\end{equation}

where $\vec{\Pi}_\mathrm{sec}$ represent the secondary ''reaction'' field generated by interaction with the ground. In case of the perfectly conducting ground we already found the solution for $\vec{\Pi}_\mathrm{sec}$ through the image charge principle. Now we assume that the ground has finite conductance $\sigma$ and an index of refraction $n_E$ given by $n_E^2 = \epsilon + i \sigma/\omega \epsilon_0$. Earlier in section \ref{SEC:reflex} we discussed reflexion and transmission for planar fields and how the reflected/transmitted part was found by considering the boundary conditions of the field at the interfaces. The same applies here. For this end one may take advantage of the cylindrical symmetry of the situation and employ Bessel functions. In terms of cylindrical coordinates $(\rho, \varphi, z)$ the distance $r$ from the antenna is given by

\[
r^2 = \rho^2 + (z-h)^2.
\]

From the cylindrical symmetry it follows that $\vec{\Pi}_\mathrm{prim}$ and $\vec{\Pi}_\mathrm{sec}$ have components only along the $z$-direction for which we also write $\Pi_\mathrm{prim}$ and $\Pi_\mathrm{sec}$. Now the primary field can be expressed in terms of the Bessel function $J_0$ as an integral \cite[p. 243]{Sommerfeld1947} (known as \textit{Sommerfeld's equation})

\begin{equation}
\label{EQ:HertzBessel}
\Pi_\mathrm{prim}(\rho, z) = \frac{e^{ikr}}{r} = 
\int_0^{\infty} J_0(\rho \zeta) \, e^{-\gamma |z - h|} \, \frac{\zeta d\zeta}{\gamma},
\end{equation}

for the region $z > 0$, and where 

\[
\gamma = \sqrt{\zeta^2 - k^2}.
\] 

(The real part of $\gamma$ is taken to be positive in order to ensure the convergence of the integral.) This suggests the ansatz 

\begin{align}
\label{EQ:Hertzground}
\Pi_\mathrm{sec}(\rho, z) &=  
\int_0^{\infty} F(\zeta) J_0(\rho \zeta) \, e^{-\gamma(z + h)} \, d\zeta,
\\
\Pi_\mathrm{earth}(\rho, z) &=  
\int_0^{\infty} F_E(\zeta) J_0(\rho \zeta) \, e^{\gamma_E z - \gamma h} \, d\zeta,
\end{align}

for the secondary field and the transmitted field in the ground. Here 

\[
\gamma_E = \sqrt{\zeta^2 - k_E^2},
\]

with $k_E = n_E k$ (the term $\gamma h$ in the exponents is added only for convenience in dealing with the boundary conditions). $\Pi_\mathrm{earth}$ represents the field transmitted in the ground. The functions $F$ and $F_E$ will be determined by the boundary conditions for the electric and magnetic field at the interface $z = 0$. The conditions are that the tangential components $E_\rho$ and $H_\varphi$ shall be continuous across the boundary $z = 0$. These components are given in terms of $\Pi$ by,

\begin{align}
\label{EQ:Hertzbound1}
E_\rho &= \frac{\partial}{\partial z}\frac{\partial \Pi}{\partial \rho},
\\
H_\varphi &= \frac{-i k^2}{\omega \mu \mu_0} \frac{\partial \Pi}{\partial \varphi}, 
\end{align}

and similarly for the earth fields ($z < 0$). The boundary conditions at $z = 0$ can thus be written,

\begin{align}
\label{EQ:Hertzbound2}
\frac{\partial}{\partial z}\frac{\partial \Pi}{\partial \rho} &= \frac{\partial}{\partial z}\frac{\partial \Pi_\mathrm{earth}}{\partial \rho},
\\
k^2 \, \frac{\partial \Pi}{\partial \varphi} &= k_E^2 \, \frac{\partial \Pi_\mathrm{earth}}{\partial \varphi}.
\end{align}

(We have assumed that the magnetic permeability is $\mu$ = 1 also for earth.) The first condition can be integrated over $\rho$ leading to 

\[
\frac{\partial \Pi}{\partial z} = \frac{\partial \Pi_\mathrm{earth}}{\partial z}.
\]

Inserting the expressions for $\Pi = \Pi_\mathrm{prim} + \Pi_\mathrm{sec}$ and $\Pi_\mathrm{earth}$ one is finally led to the following equations for the functions $F$ and $F_E$,

\begin{align}
\label{EQ:HertzFresnel}
F &= \frac{\zeta}{\gamma} \left(1 - \frac{2 \gamma_E}{n_E^2 \gamma + \gamma_E} \right),
\\
F_E &= \frac{2 \zeta}{n_E^2 \gamma + \gamma_E}.
\end{align}

These functions are related to the Fresnel coefficients $\rho$ and $\tau$ Eq.(\ref{EQ:Fresnelv}), although the relation is not that transparent due to the integral form in the present case.\footnote{As pointed out above there is an approach by H Weyl \cite{Weyl1919} from 1919 (described in \cite[sec 9.29]{Stratton1941}) that may be more transparent in this regard, since it develops $\exp({ikr})/r$ in terms of plane waves starting from the identity $\exp(ikr)/ikr = \int_{i\infty}^1 \exp({ikr\eta}) d\eta$. } However, the solution for $z > 0$ can be written \cite[p.251]{Sommerfeld1947},

\begin{equation}
\label{EQ:Hertztot}
\Pi = \frac{e^{ikr}}{r} + \frac{e^{ikr^\star}}{r^\star} -
2 \int_0^\infty J_0 (\zeta \rho) e^{-\gamma (z + h)} \, \frac{\gamma_E}{n_E^2 \gamma + \gamma_E} \, \frac{\zeta d\zeta}{\gamma}.
\end{equation}
   
The second term corresponds to the reflected field in the case of a perfectly conducting ground where	

\[
r^\star = \sqrt{\rho^2 + (z + h)^2}.
\]

Indeed, we see that the integral in Eq.(\ref{EQ:Hertztot}) vanishes in the limit of $|n_E| \rightarrow \infty$ which corresponds to the case of a perfectly conducting ground. Hence the integral term describes the deviation from the case of a perfectly conducting ground.

\subsection{Breaking point, non-smooth surface}

In the case of a conductive medium the permittivity becomes a complex quantity, $\epsilon = \epsilon' + i \epsilon''$. For a typical ground the imaginary part is negligible at the frequencies considered here (2.4 GHz). Assuming a real permittivity there is a special grazing angle, the Brewster angle $\theta_B$, as explained above, at which the reflected vertical component disappears. If the transmitter and the receiver are at the heights $h_1$ and $h_2$ above the ground, then the Brewster angle corresponds to a distance 

\begin{equation}
\label{EQ:BrewsterD}
D_B = \frac{h_1 + h_2}{\tan \theta_B } = \left(h_1 + h_2 \right) \, \sqrt{\epsilon}.
\end{equation}

This distance is of interest since for $r > D_B$ the reflexion coefficient for the vertical (V) component turns negative and approaches the value -1. (This is in contrast to a perfectly conducting ground where the V-reflexion coefficient is always 1.) From this it follows that, for vertical antennas at large distances, $r \gg D_B$ and $r \gg 2 \pi h_1 h_1/\lambda$, the power varies as

\begin{equation}
P \propto \frac{(h_1 h_2 )^2}{r^4},
\end{equation}

where $h_1$ and $h_2$ are the heights of the antennas above the ground. The ''breaking point'' where the $r^{-4}$-dependence starts to dominate is defined as

\begin{equation}
r_{BP} = \frac{4 h_1 h_2}{\lambda}.	
\end{equation}

Interesting it seems that the early investigators of radio communication, such as Sommerfeld, did not pay much attention to the drastic interference effect of the dielectric ground resulting in the $r^{-4}$ power-distance relation.

The ''mirror'' (or two ray) model discussed above assumes a ''smooth'' ground. If we measure roughness $s$ of the surface as the typical height variation of the surface, then the Rayleigh criterion for smoothness requires that

\begin{equation}
\label{EQ:Rayleigh}
s < \frac{\lambda}{16 \sin\theta}, 
\end{equation}

for the grazing angle $\theta$. This corresponds to the idea that such a height variation causes a phase difference less than $2 s \sin \theta < \lambda/8$; that is, less than 45$^\circ$. Along similar lines of thought we can obtain an estimate how much a rough surface affects the reflection in terms of a scattering loss factor $f_s$. Suppose height variation $s$ around an average height $s$ = 0 is Gaussian distributed with standard deviation $\delta_s$. 
Consider the parallel reflected rays with the grazing angle $\theta$. Due to the height variation we get an extra phase factor $\exp(ik 2 s \sin \theta )$. Summing over all the rays in the $\theta$ direction the contributions have to be weighed by the Gaussian distribution, which leads to the integral,

\begin{equation}
\label{EQ:Scatter}
f_s = \frac{1}{\sqrt{2\pi} \delta_s}
\int_{-\infty}^{\infty} \exp\left(i k 2 s \sin \theta - \frac{s^2}{2\delta_s^2} \right) ds =
\exp\left\{ - 8 \left(\frac{\pi \delta_s \sin \theta}{\lambda} \right)^2  \right\}.
\end{equation}

The corrected reflexion coefficient is thus obtained as 
$f_s \, \rho$. Thus in the case the distance between the transmitter and the receiver is 10 m, the devices are at the height 1 m (corresponding to $\theta \approx 11.3^\circ$) and $\delta_s$ = 1 cm we get $f_s$ = 0.98, and when $\delta_s$ = 5 cm we get $f_s$ = 0.62 (@ 2.4 GHz). 

\subsection{Ground wave}

The two ray model seems to account rather well for the ground effects when comparing to measurements. It is based on calculating the reflection of rays, or planar waves, from the ground which usually are good approximations in the far field region. The mirror method gives the exact solution in case of a perfectly conducting ground but for a general dielectric ground the situation is different. Near the ground the propagation may be altered since, due to the air-ground interface, the EM-field might no longer be a pure transversal TEM field. This is in analogy to the waveguide case discussed earlier. The so called ground or surface wave (Oberfl\"{a}chenwelle) may become of interest in settings where the transceivers are directly on the ground. Sommerfeld in fact originally got interested in the problem of the field of the dipole near the ground in order to find out whether the dipole solution contained the surface wave conceived earlier by Zenneck \cite{Zenneck1907}. It may be however in the interest of clarity to treat the simpler case of Zenneck. We consider a plane wave that travels in the $x$-direction over a plane ground defined by $z$ = 0. Thus we have a ground for $z < 0$ and e.g. air for $z > 0$ with the interface at $z = 0$. We assume that electrical field is polarized along $z$. However from the discussion of the skin effect we already learned that the field becomes damped in a conducting media and that $E_z$ therefore also is a function of $z$ (besides of $x$), $E_z(z)$. Then from the $\nabla \cdot \mathbf{E} = \partial E_x/\partial x + \partial E_y/\partial y + \partial E_z/\partial z = 0$ it follows that $E_x$ and $E_y$ cannot both vanish because then we would have $\partial E_z/\partial z = 0$. We may thus assume that $\mathbf{E}$ is of the form $(E_x, 0, E_z)$ and $\mathbf{H}$ is of the form $(0, H_y, 0)$. From Maxwell equations we obtain the wave equations in the harmonic case,

\begin{align}
\label{EQ:Eza}
&\frac{\partial^2 E_z}{\partial x^2} + k^2 E_z + i \sigma \mu \mu_0 \omega E_z = \frac{\partial^2 E_x}{\partial z \partial x},\\
\label{EQ:Ezb}
&\frac{\partial^2 E_x}{\partial z^2} + k^2 E_x + i \sigma \mu \mu_0 \omega E_x = \frac{\partial^2 E_z}{\partial z \partial x},
\end{align}  

where $k^2 = \mu \mu_0 \epsilon \epsilon_0 \omega^2$. For air we set $\epsilon = \mu = 1$ and $\sigma = 0$, and for the ground $\epsilon = \epsilon_E, \mu = \mu_E = 1$ and $\sigma = \sigma_E$. From Eq.(\ref{EQ:Ezb}) we can again infer that $\partial E_z/\partial z \neq 0$ leads to a non vanishing $E_x$-component. For these linear equations we have solutions of the form (due to their couplings $E_x$ and $E_z$ must have the same exponential factor)

\begin{align*}
&E_x = A \exp(i k_x x + i k_z z),\\
&E_z = B \exp(i k_x x + i k_z z),
\end{align*}

which inserted into the equations (\ref{EQ:Eza}, \ref{EQ:Ezb}) yield the relations (and their duplicates for ground) 

\begin{align}
\label{EQ:Ein}
&k_x^2 + k_z^2 = \tilde{k}^2 = k^2 + i \sigma \mu \mu_0 \omega,\\
\nonumber
&A k_x + B k_z = 0.
\end{align} 

From the boundary conditions (continuity of the tangential component $E_x$, and the normal component $D_z$) at the air-ground interface $z = 0$ we further obtain,

\begin{align*}
&k_{xE} = k_x,\\
&A = A_E,\\
&B = \tilde{\epsilon}_E B_E,
\end{align*}

with

\[
\tilde{\epsilon}_E = \epsilon_E + i \frac{\sigma_E}{\omega \epsilon_0}.
\]

The index $E$ refers to earth/ground. Summing up we have, after setting $A = 1$, altogether 7 equations for the 7 unknowns $(k_x, k_y, k_{xE}, k_{yE}, A_E, B, B_E)$. From these relations we obtain,

\begin{align*}
&k_x = k \cdot \sqrt{\frac{\tilde{\epsilon}_E}{1+\tilde{\epsilon}_E}},\\
&k_z = \pm k \cdot \sqrt{\frac{1}{1+\tilde{\epsilon}_E}},\\
&k_{zE} = \tilde{\epsilon}_E \cdot k_z. 
\end{align*}

This is nothing but a solution of the Fresnel equations in case of zero reflexion (corresponding to the Brewster angle). Zero reflexion was here enforced by the initial assumption  Eq.(\ref{EQ:Ein}). The $k_x$ and $k_z, k_{zE}$ are in general complex. In order that the field stay finite as $z \rightarrow \pm \infty$ we have to choose the negative sign of the square root in the above equation. If $\Im(\tilde{\epsilon}_E) \ll 1$ then $\exp(i k_{x} x)$ results in a damping factor that can be written as $\exp\left( - x \cdot k \sqrt{1/2} \sigma / (2\epsilon_0 \omega) \right)$ which vanishes as $x \rightarrow +\infty$. These surface waves were studied by Cohn (1900), Uller (1903) and Zenneck (1907) \cite[sec. 9.10]{Stratton1941}. Sommerfeld also derived a surface wave from his vertical dipole solution discussed above in the limit $k_x r \rightarrow \infty$. In this limit he obtains ($z \geq 0$) for the Hertz function a term \cite[p.256]{Sommerfeld1947},

\begin{equation}
\label{EQ:Somlimit}
\Pi = \frac{A}{\sqrt{r}} \cdot e^{i k_x r + k_z z},
\end{equation}

where $k_x$ and $k_z$ are as above ($A$ is a constant). This describes a sort of a two-dimensional wave in the $z$-plane since it is proportional to $1/\sqrt{r}$ instead of $1/r$ as in the case of ''space waves''. Interestingly there has been an ongoing controversy about the significance and reality of the surface waves that has lasted for over a century \cite{Zenneck1907,Sommerfeld1909,Weyl1919,Norton1935,Norton1937,Kahan1950,Wait1998,Collin2004}. One of the famous ''legends'' promulgated as recently as 1998 in a review of the field by Wait \cite{Wait1998} is that Sommerfeld made a ''sign error'' in his pioneering paper \cite{Sommerfeld1909} with serious consequences for the interpretation of the results. Sommerfeld never admitted to any such ''errors'' (see for instance \cite{Sommerfeld1947} section 32 and ch. 23 in \cite{Frank1935}), and the recent study \cite{Collin2004} by Collins indeed reaffirms ''that the famous sign error is a myth''. Collin traces the myth back to a short paper \cite{Norton1935} by K A Norton in 1935 which asserts that there is a sign error in Sommerfeld's 1909 paper -- this allegation was later uncritically repeated by numerous authors (for instance in the book \cite{Stratton1941}) although the exact location of the ''error'' was never revealed. More importantly Norton did endorse the concept of the surface waves based on his ''corrected'' version of the theory. Nevertheless, the asymptotic evaluation of the Sommerfeld solution remains a somewhat tricky business. In fact, Kahan and Eckhart have presented a series of studies \cite{Kahan1950} where they argue that no surface waves are contained in the dipole solutions, contrary to the opinions of Sommerfeld, Norton, and others. They assert\footnote{They claim to have settled the issue ''in the present paper by proving in a quite general way that this surface wave cannot be included in the said dipole radiation and by pointing out a thus far hidden error in Sommerfeld's computation'' \cite[p.807, abstract]{Kahan1950}. However, the problematics of the asymptotic evaluation of the integral involved was early recognized by e.g. F Noether and V Fock and communicated to Sommerfeld.} that a careful evaluation of the Sommerfeld solutions shows that the surface term will be canceled by an equal term of opposite sign missed by Sommerfeld. Kahan and Eckhart thus agree with the 1919 analysis by Weyl. The interesting thing is that Sommerfeld's result agrees with Weyl's result \cite[p.937]{Frank1935}.

The starting point in Sommerfeld's approach (which no one diagrees with) is to replace $J_0(\zeta\rho)$ in Eq.(\ref{EQ:Hertztot}) by 

\[
J_0(\zeta\rho) = \frac{1}{2} \left( H^{(1)}_0(\zeta\rho) + H^{(2)}_0(\zeta\rho) \right) = \frac{1}{2} \left( H^{(1)}_0(\zeta\rho) - H^{(1)}_0(-\zeta\rho) \right).
\]

This makes it possible to rewrite the integral in Eq.(\ref{EQ:Hertztot}) as an integral extending from infinity to infinity,

\begin{equation}
\label{EQ:HankelHertz}
\Pi = \int_W H^{(1)}_0(\zeta\rho) e^{-\gamma |z|} \frac{n_E^2 \zeta d\zeta}{n_E^2 \gamma + \gamma_E}.
\end{equation}

The path $W$ extends from $-\infty + i\delta$ to $0 + i \delta$, and from $0 - i\delta$ to $\infty - i \delta$ where $\delta > 0$ is an infinitesimally small real quantity. This rule is enforced in order to ensure the correct evaluations of the square roots involved in the integrand. The tricky part is the asymptotic evaluation of the integral for $r \rightarrow \infty$. The reason is that the pole of the integral may be very close to one of the branching points which is given by $k$ and which will affect the steepest descent evaluation of the integral. This was the point overlooked by Sommerfeld in his earlier works according to \cite{Collin2004,Kahan1950} though Sommerfeld mentions the problem \cite[p.258]{Sommerfeld1947} later but refers for further details to a paper by H.\ Ott \cite{Ott1942}. In \cite[p.932]{Frank1935} he also acknowledges comments from F.\ Noether and V.\ Fock on the perils of the approximation procedure which in the end means that in practice the ''surface wave'' cannot be separated from the space wave (''Es d\"urfte keine Bedienungen geben, unter denen sich der Oberfl\"achenwellentypus P rein ausbildet und den Hauptbestandteil des Wellenkomplexes darstellt''). However if the ground is covered by a dielectric layer (such as ice over the sea) then there may indeed arise trapped surface waves in the layer as pointed out by Wait \cite{Wait1998b}\footnote{The question of the presence of surface waves may be compared to similar problems of identifying particle states and resonances in quantum mechanics; investigations which too involve scrutnizing the poles in evaluating the ''transfer functions'' of the $\psi$-wave.}. The branching points mentioned above arise because of the square root expressions when the integral is evaluated using the residue calculus. In this procedure the path of integration is closed by an infinite half circle in the upper half plane inside which there will be one pole and two branching points (at $k$ and $k_E$). Cutting up the plane along lines from $k$ and $k_E$ to $k + i \infty$ and $k_E + i \infty$ will make the integrand single-valued and the residue theorem becomes thus applicable. For further discussion of this topic see the references that have been listed above. Returning to the discussion of the Zenneck waves we may envisage that they can appear if a plane wave travelling along the $x$-axis hits a dielectric/conductor at $x$ = 0, which extends to $z < 0$ and $x > 0$. Part of the wave scatters but part of it will continue along $x > 0$ in the half-space $z > 0$ and is expected to approach the form of a Zenneck wave far from $x = 0$.

\newpage

\appendix

\section{Some definitions and results from vector analysis}

Vectors are denoted by $\mathbf{A} =(A_x, A_y, A_z) = A_x \hat{\mathbf{x}} + A_y \hat{\mathbf{y}} + A_z \hat{\mathbf{z}}$~ in a Cartesian coordinate system; $\mathbf{A} =(A_{\rho}, A_{\varphi}, A_z)$ in a cylindrical coordinate system; $\mathbf{A} =(A_{r}, A_{\theta}, A_{\varphi})$ in a spherical coordinate system. Note that the $\varphi$-component appears in different order in spherical and cylindrical coordinate systems though it has the same geometrical meaning in both systems.

{\allowdisplaybreaks
\begin{align}
\label{EQ:App1}
\tag{dot product}
&\mathbf{A} \cdot \mathbf{B} = A_x B_x + A_y B_y + A_z B_z
\\ 
\tag{cross product}
&\mathbf{A} \times \mathbf{B} = (A_y B_z - A_z B_y, A_z B_x - A_x B_z , A_x B_y - A_y B_x)
\\
\tag{Lagrange identity}
&\mathbf{A} \times (\mathbf{B} \times \mathbf{C}) =
(\mathbf{A} \cdot \mathbf{C}) \mathbf{B} -
(\mathbf{A} \cdot \mathbf{B}) \mathbf{C}
\\
\tag{gradient}
&\nabla \phi = 
\left(
\frac{\partial \phi}{\partial x} , \frac{\partial \phi}{\partial y} , \frac{\partial \phi}{\partial z} 
\right)
\\
\tag{--cyl. coord.}
&\nabla \phi = 
\left(
\frac{\partial \phi}{\partial \rho} ,\frac{1}{\rho} \frac{\partial \phi}{\partial \varphi} , \frac{\partial \phi}{\partial z} 
\right)
\\
\tag{--spher. coord.}
&\nabla \phi = 
\left(
\frac{\partial \phi}{\partial r} ,\frac{1}{r} \frac{\partial \phi}{\partial \theta} , \frac{1}{r \sin\theta}\frac{\partial \phi}{\partial \varphi} 
\right)
\\
\tag{divergence}
&\nabla \cdot \mathbf{A} = \frac{\partial A_x}{\partial x} + \frac{\partial A_y}{\partial y} +
\frac{\partial A_z}{\partial z} 
\\
\tag{--cyl. coord.}
&\nabla \cdot \mathbf{A} = \frac{1}{\rho}\frac{\partial (\rho A_{\rho})}{\partial \rho} + \frac{1}{\rho}\frac{\partial A_{\varphi}}{\partial \varphi} +
\frac{\partial A_z}{\partial z} 
\\
\tag{--spher. coord.}
&\nabla \cdot \mathbf{A} = \frac{1}{r^2}\frac{\partial (r^2 A_r)}{\partial r} + \frac{1}{r \sin\theta}\frac{\partial (\sin\theta A_{\theta})}{\partial \theta} +
\frac{1}{r \sin\theta}\frac{\partial A_{\varphi}}{\partial \varphi} 
\\
\tag{rotor}
&\nabla \times \mathbf{A} = 
\left(
\frac{\partial A_z}{\partial y} - \frac{\partial A_y}{\partial z} ,
\frac{\partial A_x}{\partial z} - \frac{\partial A_z}{\partial x} ,
\frac{\partial A_y}{\partial x} - \frac{\partial A_x}{\partial y}
\right) 
\\
\tag{--cyl. coord.}
&\nabla \times \mathbf{A} = 
\left(
\frac{1}{\rho}\frac{\partial A_z}{\partial \varphi} - \frac{\partial A_{\varphi}}{\partial z} ,
\frac{\partial A_{\rho}}{\partial z} - \frac{\partial A_z}{\partial \rho} ,
\frac{1}{\rho}\frac{\partial (\rho A_{\phi})}{\partial \rho} - \frac{1}{\rho}\frac{\partial A_{\rho}}{\partial \varphi}
\right) 
\\
\nonumber
&\nabla \times \mathbf{A} = 
\left(
\frac{1}{r \sin\theta}\left(\frac{\partial (\sin\theta A_{\varphi})}{\partial \theta} - \frac{\partial A_{\varphi}}{\partial z}\right) ,
\frac{1}{r \sin\theta}\frac{\partial A_r}{\partial \varphi} - \frac{1}{r}\frac{\partial (r A_{\varphi})}{\partial r} ,
\right.
\\
\tag{--spher. coord.}
&\left. \frac{1}{r}\left(\frac{\partial (r A_{\varphi})}{\partial r} - \frac{\partial A_{r}}{\partial \theta}\right)
\right) 
\\
\tag{Laplacian operator}
&\nabla^2 = \nabla \cdot \nabla = \frac{\partial^2}{\partial x^2} +
\frac{\partial^2}{\partial y^2} + \frac{\partial^2}{\partial z^2}
\\
\tag{wave operator}
&\square = \nabla^2 - \frac{1}{c^2} \frac{\partial^2}{\partial t^2} 
\\
\nonumber
&\nabla \times (\nabla \times \mathbf{A}) = - \nabla^2 \mathbf{A} + \nabla (\nabla \cdot \mathbf{A})
\\
\nonumber
&\nabla \cdot (\mathbf{A} \times \mathbf{B}) =
\mathbf{B} \cdot (\nabla \times \mathbf{A}) -
\mathbf{A} \cdot (\nabla \times \mathbf{B})
\\
\tag{Stokes-Green theorem}
&\oint_{\partial S} \mathbf{A} \cdot d\mathsf{s} = \int_S \nabla \times \mathbf{A} \cdot d\mathsf{S} 
\\
\tag{Gauss' theorem}
&\oint_{\partial V} \mathbf{A} \cdot d\mathsf{S} = \int_V \nabla \cdot \mathbf{A} \, d\mathsf{V} 
\\
\tag{Green's theorem}
&\oint_{\partial V} (F \nabla G - G \nabla F) \cdot d\mathsf{S} = \int_V (F \nabla^2 G - G \nabla^2 F) \, d\mathsf{V} 
\\
\nonumber
&\left( \nabla^2 + k^2 \right) \frac{e^{ikr}}{4 \pi r} = -\delta(\mathbf{r})
\\
\tag{Gauss integral}
&\int_{-\infty}^{\infty} e^{-a x^2} dx = \sqrt{\frac{\pi}{a}} \quad (\Re(a) \geq 0)
\end{align}

}

The scalar Laplacian $\nabla^2 \phi$ can be calculated for cylindrical and spherical coordinates by using the identity $\nabla^2 \phi = \nabla \cdot \left(\nabla \phi \right)$. For vector functions the Laplacian operator can be evaluated using the identity $\nabla^2 \mathbf{A} = \nabla \left( \nabla \cdot \mathbf{A} \right) - \nabla \times \left( \nabla \times \mathbf{A} \right)$. The Dirac delta\footnote{It became well known through Paul Dirac's \textit{Principles of Quantum Mechanics} (1930) but the concept had been introduced in electrical engineering by Oliver Heaviside (1850-1925) from where Dirac picked it up.} $\delta(x)$ is a so called functional defined by the property

\begin{equation}
\label{EQ:dirac}
\tag{Dirac delta.}
\int_\mathbf{R} f(x) \delta(x - x_0) dx = f(x_0),
\end{equation}

for any function $f$. The Fourier representation of the Dirac delta
can be expressed as

\begin{equation}
\label{EQ:DiracFou}
\delta(x) = \frac{1}{2 \pi} \int_\mathbf{R} e^{ikx} dk.
\end{equation}

We may extend the Dirac delta to vectors by defining (in Cartesian coordinates) $\delta(\mathbf{r}) = \delta(x) \delta(y) \delta(z)$.

In the text we have often used complex notation for the EM fields. Thus, an harmonic E-field is written\footnote{We here adhere to the physics tradition using the time factor $\exp(-i \omega t)$, while many engineering texts such as \cite{Balanis2005} assume a time dependence of the form $\exp(i \omega t)$ (also often denoting $\sqrt{-1}$ by $j$ instead of $i$).}

\begin{equation}
\label{EQ:Ecplx}
\mathbf{E}(\mathbf{r},t) = \mathbf{E}(\mathbf{r}) e^{-i \omega t}.
\end{equation}

The physical field will then corresponds to the real part of the ''phasor'' (\ref{EQ:Ecplx}), 

\[
\mathbf{E}_{\text{phys}}(\mathbf{r},t) = \Re (\mathbf{E}(\mathbf{r},t)).
\] 

When computing dot-products and cross-products for complex fields, we have to use complex conjugation in order to obtain the corresponding physical quantities. For instance, $\mathbf{E}_{\text{phys}} \cdot \mathbf{E}_{\text{phys}}$ is evaluated as

\[
\frac{1}{2} \mathbf{E} \cdot \mathbf{E}^\star.
\]

The factor 1/2 comes from the fact that a time average is implied (RMS-value), which for real fields amounts to the factor

\[
\frac{1}{T}\int_0^T (\cos(\omega t))^2 dt = \frac{1}{2}. \quad (T = 2\pi/\omega.)
\]   

Similarly the Poynting vector is expressed as

\[
\mathbf{S} = \frac{1}{2} \, \Re \left(\mathbf{E} \times \mathbf{H}^\star\right) ,
\]

and the power density $\mathcal{P} = \mathbf{E} \cdot \mathbf{J}$ becomes likewise in terms of phasor quantities

\[
\mathcal{P} = \frac{1}{2} \, \Re \left(\mathbf{E} \cdot \mathbf{J}^\star \right).
\]


\section{Tables}

\begin{table}[h]
	\centering
	\begin{tabular}{l|c|c|c|}
		\hline	
		Constant & Value & Unit & Legend\\
		\hline
		\hline
		$\epsilon_0$ & 8.854187817 $\cdot 10^{-12}$  & As/Vm & Permittivity of  vacuum\\
		$\mu_0$ & 4$\pi \cdot 10^{-7}$ & Vs/Am & Permeability of vacuum\\
		$\eta_0$ & 376.7303 & $\Omega$ & Wave-impedance of vacuum\\
		$c_0$ & 2.99792458 $\cdot 10^8$ & m/s & Velocity of light in vacuum\\
		$|e|$ & 1.60217733 $\cdot 10^{-19}$ & As & Electron charge\\
		\hline
		\end{tabular}
		\label{TAB:const}
		\caption{Constants}
\end{table}

\begin{table}[h]
	\centering
	\begin{tabular}{l|l|l|c|}
		\hline	
		Material & $\epsilon$ & $\sigma$ ($\Omega$ m)$^{-1}$ & $\mu$\\
		\hline
		\hline
		Concrete & 8--12  & $10^{-5}$ & 1\\
		Plasterboard & 14 & $10^{-7}$ & 1\\
		Glass & 8 & $10^{-11}$ & 1\\
		Dry brick & 4 & 0.01 -- 0.02 (@ 4.3 GHz) & 1\\
		Limestone & 7.5 & 0.03 & 1 \\
		Wood & 3  & $10^{-13}$--$10^{-4}$ & 1\\
		Douglas fir (plywood) & 1.82 & 0.049 (@ 3 GHz) & 1\\
		Soil (sand) & 3.4 & $10^{-5}$ & 1\\
		Water (sea) & 80 & 3--5 & 1\\
		Water (lake) & 80 & $10^{-3}$ (@ 3GHz) & 1\\
		Ice & 3.2 & 5 $\cdot 10^{-4}$ (@ 3GHz) & 1 \\
		Snow & 1.5 & $10^{-3}$ & 1\\  
		Air & 1 & 0 & 1\\
		Human tissue & 70 & 0.2 & 1\\
		Iron & NA & $10^{6}$ & 5000\\
		Copper & NA & 5.8 $\cdot 10^{7}$ & 1\\
		Aluminum & NA & 3.5 $\cdot 10^{7}$ & 1\\
		\hline
		\end{tabular}
		\label{TAB:material}
		\caption{Electric properties}
\end{table}

Here $c_0$ and $\mu_0$ are exact values. These and $\eta_0$ and are connected by 

\begin{equation*}
c_0 = \frac{1}{\sqrt{\epsilon_0 \mu_0}}, \quad \eta_0 = \sqrt{\frac{\mu_0}{\epsilon_0}},
\end{equation*}

from which we obtain 

\[
\eta_0 = c_0 \mu_0 \approx \pi 120 \; \Omega \approx 377 \; \Omega.
\]

Sometimes tables list the ''loss tangent'' defined by 

\[
\tan \delta = \frac{\epsilon''}{\epsilon'} = \frac{\sigma}{ \epsilon_0 \omega}.
\]

That is, $\epsilon'' = \epsilon' \tan \delta$. Electric properties depend generally on temperature and frequency. The complex permittivity of water, for instance, can be quite well represented for $f$ < 50 GHz and 20$^\circ$ C by the Debye model (quoted by \cite{Kristensson1999})

\[
\epsilon = \epsilon_{\infty} + \frac{\epsilon_s - \epsilon_{\infty}}{1 - i \omega \tau} + i \frac{\sigma}{\omega \epsilon_0},
\]

where $\epsilon_{\infty}$ = 5.27, $\epsilon_s$ = 80.0, $\tau$ = 10$^{-11}$ s, and the final term accounts for the effect of salt if present (salt water having a conductivity around 3 -- 5 $(\Omega \text{m})^{-1}$).

About wood note that is usually anisotropic due to the fibers. Fields polarized along the direction of the fibers pass more easily through the material \cite{Perkalkis1995}. For human tissue there has been reported \cite{Schwan1971} the following expression for (real) permittivity and conductivity, 

\begin{align}
\label{EQ:dielbody}
&\epsilon' = 5 + \frac{70 - V}{1 + (1.5/\lambda)^2},
\\
&\sigma = \sigma_0 + \frac{70-V}{60 \lambda} \frac{(1.5/\lambda)^2}{1 + (1.5/\lambda)^2},
\end{align}

where $V \sim 5$ (''volume fraction occupied by macromolecular components''), $\sigma_0 \approx$ 1/70 (cm $\Omega)^{-1}$ (observe the unit) and $\lambda$ is expressed in cm. These values are to be considered as averages since the properties vary with the type of organ and tissue (blood, bone, muscle, brain, etc). 

\section{Bessel functions}
\label{SEC:ABessel}

\subsection{Bessel $J$-functions}

While $\cos$- and $\sin$-functions are associated with rotational symmetry in the plane (the circle), Bessel functions are associated with rotational (cylindrical) symmetry around an axis in the 3-dimensional space. Bessel functions come typically into play when we have to solve the Laplace equation $\nabla^2 u = 0$ in space for a system with cylindrical symmetry. The Laplace operator is given in cylindrical coordinates as,

\begin{equation}
\label{EQ:LaplCyl}
\nabla^2 u = \frac{1}{\rho}\frac{\partial}{\partial \rho} \left(
\rho \frac{\partial u}{\partial \rho}\right) + \frac{1}{\rho^2} \frac{\partial^2 u}{\partial \varphi^2} + \frac{\partial^2 u}{\partial z^2}.
\end{equation}

Consider the \textit{eigenvalue equation} $\nabla^2 u = \alpha u$ in the case $\alpha$ = -1. Apparently $u = \exp(iy) = \exp(i\rho \sin\varphi)$ is a solution. We develop it as a series in $\exp(i \varphi)$,

\begin{equation}
\label{EQ:BesselDef} 
e^{i \rho \sin \varphi} = \sum_n J_n(\rho) e^{i n \varphi}
\end{equation}

which defines the Bessel functions $J_n$ of integer order ($n = 0, \pm 1, \pm 2, \dotsm$). Inserting the expression (\ref{EQ:LaplCyl}) into $\nabla^2 u + u = 0$ we find that the $J_n$ satisfies the equation,

\begin{equation}
\label{EQ:BesselEq1}
\frac{1}{\rho}\frac{\partial}{\partial \rho} \left(
\rho \frac{\partial J_n(\rho)}{\partial \rho}\right) + \left(1 - \frac{n^2}{\rho^2} \right) J_n(\rho) = 0.
\end{equation}

From Eq.(\ref{EQ:BesselDef}) it follows also that $J_n$ can be defined via the integral,

\begin{equation}
\label{EQ:BesselEq2} 
J_n(\rho) = \frac{1}{2 \pi} \int_0^{2 \pi} e^{i \rho \sin \varphi} e^{-i n \varphi} d \varphi,
\end{equation}

and that we have a series expansion (for $n \geq 0$),

\begin{equation}
\label{EQ:BesselEq3} 
J_n(\rho) = \left( \frac{\rho}{2} \right)^n 
\sum_{k = 0}^\infty \frac{1}{k! \, \Gamma(n + k + 1)} \left( -\frac{\rho^2}{4} \right)^k \quad (\rho > 0).
\end{equation}

We have in Eq.(\ref{EQ:BesselEq3}) employed the Gamma-function\footnote{A more general definition is (L Euler 1729)

\[
\Gamma(t) = \lim_{n \rightarrow \infty} \frac{n^t n!}{t(t+1) \dotsm (t+n)}
\]

which is valid for all points $t$ in the complex plane except for 0 and the negative integers (the poles).} $\Gamma(t) = \int_0^\infty s^{t-1} e^{-s} ds$ ($\Re (t) > 0$) which for integer values $m > 0$ satisfies $\Gamma (m) = (m-1)!$. Eq.(\ref{EQ:BesselEq3}) can in fact be extended by replacing $n$ by fractional values $\nu$ to Bessel functions $J_\nu$ of \textit{fractional order} which also satisfy the Bessel equation (\ref{EQ:BesselEq1}). As a historical note we may mention that Daniel Bernoulli studied the series expansion of what is now called $J_0$ already in 1738 in connection with the problem of determining the shape of an oscillating chain (\textit{catena}) supported from both ends \cite[sec. 1.2]{Watson1944}. Using the method of steepest descent or stationary phase \cite[sec. 16.E]{Sommerfeld1947}\cite[sec. VII.6]{Courant1993} one can show that the asymptotic value of $J_n (\rho)$ for $\rho \rightarrow \infty$ and fixed $n$ is given by

\begin{equation}
\label{EQ:BesselAsymp}
J_n(\rho) \approx \sqrt{\frac{2}{\pi \rho}} \cos\left(\rho - (2n + 1)\frac{\pi}{4}\right).
\end{equation}

Thus for large $\rho$ the function $J_n(\rho)$ behaves as the $\cos$-function. One also sees from this that $J_n(\rho)$ has an infinite number of zeros, as already pointed out by Daniel Bernoulli in case of $J_0$. Below we will also allow $J_n(z)$ to take complex valued arguments $z$. From small $\rho$ we have from Eq.(\ref{EQ:BesselEq3}) the asymptotic relation,

\begin{equation}
\label{EQ:BesselAsymp0}
J_{\nu}(\rho) \approx \left( \frac{\rho}{2} \right)^{\nu} \frac{1}{\Gamma(\nu + 1)} \quad (\nu \neq -1, -2, \cdots).
\end{equation}

Since $\exp(i \zeta y) = \exp(i \zeta \rho \sin \varphi) = \sum_n J_n(\zeta \rho) \exp(i n\varphi)$ it follows that $J_n(\zeta \rho)$ satisfies the equation,

\begin{equation}
\label{EQ:BesselEq4}
\frac{1}{\rho}\frac{\partial}{\partial \rho} \left(
\rho \frac{\partial J_n(\zeta \rho)}{\partial \rho}\right) + \left(\zeta^2 - \frac{n^2}{\rho^2} \right) J_n(\zeta \rho) = 0.
\end{equation}

Especially we have in the case $n = 0$,

\begin{equation}
\label{EQ:Bessel0}
\left( \frac{\partial^2}{\partial x^2} + \frac{\partial^2}{\partial y^2} + \zeta^2 \right) J_0(\zeta \rho) = 0.
\end{equation} 
 
This implies that for any function $F(\zeta)$

\begin{equation}
\label{EQ:BesselWav}
\Pi = \int_0^\infty F(\zeta) J_0(\zeta \rho) e^{\pm \sqrt{\zeta^2 - k^2} z} d\zeta
\end{equation}

satisfies the wave equation $\nabla^2 \Pi + k^2 \Pi = 0$. This fact was employed in section \ref{SEC:Sommerfeld} in order to write $\exp(ikr)/r$ on the form (\ref{EQ:BesselWav}). One makes the ansatz that there is a function $F$ such that (\ref{EQ:BesselWav}) is satisfied in case of $\Pi = \exp(ikr)/r$. One studies first the instance $z = 0$,

\begin{equation}
\label{EQ:BesselWav2}
\frac{e^{ik \rho}}{\rho} = \int_0^\infty F(\zeta) J_0(\zeta \rho) d\zeta.
\end{equation}

Using the general orthogonality property

\begin{equation}
\label{EQ:Jortho}
\int_0^\infty J_n(\zeta \rho) J_n(\tau \rho) \rho d\rho = \tau^{-1} \delta(\zeta - \tau),
\end{equation}

of the Bessel functions one can invert Eq.(\ref{EQ:BesselWav2}) obtaining \cite[p.243]{Sommerfeld1947},

\begin{equation}
\label{EQ:BesselWav3}
\zeta^{-1} F(\zeta) = \int_0^\infty e^{i k \rho} J_0(\zeta \rho) d\rho = \frac{1}{\sqrt{\zeta^2 - k^2}}.
\end{equation}

Once we have the result for $z = 0$ (\ref{EQ:BesselWav2}) we obtain the general solution by inserting the factor $\exp(\pm \sqrt{\zeta^2 - k^2} z )$ as in Eq.(\ref{EQ:HertzBessel}).

The orthogonality relation Eq.(\ref{EQ:Jortho}) can be demonstrated by a clever use of the identity

\[
	f(x,y) = \iint f(\xi, \eta) \delta(x- \xi)\delta(y- \eta) d\xi d\eta,
\] 

inserting the representation Eq.(\ref{EQ:DiracFou}) for the Dirac deltas and then going over to polar coordinates assuming the special form $f(r,\varphi) = g(r) \exp(i n \varphi)$ for the function $f$ \cite[sec. 21]{Sommerfeld1947}. That $\int_0^\infty J_n(\tau \rho) J_n(\zeta \rho) \rho d\rho = 0$ for $\tau \neq \zeta$ can be seen also directly by multiplying the Bessel differential equation for $J_n(\tau \rho)$ by $J_n(\zeta \rho)$ and vice versa and subtracting the expressions.

\subsection{Hankel functions}

As we found above the Bessel function $J_n$ resembles the $\cos$-function. It is often more useful to employ the complex $\exp$-function than the $\cos$-function, for instance when dealing with time-harmonic fields. It likewise exists a Bessel counterpart to the $\exp$-function that may be preferable to the $J$-functions in some circumstances. These are the so called Hankel functions $H^{(1)}_{\nu}$ and $H^{(2)}_{\nu}$. In order to arrive at these one may start by generalizing the integral representation (\ref{EQ:BesselEq3}) to one that covers also the case of non-integral orders (Sch\"{a}fli 1871),

\begin{equation}
\label{EQ:Schafli}
J_{\nu}(\rho) = \frac{1}{2 \pi} \int_{W_0} e^{i \rho \cos z + i \nu (z- \pi/2)} dz.
\end{equation} 

\begin{wrapfigure}{r}{50mm}
\begin{flushleft}
\setlength{\unitlength}{1mm}
\begin{picture}(50,50)(0,0)
\Thicklines
\drawline(6,50)(6,22)(44,22)(44,50)
\drawline(4,50)(4,18)(24,18)(24,0)
\drawline(46,50)(46,18)(26,18)(26,0)
\put(6,45){\vector(0,-1){4}}
\put(4,40){\vector(0,-1){4}}
\put(26,10){\vector(0,1){4}}
\put(24,10){\vector(0,-1){4}}
\put(7,40){\makebox(6,6)[l]{$W_0$}}
\put(17,5){\makebox(6,6)[l]{$W_1$}}
\put(28,5){\makebox(6,6)[l]{$W_2$}}
\put(2,11){\makebox(6,6)[l]{-$\frac{\pi}{2}$}}
\put(44,11){\makebox(6,6)[l]{$\frac{3\pi}{2}$}}
\thicklines
\drawline(0,20)(50,20)
\drawline(25,0)(25,50)
\multiput(5,20)(10,0){5}{\drawline(0,-1)(0,1)}
\end{picture}
\end{flushleft}
\label{FIGHankel}
\end{wrapfigure}

The integration is along an infinite path $W_0$ in the complex $z$-plane, chosen such as to make the integral convergent for real positive $\rho$. The standard choice for $W_0$ is the path $-\frac{\pi}{2} + i \infty \rightarrow -\frac{\pi}{2} + i 0 \rightarrow \frac{3\pi}{2} + i0 \rightarrow \frac{3\pi}{2} + i\infty$. In case of an integer order $\nu= n$ the expression (\ref{EQ:Schafli}) reduces to the earlier one (\ref{EQ:BesselEq3}), since the integrals along the vertical sections of the path $W_0$ cancel each other in this case. In the adjoining figure we introduced two other infinite paths $W_1$ and $W_2$ which also define two convergent integrals (introduced by Sommerfeld in 1896) and the corresponding functions called the first and second \textit{Hankel functions},

\begin{align}
\label{EQ:Hankel1}
&H^{(1)}_{\nu}(\rho) = \frac{1}{\pi} \int_{W_1} e^{i \rho \cos z + i\nu (z - \pi/2)} dz , \\
\nonumber 
&H^{(2)}_{\nu}(\rho) = \frac{1}{\pi} \int_{W_2} e^{i \rho \cos z + i\nu (z - \pi/2)} dz.
\end{align}

Now we have formally $W_0 = W_1 + W_2$ since the integrals along the negative imaginary axis cancel each other. We therefore obtain,

\begin{equation}
\label{EQ:Hankel2}
J_{\nu}(\rho) = \frac{1}{2} \left(H^{(1)}_{\nu}(\rho) + H^{(2)}_{\nu}(\rho) \right).
\end{equation}  

This can be compared to $\cos\varphi = (\exp(i\varphi)+\exp(-i\varphi))/2$. Indeed the asymptotic form as $\rho \rightarrow \infty$ is given by

\begin{align}
\label{EQ:Hankel3}
&H^{(1)}_{\nu}(\rho) \approx \sqrt{\frac{2}{\rho \pi}} \cdot e^{i(\rho - (\nu + 1/2)\pi/2)},\\
\nonumber
&H^{(2)}_{\nu}(\rho) \approx \sqrt{\frac{2}{\rho \pi}} \cdot e^{-i(\rho - (\nu + 1/2)\pi/2)}.
\end{align}

For $\rho \rightarrow 0$ we have instead

\begin{align}
\label{EQ:Hankel0}
&H^{(2)}_0(\rho) \approx -i \frac{2}{\pi} \ln\rho, \\
\nonumber
&H^{(2)}_{\nu}(\rho) \approx i \frac{\Gamma(\nu)}{\pi} \left(\frac{\rho}{2}\right)^{-\nu} \quad (\Re\nu > 0).
\end{align}

One property following from the definitions (\ref{EQ:Hankel3}) is that for real $\rho$ and $\nu$

\begin{equation}
\label{EQ:Hankel4} 
H^{(1)}_{\nu}(\rho) = \overline{H^{(2)}_{\nu}(\rho)},
\end{equation}

where the bar denotes complex conjugation. The Bessel function corresponding to the $\sin$-function is called the \textit{Neumann function} and is defined by

\begin{equation}
\label{EQ:Neumann}
N_{\nu}(\rho) = \frac{1}{2i}\left(H^{(1)}_{\nu}(\rho) - H^{(2)}_{\nu}(\rho) \right).
\end{equation} 

The Hankel functions play an important role in Sommerfeld's theory of the dipole over ground and e.g. in the treatment of EM-fields along circular conductors. Using the decomposition (\ref{EQ:Hankel2}) of the Bessel function $J_0$ in Eq.(\ref{EQ:Hertztot}) one can transform the $\int_0^{\infty}$-integral into an $\int_{-\infty}^{\infty}$-integral. 

\begin{wrapfigure}{r}{60mm}
\begin{flushleft}
\setlength{\unitlength}{1mm}
\begin{picture}(60,50)(0,0)
\thicklines
\multiput(0,0)(10,0){7}{\drawline(0,0)(0,40)}
\multiput(0,20)(20,0){3}{\multiput(0,0)(0,1){20}{\drawline(0,0)(10,0)}}
\multiput(10,0)(20,0){3}{\multiput(0,0)(0,1){20}{\drawline(0,0)(10,0)}}
\put(1,15){\makebox(6,6)[l]{$-3\pi$}}
\put(21,15){\makebox(6,6)[l]{$-\pi$}}
\put(41,15){\makebox(6,6)[l]{$\pi$}}
\put(11,20){\makebox(6,6)[l]{$-2\pi$}}
\put(31,20){\makebox(6,6)[l]{0}}
\put(51,20){\makebox(6,6)[l]{$2\pi$}}
\put(40,40){\makebox(6,6)[l]{$\widetilde{W}_2$}}
\drawline(0,20)(60,20)
\Thicklines
\spline(45,40)(44,30)(40,20)(36,20)(35,0)
\end{picture}
\end{flushleft}
\label{FIGHankel2}
\end{wrapfigure}

This makes it possible to apply the calculus of residues and evaluate part of the integral in terms of a pole of the integrand. An important step is the relation

\begin{align}
\label{EQ:Hankel5}
&H^{(2)}_0(-\rho) = - H^{(1)}_0(\rho), \\
\nonumber
&H^{(1)}_0(-\rho) = - H^{(2)}_0(\rho),
\end{align} 

which is a special case of the \textit{Umlaufrelationen} for the Hankel functions \cite[p.314]{Sommerfeld1947}. Note that the integration paths used in order to define Bessel functions need to stay in the limits of $|z| \rightarrow \infty$ in the checker board patterns shown in the adjoining figure in order to ensure convergence for $\rho > 0$; for $\rho < 0$ the checker pattern is shifted by $\pi$ along the real axis). The paths may though be deformed as long as they do not cross borders and still define the same functions. Thus $\widetilde{W}_2$ is an allowed version of $W_2$ described earlier. Now $H^{(1)}_0(-\rho)$ ($\rho > 0$) is defined using the path $W_1$ translated by $\pi$ along the real axis, which we denote $W_1 +\pi$. Since $-\rho \cos z = \rho \cos(z \pm \pi)$ we get $H^{(1)}_0(-\rho) = (1/\pi) \int_W \exp(i \rho \cos(z)) dz$ where $W_1 +\pi$ is shifted by $\pi$ for $z > 0$ and by $-\pi$ for $z < 0$. This gives a path equivalent to $W_2$ except that it is traversed in the opposite direction.

\section{The cable equation and impedance}
\label{SEC:Acab}

Waves do not travel only in open space, but also along cables. Thus, for high frequency currents, circuit characteristics can no longer be treated according to the usual point-to-point models. For instance, the lengths of the connections may affect the circuit properties. One central property of circuit elements, cables and loads, is that of \textit{impedance} $Z$. Generally speaking, if we feed a voltage $ V = V_0 \cdot \exp(- i \omega t)$ into a load and measure a resulting current $I = I_0 \cdot \exp(- i \omega t)$, then the impedance (at the given frequency) of the load is defined by the quotient

\[
Z = \frac{V}{I}.
\]

We will consider a discrete model of a cable made up of a series of inductors $L$ and capacitors $C$ (in parallel). 

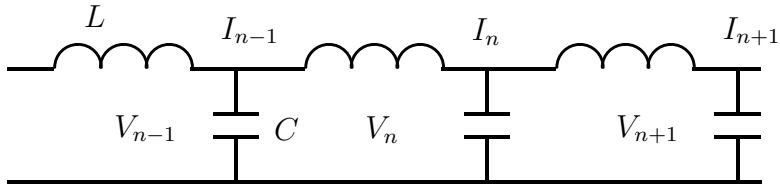
\begin{figure}[h]
\centering
\unitlength = 1mm
\begin{picture}(160,35)(0,0)
\Thicklines
\multiput(5,5)(33,0){3}{\LCfilt{0,0}}
\put(19,9){\makebox(6,6)[l]{$V_{n-1}$}}
\put(52,9){\makebox(6,6)[l]{$V_{n}$}}
\put(85,9){\makebox(6,6)[l]{$V_{n+1}$}}
\put(33,22){\makebox(6,6)[l]{$I_{n-1}$}}
\put(66,22){\makebox(6,6)[l]{$I_{n}$}}
\put(99,22){\makebox(6,6)[l]{$I_{n+1}$}}
\put(15,24){\makebox(6,6)[l]{$L$}}
\put(40,9){\makebox(6,6)[l]{$C$}}
\end{picture}
\caption{Discrete model of a transmission line as a series of inductors $L$ and capacitors $C$.}
\label{FIGcable}
\end{figure}

These elements are defined by their impedances $Z_L = -i \omega L$ and $Z_C = 1/i \omega C$; that is, the voltage over an inductor is given by $V = L \; \partial I/\partial t$ and over a capacitor by $V = Q/C$. If one applies the laws of Kirchoff and Ohm to the currents and voltages over the $n-1$, $n$, and $n+1$:th elements one obtains the equations

\begin{align}
\label{EQ:cab1}
 -C {\dot V}_n &= I_n - I_{n-1},
\\
-L {\dot I}_n &=  V_{n+1} - V_n. 
\end{align}

Letting the spacing $\Delta x$ between the units go to zero we may replace the difference by derivatives,

\begin{align}
\label{EQ:cab2}
 -\mathcal{C} {\dot V} &= \frac{\partial I}{\partial x},
\\
-\mathcal{L} {\dot I} &=  \frac{\partial V}{\partial x}, 
\end{align}

where $\mathcal{C} = C/{\Delta x}$ is the capacitance per unit length, and
$\mathcal{L} = L/{\Delta x}$ the inductance per unit length of the cable. This leads to the ''cable equation''

\begin{equation}
\label{EQ:cabwav}
\frac{\partial^2 V}{\partial x^2} - \frac{1}{u^2} \frac{\partial^2 V}{\partial t^2}  = 0 \quad \mbox{with} \quad u = {\frac{1}{\sqrt{\mathcal{C}\mathcal{L}}}}.
\end{equation}

This is a wave equation for EM-waves propagating along the cable with the velocity $u$. Inserting solutions of the form $V = V_0 \cdot \exp(i(kx - \omega t))$, $I = I_0 \cdot \exp(i(kx - \omega t))$ into (\ref{EQ:cab2}), where $k = \omega /u$, we obtain for the impedance,

\begin{equation}
\label{EQ:Z0}
Z = \frac{V}{I} = \frac{k}{\mathcal{C} \omega} =  \sqrt{\frac{\mathcal{L}}{\mathcal{C}}}.
\end{equation}

Thus, the quantity $Z_0 = \sqrt{\frac{\mathcal{L}}{\mathcal{C}}}$ gives the cable impedance for this model. One can also add resistive elements to the model, but the general features are already apparent in this simple model. Standard coaxial cables have impedances of 50 $\Omega$ or 75 $\Omega$, but one may note that these values apply only in some restricted frequency range.

In the above case we considered an infinite long cable. Suppose instead we have a cable with impedance $Z_0$ of length $l$ which is ended by a load $Z_L$. What will then be the impedance of this system? Assume the feeding point is at $x$ = 0, and the load is at $x = l$. If we feed a voltage $V_0 \cdot \exp(-i \omega t)$ at the point $x$ = 0, then the solution for the system is of the form

\begin{align}
\label{EQ:cab3}
&V(x,t) = a e^{i(kx - \omega t)} + b e^{i(-kx - \omega t)},
\\
\nonumber
&I(x,t) = c e^{i(kx - \omega t)} + d e^{i(-kx - \omega t)}.
\end{align}

The terms with $- i k x$ in the exponents correspond to the reflected part of the wave. The impedance will thus be given by the quotient

\[
Z = \frac{V(0,t)}{I(0,t)} = \frac{a+b}{c+d}.
\]
 
\begin{wrapfigure}{r}{50mm}
\begin{flushleft}
\unitlength = 1mm
\begin{picture}(50,40)(0,0)
\Thicklines
\multiput(5,10)(0,20){2}{\drawline(0,0)(30,0)}
\multiput(35,10)(0,15){2}{\drawline(0,0)(0,5)}
\multiput(33,15)(0,10){2}{\drawline(0,0)(4,0)}
\multiput(33,15)(4,0){2}{\drawline(0,0)(0,10)}
\put(4,10){\circle{2}}
\put(4,30){\circle{2}}
\put(38,17){\makebox(6,6)[l]{$Z_L$}}
\put(17,31){\makebox(6,6)[l]{$Z_0$}}
\put(15,5){\vector(-1,0){10}}
\put(25,5){\vector(1,0){10}}
\put(20,3){\makebox(6,6)[l]{$l$}}
\end{picture}
\end{flushleft}
\label{FIGimp}
\end{wrapfigure}

The coefficients $a, b, c, d$ are determined by the condition $V(l,t) = Z_L \cdot I(l,t)$, and using (\ref{EQ:cab2}) (which yields the relations $a = Z_0 c, b = - Z_0 d$). After some algebra one obtains finally

\begin{equation}
\label{EQ:ZZ0}
Z = Z_0 \frac{Z_L - i Z_0 \tan (kl)}{Z_0 - i Z_L \tan (kl)}.
\end{equation}

We have the interesting result that the impedance depends on the cable length $l$ and the wavelength $\lambda$ through the term $\tan (2 \pi l/\lambda)$. For instance, for a quarter wave cable, $l = \lambda/4$, we get $Z = Z_0^2/Z_L$. In the short-circuited case ($Z_L = 0$) we obtain

\[
Z = -i Z_0 \tan (kl). \qquad \mbox{(Short circuited case.)}
\]

In this case we have infinite impedance for quarter wave cables. This property is used in microwave chokers which are quarter wave cavities designed to trap unwanted leakage of EM-waves e.g.\ through the door slits microwave ovens.

For an open ended cable ($Z_L = \infty$) we have instead

\[
Z = i Z_0 \cot (kl), \qquad \mbox{(Open ended case.)}
\]  

and this corresponds to an infinite impedance for a half wave cable. We can calculate the rms power dissipated by the load $Z_L$ from (assuming $Z_0$ is real)

\begin{align*}
P = \frac{1}{2} \Re (I^\star (l) V(l)) = \frac{1}{2 Z_0} (|a|^2 - |b|^2) =
\\
\frac{V_0^2}{2Z_0} \frac{\left| 1 - \left|\frac{Z_L - Z_0}{Z_L + Z_0}\right|^2 \right|}{\left| 1 + \frac{Z_L - Z_0}{Z_L + Z_0}e^{i2kl}\right|^2}.
\end{align*}

The quotient $|b/a|$ describes the fraction that is reflected, and it is obtained from

\begin{equation}
\label{EQ:zba}
\frac{b}{a} = \frac{Z_L - Z_0}{Z_L + Z_0} \cdot e^{i2kl}.
\end{equation}

From this we infer that there is no reflexion when we have \textit{impedance matching}, $Z_L = Z_0$. In this case the energy transferred to the load is simply $V^2/2Z_0$. The term impedance matching is also used for point-to-point circuits where reflexion plays no part. If we have resistances $R_0$ and $R_L$ in series and apply the voltage $V$, then the power dissipated by the resistance $R_L$ becomes

\[
P_L = R_L \cdot I^2 = R_L \cdot \left(\frac{V}{R_0 + R_L}\right)^2, 
\]

which for fixed $V$ and $R_0$ attains its maximum when $R_L = R_0$. Thus, for this case, in order to deliver maximum of power to the load, its impedance must be matched to the circuit impedance $R_0$. If we repeat this exercise with the complex impedances $Z_0$ and $Z_L$ we obtain the matching condition $Z_L = Z_0^\star$. These results can be applied to the case where we have an RF-circuit with impedance $Z_0$ coupled to an antenna with the impedance $Z_L$. From (\ref{EQ:zba}) we get for the magnitude of the reflexion coefficient,

\begin{equation}
\label{EQ:antr} 
|\rho| = \left|\frac{Z_L - Z_0}{Z_L + Z_0} \right|.
\end{equation}

Two antenna design parameters are defined in terms of $r$: the Voltage Standing Wave Ratio,

\begin{equation}
\label{EQ:VSWR}
\text{VSWR} = \frac{1 - |\rho|}{1 + |\rho|},
\end{equation}

and the Return Loss,

\begin{equation}
\label{EQ:RL}
\mathcal{R} = 20 \, \log (|\rho|).
\end{equation}

\begin{wrapfigure}{r}{50mm}
\begin{flushleft}
\setlength{\unitlength}{1mm}
\begin{picture}(50,50)(0,0)
\Thicklines
\put(25,25){\circle{4}}
\put(25,25){\circle{40}}
\put(25,25){\circle{36}}
\put(25,27){\drawline(0,0)(0,7)}
\put(25,34){\vector(-1,0){10}}
\put(25,34){\vector(0,1){9}}
\put(27,35){\makebox(6,6)[l]{$E$}}
\put(17,27){\makebox(6,6)[l]{$H$}}
\end{picture}
\label{FIGCoax}
\end{flushleft}
\end{wrapfigure}

Designing the antenna conditioning one has to make a choice between minimizing the reflection ($Z_0 = Z_L$) and maximizing the radiation power ($Z_L = Z_0^\star$). One wants to get rid of the reflexions because they can interfere with RF-circuit operations. Above we considered a discrete model of a transmission line. The straight coaxial cable can be easily solved too because of the cylindrical symmetry. Thus assume the center wire has the radius $a$ and outer conductor an inner radius $b$. The transversal electric and magnetic mode (TEM) correspond to the case where the electrical field $\mathbf{E} = (E_r, 0, 0)$ is radial while the magnetic field is azimutal, $\mathbf{H} = (0, H_{\varphi}, 0)$. The Maxwell equations of interest in the region $a < r < b$ become,

\begin{align}
\label{EQ:Maxcyl}
\frac{\partial E_r}{\partial z} = i \omega B_{\varphi},\\
\frac{\partial H_{\varphi}}{\partial z} = i \omega B_{\varphi},\\
\frac{1}{r}\frac{\partial (r H_{\varphi})}{\partial r} = 0.
\end{align}

From this follows that

\[
\frac{\partial^2 E_r}{\partial z^2} + k^2 E_r = 0,
\]

where $k = \omega \sqrt{\epsilon \epsilon_0 \mu \mu_0}$. It follows from the third equation that $H_{\varphi} \propto 1/r$. More precisely we have $\oint \mathbf{H} \cdot d\mathbf{r} = I$ where the integral is along a circle of radius $a < r < b$ encircling the inner wire which carries a current $I$. Thus

\[
H_{\varphi}(r) = \frac{I}{2 \pi r},
\]

which in combination with the equation (\ref{EQ:Maxcyl}) gives

\[
E_r = \sqrt{\frac{\mu \mu_0}{\epsilon \epsilon_0}} \frac{I}{2 \pi r} =
\eta \frac{I}{2 \pi r}.
\]

The potential $V$ between between the inner and outer conductor becomes therefore

\[
\int_a^b E_r dr =  \frac{\eta I}{2 \pi} \ln\left(\frac{a}{b}\right).
\]

Finally the characteristic impedance $Z = V/I$ of the cable becomes 

\[
Z = \frac{\eta}{2 \pi} \ln\left(\frac{a}{b}\right).
\]

\section{Chipcon WCR2400}

The Chipcon device CC2420 \citep{Chipcon2005} operates in the 2.4 frequency band ($\lambda$ = 12.5 cm). IEEE 802.15.4 defines 16 channels in 5 MHz with the frequencies $f_k$ = 2405 + $5 \cdot (k - 11)$, with $k$ = 11, 12, $\dots$ , 26. The effective data rate is 250 kbps (2 MChips s$^{-1}$) and the device uses a coding with 4 bit symbols in 32 chip spread sequences. The RSSI values are determined from the average over 8 symbols corresponding to a time interval $\Delta t$ = 128 $\mu$s. The signal strength is determined from the automatic gain control factor in the signal amplifier part (variable gain amplifier, VAG). From the bit rate we gather that every bit has a 4 $\mu$s time window. Since the velocity of light $c$ is about 3 $\times$ 10$^8$ ms$^{-1}$ we get $c$ $\times$ 4 $\mu$s $\approx$ 1200 m for the distance that the radiation can travel during the time window which means that the reflected and refracted parts (from objects within 600 m which by far exceeds the normal range of the devices) can contribute to the bit-signal, either by constructive or destructive interference. The corresponding chip time window is 1 $\mu$s which translates into a 300 m distance. In our measurements we used a standard coaxial $\lambda/2$-dipole antennas model WCR-2400-SMA. The dipole antennas approximate well a circular radiation pattern in the horizontal plane \citep{WCR2400}.

\begin{figure}[h]
\centering
\includegraphics[width = 0.6\textwidth, angle = -90]{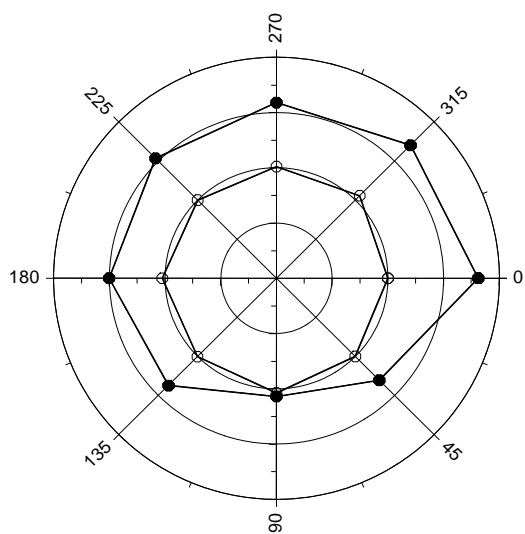}
\caption{Variation of RSSI (radius corresponds to absolute RSSI value) with angular orientation of the receiver with shielding (circle) and no shielding (filled circle) in outdoor setting. The receiver-transmitter separation was 10 m. Origin of the diagram corresponds to RSSI = -50, and the radial intervals to 10 RSSI units.}
\label{FIG3}
\end{figure}

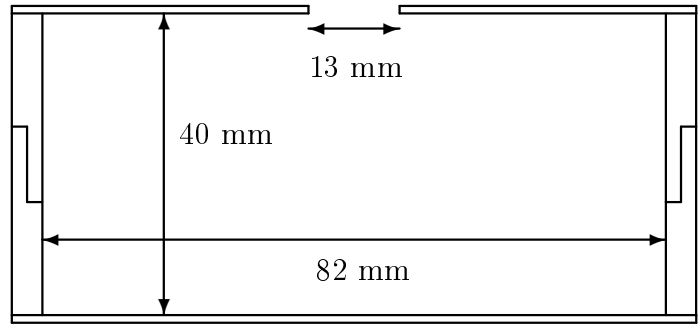
\begin{figure}[h]
\begin{center}
\setlength{\unitlength}{1mm}
\begin{picture}(90,50)(0,0)
\thicklines
\multiput(0,5)(86,0){2}{\drawline(0,0)(0,40)}
\multiput(4,5)(86,0){2}{\drawline(0,0)(0,40)}
\drawline(0,4)(0,5)(90,5)(90,4)(0,4)
\drawline(0,45)(0,46)(39,46)(39,45)(0,45)
\drawline(51,45)(51,46)(90,46)(90,45)(51,45)
\drawline(4,20)(2,20)(2,30)(0,30)
\drawline(86,20)(88,20)(88,30)(90,30)
\drawline(20,5)(20,45)
\put(20,5){\vector(0,-1){0}}
\put(20,45){\vector(0,1){0}}
\drawline(4,15)(86,15)
\put(4,15){\vector(-1,0){0}}
\put(86,15){\vector(1,0){0}}
\drawline(39,43)(51,43)
\put(39,43){\vector(-1,0){0}}
\put(51,43){\vector(1,0){0}}
\put(40,8){\makebox(6,6)[l]{82 mm}}
\put(22,26){\makebox(6,6)[l]{40 mm}}
\put(39,35){\makebox(6,6)[l]{13 mm}}
\end{picture}
\end{center}
\caption{Shielding box.}
\label{FIGcylind}
\end{figure}

However, the antenna + RF device results in orientation effects apparently because of the PCB affects the radiation pattern of the antenna in the horizontal plane. Shielding the RF device using an appropriate metallic enclosure (''Faraday cage'') reduces the orientation effect significantly as shown by Fig.\ref{FIG3}. The orientation effect is similar in the transmitting and receiving mode. One may also note that we get a better RSSI value in the shielded case. Cylindrical enclosures of diameter 90 mm and height 40 mm were custom made in 4 mm aluminum. The enclosure housed the PCB plus the battery, with the antenna fitted to the center of the top surface (top and bottom plates 1 mm thick).

The enclosure + antenna naturally will be expected to have a different input impedance than the antenna alone. However, there seems not be any readily available treatments of how such a cylindrical enclosure modifies the impedance. This problem as well as impedance measurements using a network analyzer will be left to a separate study.    

\section{Reciprocity theorem}

A basic property used in antenna measurement \cite{Kummer1978} is the so called \textit{reciprocity theorem} which says that an antenna has the same field pattern both in the receiving and transmitting mode. Therefore one needs to test it only as either receiver (usually the simpler alternative) or as a transmitter. More generally, suppose an antenna T$_1$ at $\mathbf{r}_1$ is transmitting radiation at the frequency $f$ and is intercepted at $\mathbf{r}_2$ by an antenna T$_2$, then field strength measured with T$_2$ is the same as would be measured T$_1$ if T$_2$ is transmitting with the same energy and at the same frequency $f$. If we imagine the antennas as part of of circuit \cite{Carter1932} with voltages and currents related by

\begin{align}
\label{EQ:antennaport}  
V_1 = Z_{11} I_1 + Z_{12} I_2 \\
\nonumber
V_2 = Z_{21} I_1 + Z_{22} I_2
\end{align}

then reciprocity boils down to the symmetry of the mutual impedance, $Z_{12} = Z_{21}$: A given current in one antenna causes the same (open circuit) voltage in the other antenna irrespective which of them acts as the transmitter or receiver.

\begin{figure}[H]
\centering
\setlength{\unitlength}{1mm}
\begin{picture}(100,35)(0,0)
\Thicklines
\put(15,5){\drawline(0,0)(0,20)(20,20)(20,0)(0,0)}
\put(10,10){\vector(1,0){0}}
\put(40,10){\vector(-1,0){0}}
\put(8,3){\makebox(6,6)[l]{$I_1$}}
\put(8,12){\makebox(6,6)[l]{$V_1$}}
\put(38,3){\makebox(6,6)[l]{$I_2$}}
\put(38,12){\makebox(6,6)[l]{$V_2$}}
\multiput(5,10)(0,10){2}{\drawline(0,0)(10,0)}
\multiput(35,10)(0,10){2}{\drawline(0,0)(10,0)}
\matrixput(4,10)(42,0){2}(0,10){2}{\circle{2}}

\put(55,5){\drawline(0,0)(40,0)}
\put(75,5){\drawline(0,0)(0,5)}
\put(74,10){\drawline(0,0)(0,10)(2,10)(2,0)(0,0)}
\put(75,20){\drawline(0,0)(0,5)}
\multiput(55,25)(20,0){2}{\drawline(0,0)(5,0)(5,1)(15,1)(15,-1)(5,-1)(5,0)}
\multiput(70,25)(20,0){2}{\drawline(0,0)(5,0)}
\matrixput(54,5)(42,0){2}(0,20){2}{\circle{2}}
\put(78,12){\makebox(6,6)[l]{$Z_{12}$}}
\put(57,27){\makebox(6,6)[l]{$Z_{11} - Z_{12}$}}
\put(77,27){\makebox(6,6)[l]{$Z_{22} - Z_{12}$}}
\end{picture}
\caption{Antenna system as a two-port network}
\label{FIGTwoPort}
\end{figure}
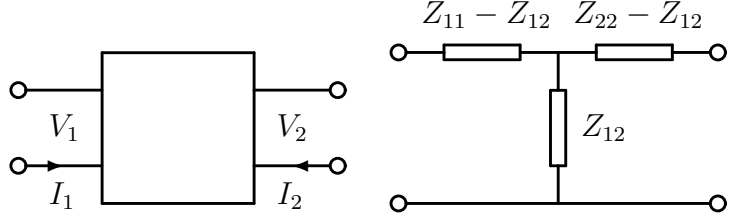

The electromagnetic reciprocity theorem goes back to the work by H A Lorentz in 1895-1896 and an account is given by Sommerfeld in \cite[p.653-663]{Frank1935} (which however rests on the assumption that the antenna length $L$ is insignificant compared to the wavelength $\lambda$). It is built on the fundamental equation 

\begin{equation}
\label{EQ:Lorentzrel}
\nabla \cdot \mathbf{E}_1 \times \mathbf{H}_2 - 
\nabla \cdot \mathbf{E}_2 \times \mathbf{H}_1 =
\mathbf{E}_2 \cdot \mathbf{J}_1 -
\mathbf{E}_1 \cdot \mathbf{J}_2
\end{equation} 

which is obtained from Maxwell's equations for $(\mathbf{E}_1,\mathbf{H}_1,\mathbf{J}_1)$ and $(\mathbf{E}_2,\mathbf{H}_2,\mathbf{J}_2)$ in the harmonic case (time derivatives of the fields are replaced by multiplication by $-i \omega$) assuming we have a linear and isotropic medium ($\mathbf{D}_k = \epsilon \epsilon_0 \mathbf{E}_k$; $\mathbf{B}_k = \mu \mu_0 \mathbf{H}_k$; $k$ = 1, 2), 

\begin{align}
\label{EQ:MaxH}
\nabla \times \mathbf{E}_k &= i \omega \mathbf{B}_k ,\\
\nonumber
\nabla \times \mathbf{H}_k &= -i \omega \mathbf{D}_k + \mathbf{J}_k.
\end{align}

We may take the index 1 to refer to the case when device 1 acts as sender, and the index 2 to refer to the case when device 2 acts as sender, everything else remaining similar. The fields have to satisfy the boundary conditions on the conductor surfaces which is why we have in general $Z_{12} \neq 0$.\footnote{If the indexes 1 and 2 refer to different instances, then $\mathbf{E}_k$ will be the total electric field and the boundary condition has to be applied to it. In this case $\mathbf{J}_k$ will differ from zero not just for the sending antenna but also for the receiving antenna because of the induced current. If the indexes refer to the same instance but where $\mathbf{J}_1$ is the current of antenna 1 alone (and vice versa for $\mathbf{J}_2$) then the boundary condition has to be applied to the total field $\mathbf{E}_1 + \mathbf{E}_2$.} If one integrates Eq.(\ref{EQ:Lorentzrel}) over a spherical volume containing the antennas and the devices we get for the left hand side, by using Stokes' theorem, the surface integral

\begin{equation}
\label{EQ:Lorentzrel2}
\int_{S_r} \mathbf{E}_1 \times \mathbf{H}_2 \cdot d\mathbf{S} -
\int_{S_r} \mathbf{E}_2 \times \mathbf{H}_1 \cdot d\mathbf{S}
\end{equation}   
 
which goes to zero as the radius $r$ of the sphere goes to infinity. The reason for this is that the field at a far away distance approaches locally a planar electromagnetic radiation for which $\mathbf{H} = \pm \mathbf{k} \times \mathbf{E}/(k \eta)$. Inserting this in Eq.(\ref{EQ:Lorentzrel2}) we see that the difference vanishes. Thus we get from Eq.(\ref{EQ:Lorentzrel2})

\begin{equation}
\label{EQ:Lorentzrel3}
\int \mathbf{E}_2 \cdot \mathbf{J}_1 dV =
\int \mathbf{E}_1 \cdot \mathbf{J}_2 dV.
\end{equation}

If the antenna 1 acts as a transmitter it emits a power $P_T = 1/2 \cdot \Re (V_1 I_1^\star) = 1/2 \cdot \Re (Z_{11} |I_1|^2)$ from which we deduce that $Z_{11} = R_{1s} + i X_1$ where $R_{1s}$ is the radiation resistance of the antenna while $X_1$ is called its reactance, and \textit{mutatis mutandis} for $Z_{22}$. If the receiving antenna 2 is furnished with a load $Z_L$ the current $I_2$ will be\footnote{Indeed, the antenna resistance has to be included in the total impedance because the antenna reradiates also in the receiving mode. The present analysis assumes that the distance between the antennas is large compared with the wavelength $\lambda$ and thus $Z_{12}$ small in comparison with $Z_{11}$ and $Z_{22}$; that is, the coupling between the antennas is weak -- see further \cite[sec. 4.2]{Bertoni2000}.} $- Z_{21} I_1/(Z_L + Z_{22})$ and the received power hence given by $P_R = 1/2 \cdot |Z_{21} I_1/(Z_L + Z_{22})|^2 \Re(Z_L)$. This attains the maximum for a matched load $Z_L = Z_{22}^\star$ in case which we obtain the ratio of received to transmitted power as,

\[
\frac{P_R}{P_T} = \frac{|Z_{12}|^2}{4 R_{1s} R_{2s}},
\]    
 
which is, as seen, symmetrical in the in the indexes 1 and 2 if $Z_{12} = Z_{21}$. 

\begin{figure}[H]
\centering
\includegraphics[width = 0.8\textwidth]{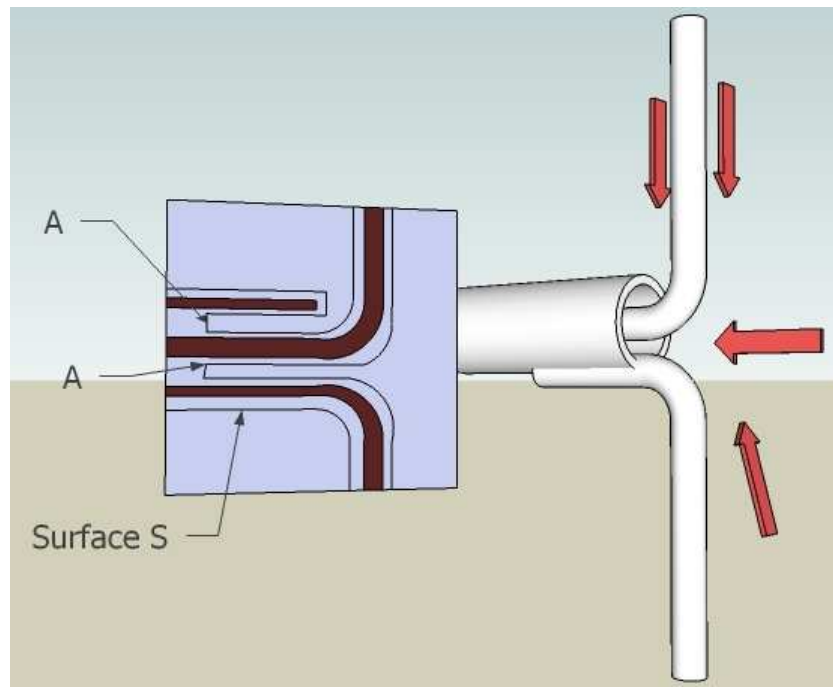}
\caption{Sketch of the dipole antenna link to the coaxial cable. On the right the arrows indicate the flow of field energy in the receiving mode. The dipole conductors steers the flow along the conductor surface into the coaxial cable. In the transmitting mode the energy flow is outwards from the coaxial cable and directed to the surrounding space by the antenna surface. On the left there is a schematic outline of the surface $S$ of integration. The surface $S$ follows the dipole antenna and the receiver/sender along the conducting surfaces where the tangential fields are zero, except where it forms the cross section $A$.}
\label{FIG6}
\end{figure}

We will consider the reciprocity relation in a bit more detail using as surfaces of integration the surface $S_1$ which follows the surface of the antenna 1+device 1 except where it crosses a section of the the coaxial cable (see the right part of Fig.(\ref{FIG6}) and \cite{Benumof1984}), and the surface $S_2$ defined mutatis mutandis for the second antenna + device. One thus integrates over a volume $V$ with the boundary $S_1 + S_2$ which excludes the interior of the antennas+devices and where we therefore have $\mathbf{J}_1 = \mathbf{J}_2 = 0$. One obtains then the reciprocity theorem in the form    
   
\begin{equation}
\label{EQ:Lorentzrel4}
\int_{S_1+S_2} \mathbf{E}_1 \times \mathbf{H}_2 \cdot d\mathbf{S} -
\int_{S_1+S_2} \mathbf{E}_2 \times \mathbf{H}_1 \cdot d\mathbf{S} = 0.
\end{equation}   

Here, for instance, the integral $\int_{S_2} \mathbf{E}_1 \times \mathbf{H}_2 \cdot d\mathbf{S}$ becomes $\int_{A_2} \mathbf{E}_1 \times \mathbf{H}_2 \cdot d\mathbf{S}$ where $A_2$ is a cross section of the coaxial cable interfacing the antenna 2, because elsewhere the electric field is orthogonal to the surface meaning that $\mathbf{E}_1 \times \mathbf{H}_2 \cdot d\mathbf{S} = - \mathbf{H}_2 \cdot \mathbf{E}_1 \times d\mathbf{S} = 0$ . In this integral $\mathbf{H}_2$ is the field in the coaxial cable generated by the impressed current $I_2$ while $\mathbf{E}_1$ refers to the field in the coaxial cable induced by the impressed current $I_1$ in the antenna 1 via the transmitted field impinging on antenna 2. In \cite{Benumof1984} the integration is only over $S_2$ and the volume $V$ thus contains the antenna 1 and we will therefore have a current term $\int_V \mathbf{E}_2 \cdot \mathbf{J}_1 dV$ on the right hand side of Eq.(\ref{EQ:Lorentzrel4}). In \cite[Eq.(15) and the preceding one]{Benumof1984} $I_1$ apparently refers to the current induced by the field $\mathbf{E}_1$ in antenna 2 (a point left somewhat unclear in \cite{Benumof1984}).

For a TEM wave in a coaxial cable we have the solution

\begin{align}
\label{EQ:coax}
&H_{\varphi}(r) = \pm \frac{ I}{2 \pi r},\\
\nonumber
&E_r(r) = \eta \frac{ I}{2 \pi r},
\end{align}

where $I$ is the current of the coaxial cable. The electric field is thus radial and orthogonal to the magnetic field which is along the angular direction. Inserting this solution into $\int_{A_2} \mathbf{E}_1 \times \mathbf{H}_2 \cdot d\mathbf{S}$ we obtain

\begin{align*}
&\int_{A_2} \mathbf{E}_1 \times \mathbf{H}_2 \cdot d\mathbf{S} =
\frac{\eta}{(2 \pi)^2} I_{1 \rightarrow 2} I_2 \int_a^b \frac{1}{r^2} 2 \pi r dr = \\
& \frac{\eta}{2 \pi} I_{1 \rightarrow 2} I_2 \ln\left(\frac{b}{a}\right) = Z_c I_{1 \rightarrow 2} I_2.
\end{align*}

Here $a$ is radius of the inner wire of the coaxial cable and $b$ the radius of the outer conductor, while $Z_c$ defines the impedance of the cable. We use $I_{1 \rightarrow 2}$ to denote the current induced in antenna 2 by the field radiated by antenna 1 with the impressed current $I_1$. The evaluation of $\int_{A_2} \mathbf{E}_2 \times \mathbf{H}_1 \cdot d\mathbf{S}$ will give the same result except being of the opposite sign (this has to do with the orientation of the fields and the fact that the impressed and induced fields go in opposite directions in the coaxial cable). This leads finally to the relation 

\[
2 Z_c I_{1 \rightarrow 2} I_2 = \int_V \mathbf{E}_2 \cdot \mathbf{J}_1 dV.
\]

If the receiver circuit is coupled to a matched load $R_L = Z_c$ then $2 Z_c I_{1 \rightarrow 2}$ will correspond to the induced voltage $V_2 = Z_{21} I_1$. From Eq.(\ref{EQ:Lorentzrel3}) and repeating the above considerations for antenna 1 we obtain $Z_{12} = Z_{21}$. The induced voltage in antenna 2 can, according to these results, be written as\footnote{When $\mathbf{E}_1$ refers to the total field then the result corresponds to closed circuit voltage, when $\mathbf{E}_1$ refers only to the incoming field then the result corresponds to the open circuit voltage.}

\begin{equation}
\label{EQ:Volt1}
V_2 = \frac{1}{I_2} \int_V \mathbf{E}_1 \cdot \mathbf{J}_2 dV.
\end{equation}

\section{Measurements of the reflexions from the ground}

The measurement procedures were designed such as to obtain knowledge how the RSSI depends on the distance the receiver and transmitter, and on the environment. The devices were typically attached to wooden supports (poles) of height 1 m. Power sources were 9V batteries. The transmitter was configured to send 20 packets per transmission. The transmissions were intercepted using a ''sniffer'' device (Chipcon Packet Sniffer) connected to a laptop computer. The data was saved on the computer for later extraction of the RSSI values, packet information and other data. For a majority of the tests the channel nr 20 was used and the power set to level 11 (0xA0B) corresponding to -11 dBm (see Tab.(\ref{TAB:0})). When calculating the average RSSI value the maximum and minimum values were dropped in order to eliminate possible outliers. Alternatively we used the median value. The RF-circuit was shielded by putting the whole PCB plus battery into a metallic can of 90 mm diameter and 40 mm height.

\begin{table}[htbp]
	\centering
		\begin{tabular}{l|c|c|}
		\hline	
		PA level & Output power (dBm) & Register\\
		\hline
		\hline
		31 & 0   & 0xA0FF \\
		\hline
		27 & -1  & 0xA0FB \\
		\hline
		23 & -3  & 0xA0F7 \\
		\hline
		19 & -5  & 0xA0F3 \\
		\hline
		15 & -7  & 0xA0EF \\
		\hline
		11 & -10 & 0xA0EB \\
		\hline
		7  & -15 & 0xA0E7 \\
		\hline
		3  & -25 & 0xA0E3 \\
		\hline
		\end{tabular}
	\caption{Power level assignments}
	\label{TAB:0}
\end{table}

\begin{figure}[H]
\centering
\includegraphics[width = 0.8\textwidth, angle = -90]{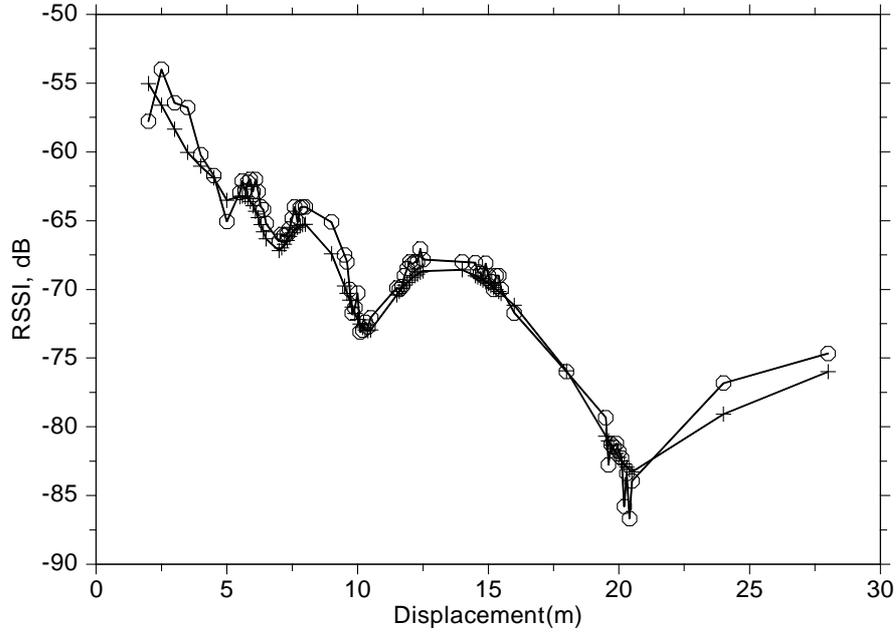}
\caption{RSSI and power dB vs distance, measured data (circle) and data computed using the two ray model (+) with $\epsilon$ = 3. The tops of the antennas were about 1.14 m above the ground (sand). At the time of measurement the temperature was -5$^\circ$C and the ground was covered with a ca 5 cm layer of fresh snow.}
\label{FIG1}
\end{figure}

In case of a smooth ground the two-ray (mirror) model can be fitted quite well to the data, see Fig.\ref{FIG1}. In the figure we compare measured data (RSSI values) with data (power in dB) computed using the two-ray model described above with the dielectric permittivity $\epsilon$ set to 3 providing a good fit. Measurement points were sampled more densely ($\Delta r$ = 10 cm) where we expected the changes to be largest. In this case the Brewster angle $\theta_B$ will be 30$^\circ$. The Brewster distance is about 3.9 m, and the breaking point is at $r \approx$ 42 m.

\begin{table}[htbp]
	\centering
		\begin{tabular}{l|c|c|c|}
		\hline	
		$h_1 (m)$ & Correlation & Error (stdev) & Offset (mean, $\mu$)\\
		\hline
		\hline
		0.64 & 0.99 & 0.95 & 49.4\\
		1.14 & 0.99 & 1.15 & 48.3\\
		1.64 & 0.98 & 1.30 & 49.3\\
		\hline
		\end{tabular}
	\caption{Statistics of model-data comparison}
	\label{TAB:1}
\end{table}

Similar measurements were made with $h_1$ = 0.64 m, 1.64 m, while $h_2$ remained 1.14 m, with equally good fits between model and data. Tab.(\ref{TAB:1}) shows the statistics of fitting the measured RSSI values to $10 \cdot \log(P) - \mu$, where $P$ is the power calculated from the model, and $\mu$ is the ''offset'' value between measured RSSI and model power (in dB). The difference (error) between the model and the data are seen to correspond to about 1 RSSI unit when measured in terms of the standard deviation. Note that the data points are not evenly distributed but were chosen such as to best cover the places where the changes in the RSSI were expected to be largest.

\begin{figure}[H]
\centering
\includegraphics[width = 0.8\textwidth, angle = -90]{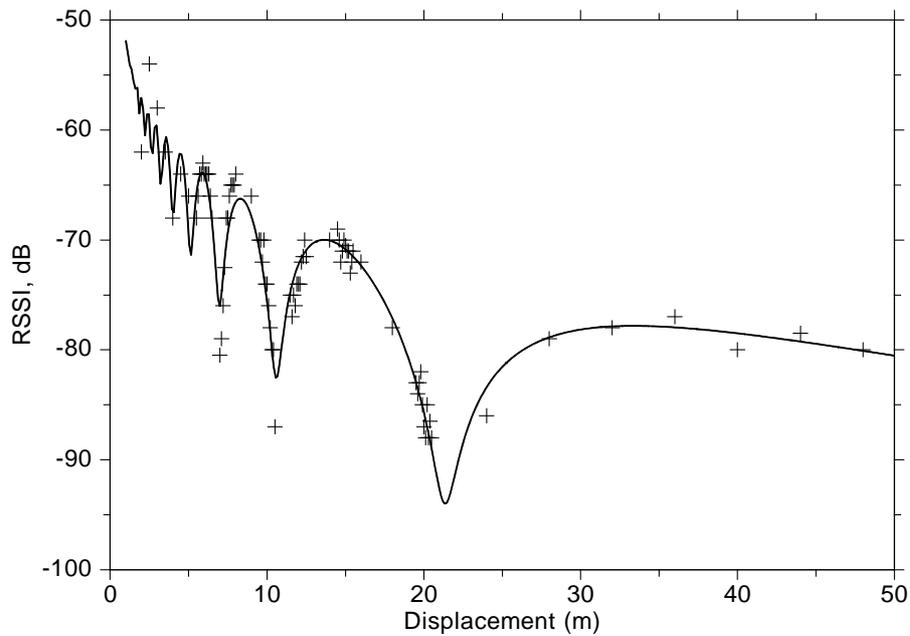}
\caption{RSSI and power dB vs distance measured on a frozen river. Measured data (+) and data computed using the two ray model (continuous) with two reflecting interfaces, with $\epsilon$ = 3 for ice (10 cm thick) and $\epsilon$ = 80 for water. The tops of the antennas were about 1.14 m and 1.16 m above the ice.}
\label{FIG4}
\end{figure}

Measurements were also made on a frozen river with a smooth ice of thickness around 10 cm. The results were modeled using the dielectric layer model treated in section \ref{SEC:reflex}, see Fig.(\ref{FIG4}). A characteristic difference when compared with Fig.(\ref{FIG1}) are the much deeper troughs of the interference pattern in case of sandwiched structure air-ice-water. In the model this feature is sensitive to the thickness of the ice.

\section{C-code for extracting RSSI values}

Below is an example of the C-implementation of the function \texttt{extract()} which has been used to extract the RSSI values from the psd-files obtained with the Packet Sniffer. The C-code will depend on what sort of information the devices are programmed to relay besides RSSI (e.g. battery power level, sensor data).

\lstset{language=C, basicstyle=\small}
\begin{lstlisting}


#include <stdio.h>
#include <stdlib.h>
#include <ansi_c.h>

/* Structure used */

struct sRSSIval {
	
	int n;       // size of RSSI-array 
	int *RSSI;   // array of RSSI-values 
		
};

typedef struct sRSSIval tRSSIval;


/* extract defines a function which opens
a binary psd-file and returns the RSSI values */ 

tRSSIval extract(char *fname)
{
	int i, j = 0;
	unsigned char temp;
	unsigned char dummy[121];
	int x=0, y=0;
	int temppu;
	tRSSIval data, err;

	err.n = 0;
	err.RSSI = 0;
			
	fp = fopen(fname, "r+b");
	data.n = 0;
	
	if (fp == NULL) {
		err.n = -1;
		return err;
	}

	fread(&temp, sizeof(char), 1, fp);
	x=temp;
	fread(&temp, sizeof(char), 1, fp);
	y=temp;
	y=y<<8;
	y=y+x;   //number of packets
	
	if(y < 1){
		err.n = -2;
		return err;	
	}
	
	data.RSSI = (int*) calloc(y, sizeof(int));
	
	for (i = 0; i < y; i++)
	{
		fread(dummy, sizeof(char), 26, fp);
		fread(&temp, sizeof(char), 1, fp);
		temppu=temp;
		if(temppu!=0) {
			if((temppu&0x40)>0) { 
				temppu=temppu&0x7f;
				temppu=temppu^0x7f;
				temppu=temppu+1;
				temppu=-45-(temppu);
				data.RSSI[j]=temppu;
				j++; 
				// increments true 
				// number of RSSI vals
			}
			else {
				temppu=temppu&0x7f;
				temppu=-45+(temppu);
				data.RSSI[j]=temppu;
				j++; 
				// increments true 
				// number of RSSI vals
				
			}
		}	
	
		fread(dummy, sizeof(char), 105, fp);
	}

	fclose(fp);
	data.n = j;
	return data;
} // extract end

\end{lstlisting}

\bibliography{radiofys12_07}
\bibliographystyle{plain}

\end{document}